\documentclass[11pt]{article}
\usepackage{fullpage,graphicx,psfrag,amsmath,amsfonts,verbatim}
\usepackage{xcolor}
\usepackage{amsthm}
\usepackage[small,bf]{caption}
\usepackage{authblk}

\usepackage{microtype}
\usepackage{graphicx}
\usepackage{booktabs} 

\usepackage{hyperref}
\usepackage{url}
\usepackage{caption}
\usepackage{subcaption}
\usepackage{amsthm}
\usepackage{amssymb}
\usepackage{stmaryrd}
\usepackage{multirow}
\usepackage{epigraph}
\usepackage[most]{tcolorbox}
\usepackage{enumitem}
\usepackage{algorithm}
\usepackage{algorithmic}
\usepackage{natbib}

\usepackage{xcolor}
\usepackage{listings}

\definecolor{mintbg}{gray}{0.95}
\definecolor{vsblue}{RGB}{0,0,255}
\definecolor{vscomment}{RGB}{0,128,0}
\definecolor{vsstring}{RGB}{163,21,21}

\usepackage[toc,header]{appendix}
\usepackage{minitoc}

\usepackage{amsmath}
\usepackage{mathtools}

\usepackage[capitalize,noabbrev]{cleveref}

\theoremstyle{plain}
\newtheorem{theorem}{Theorem}[section]
\newtheorem{proposition}[theorem]{Proposition}

\theoremstyle{definition}
\newtheorem{definition}[theorem]{Definition}

\theoremstyle{remark}

\usepackage[textsize=tiny]{todonotes}

\newenvironment{longlisting}{\captionsetup{type=listing}}{}

\hypersetup{
    colorlinks = true,
    allcolors = {gray},
    linkbordercolor = {white},
}

\allowdisplaybreaks

\title{Verbalized Bayesian Persuasion}

\author{Wenhao Li\textsuperscript{1}, 
Yue Lin\textsuperscript{2}, 
Xiangfeng Wang\textsuperscript{3}, 
Bo Jin\textsuperscript{1}, 
Hongyuan Zha\textsuperscript{2}, 
Baoxiang Wang\textsuperscript{2,4}}

\affil{\textsuperscript{1}Tongji University \textsuperscript{2}The Chinese University of Hong Kong, Shenzhen \\ \textsuperscript{3}East China Normal University \textsuperscript{4}Vector Institute \\ \texttt{\{whli,bjin\}@tongji.edu.cn, linyue3h1@gmail.com, \ xfwang@cs.ecnu.edu.cn,\{zhahy,bxiangwang\}@cuhk.edu.cn}}

\date{}

\begin{document}

\doparttoc 
\faketableofcontents 


\maketitle

\begin{abstract}
Information design (ID) explores how a sender influence the optimal behavior of receivers to achieve specific objectives.
While ID originates from everyday human communication, existing game-theoretic and machine learning methods often model information structures as numbers, which limits many applications to toy games.
This work leverages LLMs and proposes a verbalized framework in Bayesian persuasion (BP), which extends classic BP to real-world games involving human dialogues for the first time.
Specifically, we map the BP to a verbalized mediator-augmented extensive-form game, where LLMs instantiate the sender and receiver.
To efficiently solve the verbalized game, we propose a generalized equilibrium-finding algorithm combining LLM and game solver. 
The algorithm is reinforced with techniques including verbalized commitment assumptions, verbalized obedience constraints, and information obfuscation.
Numerical experiments in dialogue scenarios, such as recommendation letters, courtroom interactions, and law enforcement, validate that our framework can both reproduce theoretical results in classic BP and discover effective persuasion strategies in more complex natural language and multi-stage scenarios.
\end{abstract}


\maketitle

\setlength{\epigraphwidth}{.9\linewidth}
\epigraph{\textit{You can fool some of the people all of the time, and all of the people some of the time, but you can not fool all of the people all of the time.}}{Abraham Lincoln}

\section{Introduction}\label{sec:intro}

Persuasion plays a significant role in modern economies, with estimates suggesting that up to one-quarter~\citep{mccloskey1995one}, or even $30$\%~\citep{antioch2013persuasion} of GDP, is persuasion. 
The study of BP has deep roots in economics, with numerous applications across fields such as school grading~\citep{boleslavsky2015grading}, law enforcement deployment~\citep{lazear2006speeding}, research procurement~\citep{yoder2022designing}, matching platforms~\citep{romanyuk2019cream}, and routing systems~\citep{das2017reducing}. 
Various lines of theory have been proposed to explore the power of persuasion in different contexts~\citep{kamenica2019bayesian}.

This work investigates the Bayesian persuasion (BP) problem~\citep{kamenica2011bayesian,kamenica2019bayesian} between two agents: a sender and a receiver. 
Unlike cheap talk~\citep{lo2023cheap}, which is often employed in communication learning~\citep{foerster2016learning,sheng2022learning,zhu2022survey}, BP requires the sender to commit to an information disclosure mechanism publicly.
The focus, therefore, is on rational (Bayesian) decision-makers who understand and optimally react to the disclosed information. 
Given a specific utility function, BP is equivalent to finding an optimal Bayes-correlated equilibrium in an extensive-form game~\citep{bergemann2013robust,bergemann2019information}.

When the information space and the action space are both discrete and small, BP can often be solved analytically using optimization techniques~\citep{kolotilin2018optimal,dworczak2019simple,makris2023information,koessler2023informed}. This includes more complicated variants such as informed BP or multistage BP. 
Some research has also explored the use of multi-agent reinforcement learning to approximate solutions for more complicated problem structures~\citep{wu2022sequential,lin2023information,bacchiocchi2024markov}.

Despite these successes, most applications remain limited to games in the colloquial sense, where real-world complexity is often oversimplified. 
In fact, applying these methods to real-world settings requires constructing a model of the game in question, which involves defining the appropriate state space, action space, and transition dynamics.
For instance, a classic example in BP is the recommendation letter problem, where a professor must write a letter that conveys nuanced information about a student's background \citep{dughmi2017algorithmic}. 
In the example, the student's quality is reduced to a binary classification of either weak or strong, and the professor's decision is restricted to either recommending or not. 
This abstraction strips away much of the meaningful information inherent in the actual task. 
Consequently, it prevents BP from being generalized to the rich spectrum of real-world problems.

\begin{figure*}[t]
    \centering
    \includegraphics[width=\linewidth]{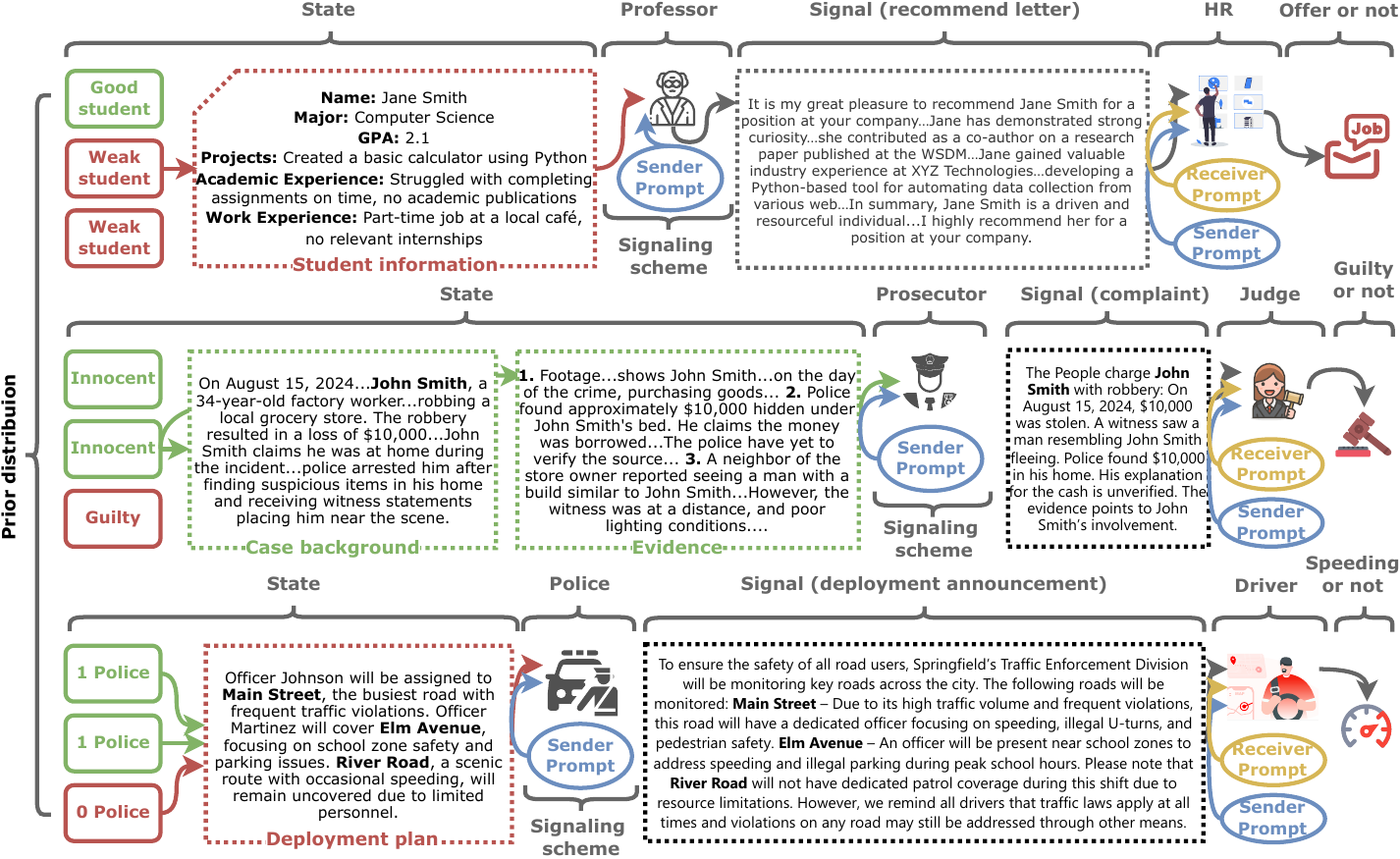}
    \caption{Extending classic BP examples to verbalized mediator-augmented, extensive-form games.}
    \label{fig:scenarios}
\end{figure*}

This work aims to leverage game-theoretic methods, enhanced by large language model (LLMs)~\citep{zhao2023survey}, to solve the original BP problem directly in the natural language domain. 
Specifically, we model the BP problem as a verbalized mediator-augmented, extensive-form game~\citep{zhang2022polynomial}, where states and actions (or signals for the sender) are all represented as text, as shown in Figure~\ref{fig:scenarios}.
For example, in the recommendation letter problem, the sender (professor) has a state that is a detailed, verbal description of the student's background, the signal is the content of the recommendation letter, while the reward remains numerical. 
The receiver (HR) observes this letter and must decide whether to accept the student. 
To enable the sender and receiver to process, understand, and generate this text, we parameterize both agents using LLMs.


Before introducing the proposed verbalized game solver, we need to address two fundamental challenges that have hindered the application of BP to real-world scenarios.
The first challenge lies in the theoretical foundation of signal space design. 
Unlike traditional BP problems where signals are discrete and finite, real-world persuasion involves natural language that carries nuanced information and contextual dependencies, creating an infinite-dimensional signal space where classic BP solution concepts break down~\citep{myerson1979incentive,kamenica2011bayesian}.
The second challenge concerns the fundamental difficulty in strategy optimization when using LLMs as agents. 
While LLMs provide a powerful way to process natural language, their parameter spaces are extremely high-dimensional and non-convex, making it theoretically impossible to guarantee the existence of Nash equilibria in such spaces~\citep{gemp2024states} and computationally intractable for direct optimization.

To address these challenges, we draw on the prompt-space response oracle (Prompt-PSRO)~\citep{gemp2024states}, which models strategy optimization for the sender and receiver as prompt optimization for their respective LLMs. 
This approach not only mitigates the challenge of optimization inefficiency but also reduces the action space from lengthy, complex text to compact, low-dimensional, discrete prompts. 
For instance, by adjusting the prompt given to the sender's LLM, we can control the ``level of details in the student's background description'' in the recommendation letter.

Building on the Prompt-PSRO, our solver is composed of several components that improve performance, efficiency, and stability.
These include verbalizing commitment assumptions, obedience constraints, and information obfuscation.
Technically, we extend Prompt-PSRO to multistage games by proposing conditional prompt optimization and providing a convergence guarantee to the equilibrium solution.
Together, these components form a comprehensive verbalized game solver tailored for BP, which we refer to as verbalized Bayesian persuasion (VBP).
To the best of our knowledge, VBP is the first general framework that attempts to solve real, non-abstract BP problems.

Our main contributions include:
1) Transforming real-world BP problems into verbalized mediator-augmented, extensive-form games, which provides a unified interface for game-theoretic solvers;
2) Proposing a general game-theoretic solver for verbalized BP problems based on the Prompt-PSRO framework, with a convergence guarantee to equilibrium solutions. Several components including verbalized commitment assumptions and obedience constraints, information obfuscation, and conditional prompt optimization enhance the solver's performance, efficiency, and stability;
3) Reproducing results in classic BP problems consistent with existing optimization and learning methods, while efficiently solving natural language and multistage BP problems. This potentially opens up a new line of studies of persuasion in real-world scenarios, which is helpful in understanding the interaction of multiple agents in both economic and societal applications.


\section{Preliminaries}\label{sec:pre}

This section will introduce research areas related to BP as well as tools and methods relevant to the techniques covered in the VBP framework. 

\subsection{Bayesian Persuasion}
The canonical BP model is structured as follows~\citep{kamenica2019bayesian}. 
A receiver, an agent, has a utility function $u_1(a, \omega)$, which depends on its action $a \in \mathcal{A}$ and the state of the world $\omega \in \Omega$. 
Another agent, the sender (also known as the information designer), has a utility function $u_0(a, \omega)$.
Both the sender and receiver share a common prior $\mu_0$ over $\Omega$.
The sender’s key decision is the choice of a signaling scheme, which is a mapping from the state to a distribution over signals, $\pi: \Omega \rightarrow \Delta(\mathcal{S})$. 
Here $\mathcal{S}$ represents a sufficiently large set of signals, which is typically enough with $|\mathcal{S}| \geq \min \{|\mathcal{A}|,|\Omega|\}$, known as the revelation principal.


%


Given knowledge of $\pi$ (i.e., under the commitment assumption~\citep{kamenica2011bayesian}), the receiver updates its belief from the prior $\mu_0$ to the posterior $\mu_\pi(\omega \mid s)$ using Bayes' rule. 
The receiver then selects an action $a^*$ that maximizes $\mathbb{E}_{\omega \sim \mu_\pi(\mid s)} u_1(a, \omega)$. 
Given this response mechanism from the receiver, the sender's objective is to solve the following maximization problem: $\max _{\pi \in \Pi} \mathbb{E}_{\omega \sim \mu_0} \mathbb{E}_{s \sim \pi(\omega)} u_0\left(a^*, \omega\right)$, where $\Pi$ denote the set of all possible signaling schemes. 
Here the revelation principle applies that an optimal signaling scheme exists that requires no more signals than there are actions available to the receiver. 
From the receiver's perspective, as long as it believes that the recommended actions are optimal according to its posterior belief, it will follow the sender's advice. 
These constraints on the sender's signaling scheme are referred to as obedience constraints~\citep{myerson1979incentive,kamenica2011bayesian}. 
Then, BP can be reduced to a simplified linear programming~\citep{lin2023information},
\begin{equation}\label{eq:bp-co}
\begin{aligned}
\max _{\pi} \mathbb{E}_{\pi}\left[u_0(a,w)\right], &\text { s.t. } \sum_w P(w) \cdot \pi(a \mid w) \cdot\left[u_1(a, w) -u_1\left(a', w\right)\right] \geq 0, \forall a, a'.
\end{aligned}
\end{equation}

\subsection{Learning Methods for BP}
The problem of BP can be approximately solved by multi-agent reinforcement learning (MARL).
In mixed-motive MARL, agents aim to advance their interests by shaping others~\citep{leibo2017multi,mckee2020social,dafoe2020open,leibo2021scalable}. 
Existing methods achieve this through either mechanism (modifying rewards)~\citep{yang2020learning,zheng2022ai,hua2023learning,wang2024carbon} or information design (modifying observations)~\citep{wu2022sequential,bernasconi2022sequential,lin2023information}, the latter of which could be used to solve BP. 
An sender can commit to a strategy for providing state information to the agents, effectively altering the observation function of the receiver. 
So the sender can influence agents' behavior by strategically providing information and guiding them toward desired outcomes~\citep{bergemann2019information}. 
In a long-term interaction, the sender and the receiver become aware of each other's strategy, which creates the game dynamics that resemble the commitment assumption and therefore the BP problem setting.
Then, the outcome of the MARL algorithm becomes an equilibrium of BP.

\begin{figure}[htb!]
    \centering
    \includegraphics[width=\linewidth]{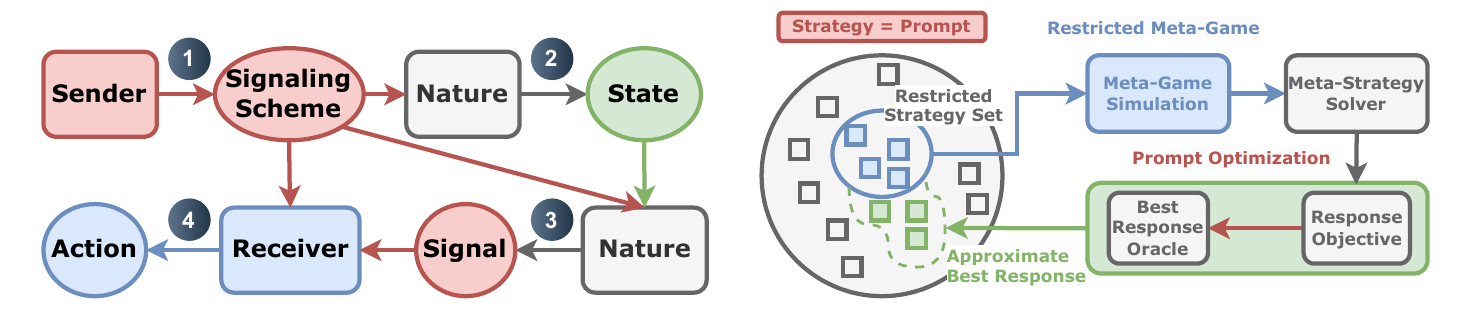}
    \caption{\textbf{Left:} BP timing in the EFG; \textbf{Right:} Illustration of the Prompt-PSRO.} 
    \vspace{-5pt}
    \label{fig:bp-prso}
\end{figure}

\subsection{Policy- and Prompt-Space Response Oracle}
While game theory offers a mathematical framework to study interactions between multiple agents~\citep{bighashdel2024policy}, classical analysis struggles with scalability due to the sheer number of strategies.
To address this, a wide range of learning methods have been applied to large-scale games, with MARL~\citep{yang2020overview,zhang2021multi} being one of the most prominent approaches. 
Unlike traditional methods, learning-based approaches do not require full representation of the game and instead create agents that explore and adapt by interacting with the environment. 
However, learning methods face inherent challenges in games, such as non-stationarity~\citep{tuyls2012multiagent} and non-transitivity~\citep{czarnecki2020real,sanjaya2022measuring}.

The PSRO framework~\citep{psro} emerged as a hybrid approach, combining traditional equilibrium computation with learning techniques. 
\textcolor{black}{It improves scalability by focusing on relevant subsets of strategies~\citep{wellman2006methods,bighashdel2024policy}.}
\citet{assos2023online} demonstrate that PSRO-like approaches lead to tractable notions of approximate local Nash equilibria. 
As illustrated in Figure~\ref{fig:bp-prso}, PSRO algorithms begin with an initial set of strategies for each agent and proceed through two alternating steps.
First, a normal-form meta-game (e.g., matrix game) is constructed, where each agent selects a meta-strategy to represent their overall behavior in the game. 
A meta-solver (e.g., Nash equilibrium solver) then computes a solution (e.g., Nash equilibrium) for this meta-game. 
In the second step, each agent computes an approximate best response to the meta-strategy, aiming to improve their reward assuming the other agents play according to the meta-strategy. 
This process repeats until no agent can benefit by deviating from their strategies~\citep{bighashdel2024policy}.

The prompt-space response oracle~\citep{gemp2024states} (Prompt-PSRO) is a verbalized adaptation of the standard PSRO framework, where strategies are parameterized by LLMs and represented as prompts. 
The approximate best response is generated by optimizing and sampling prompt strings, as opposed to the standard PSRO protocol where best responses are typically computed using MARL or gradient-based optimization. 

\section{Related Works}\label{sec:related}

The related fields primarily include the broader research area of deceptive behaviors in multi-agent learning, the persuasive or persuadable capabilities of LLMs themselves, and the LLMs in strategic interactions. 

\subsection{Game-Theoretic Solvers with LLMs}
The combination of a game-theoretic solver with prompt optimization, which we use in this work, is not the only paradigm for utilizing LLMs to solve games. 
Widely adopted parameter-efficient fine-tuning~\citep{xu2023parameter,han2024parameter}, agentic workflow~\citep{mao2023alympics,hua2024game,guo2024can,fan2024can,lore2024strategic,duan2024gtbench}, as well as the recent trend of improving reasoning and problem-solving capabilities for complex and mathematical problems by having LLMs generate longer chains of thought prior to making decisions~\citep{zelikman2022star,zelikman2024quiet,o1,guo2025deepseek}, are also very promising directions. 
The former allows for more fine-grained control of LLM outputs through in-weight updates, compared to in-context updates like prompt optimization, while the latter two may enable LLMs to discover novel game solvers.
VBP is orthogonal to these approaches.
Its primary goal is to leverage the rich foundation of game theory by incorporating various game-theoretic solvers that have already been proposed, and to extend the solid theoretical results established in classical games for solving verbalized games.





\subsection{Deception in Multi-Agent Learning}\label{sec:deception}


\citet{bond1988evolution} defines deception as false communication that benefits the communicator. 
In social learning, deception can be viewed as a means for the communicator to establish a cooperative equilibrium that is suboptimal for overall population welfare. 
Previous studies have explored deception within multi-agent reinforcement learning (MARL) settings~\citep{asgharnia2020deception, bontrager2019superstition, li2020effective, ghiya2020learning}, but these efforts typically focus on environments where agents have limited capacity to influence one another. 
More recent work~\citep{chelarescu2021deception} highlights the vulnerability of agents dependent on signals from others to guide their learning processes, pointing to the potential risks inherent in such scenarios. 
While much research focuses on the positive outcomes of mechanism design, it also reveals unforeseen risks, such as the emergence of deceptive behaviors~\citep{hughes2018inequity, jaques2019social, yang2020learning, lupu2020gifting, ndousse2021emergent}. 
Unlike these prior studies, which primarily examine how reward modifications influence deception through mechanisms like mechanism design, our work emphasizes the role of information manipulation in shaping deceptive behavior.

Game-theoretic models traditionally frame deception using signaling~\citep{ho1978teams}, where one player can send costly signals to convey false information. 
In network security, for instance, \citet{carroll2011game} examined how defenders can deceive attackers by masking honeypots as regular computers. 
Other research has studied the evolution of deceptive signaling in mixed environments. 
\citet{floreano2007evolutionary} demonstrated that, in competitive food-gathering tasks, teams of robots spontaneously developed deceptive strategies, misleading competitors to reduce resource competition.
An extension to classical game theory, known as hypergame theory~\citep{bennett1980hypergames}, accounts for players' uncertainty about others' strategies or preferences, leading to disagreements about the underlying game being played. 
By incorporating agents' differing perceptions, hypergame theory provides a natural framework to model misperception, false beliefs, and deception~\citep{kovach2015hypergame}. 
Applications of hypergame theory include \citet{vane2002using}, who analyzed deception in normal-form hypergames, and \citet{gharesifard2013stealthy}, who modeled deception based on player preferences when the deceiver has full knowledge of the target. 
Additionally, \citet{ettinger2010theory}, \citet{strouse2018learning}, and \citet{aitchison2021learning} show how agents can manage information about their roles to achieve deception by regularizing mutual information between goals and states.
In contrast to these works, which model deception as discrete, explicit signaling actions, our study explores how deception can be realized through natural language interaction.


Finally, \citet{macnally2018action} addresses the broader question of how agents can communicate intent without explicit signaling, using an online planner to select actions that implicitly reveal intent to an observer. 
\citet{masters2017deceptive} extended this approach to deception by maximizing the divergence between the agent's and observer's beliefs. 
However, these methods assume full observability and rely on environmental models for forward planning, whereas our work focuses on achieving deception through natural language in more complex, partially observable environments.

\subsection{Conversational Persuasiveness of LLMs}\label{sec:persuasion}

Recent advancements in LLMs have shown their impressive potential in the realm of persuasion. 
A growing body of research highlights how these models can enhance human communicative abilities and even autonomously generate persuasive content across various contexts.

For instance, \citet{shin2024large} demonstrated that refining complaint narratives with ChatGPT significantly improved consumers' chances of obtaining redress from financial institutions, showcasing the role of LLMs in boosting human persuasive efforts. 
Similarly, \citet{carrasco2024large} showed that LLMs outperform humans in utilizing cognitive load and moral or emotional language when crafting persuasive messages, prompting the need for ethical guidelines governing their use. 
\citet{breum2024persuasive} further explored LLMs' capacity to simulate persuasive dynamics, revealing that LLMs can influence opinion changes in other LLMs with predefined personas. 
Building on this, \citet{ramani2024persuasion} introduced a multi-agent framework in which a primary agent engages users through persuasive dialogue, while auxiliary agents handle tasks such as information retrieval, response analysis, and strategy development.
These studies illustrate that LLMs are not only capable of enhancing human persuasion but also of autonomously refining and executing persuasive strategies.

The impact of LLM-generated persuasive text on human behavior has been demonstrated across a diverse range of domains. 
For example, \citet{bai_voelkel_eichstaedt_willer_2023} showed that GPT-3.5 could influence political attitudes, while \citet{karinshak2023working} found that GPT-3’s vaccine campaign messages were more effective than those created by professionals. 
Additionally, LLM-powered romantic chatbots have been shown to sustain human engagement longer than human-to-human conversations~\citep{zhou2020design}. 
In strategic contexts, LLMs have achieved human-level negotiation capabilities in games like Diplomacy~\citep{meta2022human}, and algorithmic suggestions have been shown to shape emotional language in messaging~\citep{hohenstein2023artificial}. 
These examples collectively highlight the broad applicability of LLMs in persuasive tasks and their significant influence on human decision-making.

However, the increasing persuasive power of LLMs also raises concerns about potential misuse. 
\citet{salvi2024conversational} found that LLMs outperform humans in personalized debates, achieving a higher rate of belief change in one-on-one discussions. 
This raises ethical concerns, particularly regarding the risks of misinformation and manipulation. 
For instance, \citet{majovsky2023artificial} demonstrated that LLMs can convincingly fabricate medical facts, further complicating the ethical landscape. 
The ability of LLMs to produce persuasive yet misleading content underscores the need for stronger oversight, especially in high-stakes domains such as healthcare, politics, and public discourse.
Recent studies have thus emphasized the necessity of ethical frameworks as LLMs become more adept at persuasion. 
While LLMs have shown persuasive power across various tasks and domains~\citep{matz2024potential,durmus2024persuasion,burtell2023artificial,shin2023enhancing}, they also pose risks, particularly for vulnerable populations. 
\citet{bar2023algorithmic} highlighted that characteristics such as race, gender, and sexual identity may subject certain groups to greater risks of algorithmic persuasion and bias, potentially exacerbating existing social inequalities. 

From a computational standpoint, \citet{wojtowicz2024and} provided a novel proof showing that discovering persuasive messages is NP-hard, while adopting persuasive strategies provided by others is NP-easy. 
This insight adds to our understanding of the complexity involved in generating persuasive content and demonstrates why LLMs, with their vast data-processing capabilities, are particularly adept at these tasks.
Building on these insights, our work explores how game-theoretic methods can be leveraged to enhance the persuasive capabilities of LLMs in purely multi-agent LLM systems. 
Unlike previous studies that primarily measure the impact of LLM-generated persuasive text on humans, we investigate how multiple LLMs can engage in persuasive interactions with one another, optimizing their strategies using game-theoretic approaches.

\subsection{LLMs in Strategic Interactions}

Recent advances in large language models (LLMs) have showcased their potential in reasoning and planning, particularly in strategic interactions. 
LLMs have demonstrated strong capabilities in in-context learning, allowing them to reason about possible outcomes~\citep{kojima2022large} and plan their actions to achieve strategic objectives~\citep{liu2023llm+}. 
However, their performance in game environments can vary significantly depending on the type of game, as shown by \citet{lore2023strategic}, where LLMs struggled in different ways across various games. 
To address these challenges, \citet{gandhi2023strategic} introduced an automated ``prompt compiler'' that facilitates strategic reasoning by constructing demonstrations, enabling LLMs to solve games through in-context learning. 
Similarly, \citet{meta2022human} designed an action space of ``intents'' to control a generative language model, also leveraging in-context learning, which aligns closely with the approach taken in our work here. 
Additionally, game-theoretic models have been employed to improve the factual accuracy of LLMs~\citep{jacob2024the} and enhance their security~\citep{ma2023red}. 
For a broader overview of LLMs in strategic reasoning, \citet{zhang2024llm} provides a comprehensive survey.

The BP problem, however, goes beyond mere reasoning or planning. 
It requires the ability to anticipate and account for the intentions, beliefs, and goals of other participants-a hallmark of game-theoretic settings. 
While some initial studies have begun to explore how LLMs perform in game environments, most of this work focuses on leveraging in-context learning. 
For example, research has examined LLMs' behavior in matrix games~\citep{xu2023magic,fan2024can}, repeated games~\citep{akata2023playing,zhang2024k,huang2024far,silva2024large}, economic mechanisms like auctions~\citep{chen2023put,mao2023alympics}, and collective decision-making scenarios~\citep{jarrett2023language}. 
These studies collectively illustrate the potential of LLMs to navigate complex environments that require both strategic thinking and interaction with other agents.

In contrast to prior work that primarily evaluates LLMs' reasoning or game-playing capabilities through in-context learning or agentic workflows, our approach focuses specifically on solving the BP problem. 
Our key contribution lies in providing a general interface that integrates LLMs with game-theoretic solvers to address BP problems effectively. 
Based on this interface, we propose a solution framework called VBP, which combines prompt optimization with game-theoretic methods. 
This framework offers a convergence guarantee to equilibrium solutions, ensuring robust performance.

\paragraph{Remark 1}  

While both our work and~\citet{bai2024efficient} leverage BP, they address fundamentally different problem spaces. 
\citet{bai2024efficient} apply classic BP as a tool for model alignment, optimizing signaling strategies between a smaller ``Advisor'' model and a larger ``Receiver'' model to improve downstream task performance in areas like mathematical reasoning and code generation. 
In contrast, our work extends BP into natural language settings by introducing a verbalized BP framework, enabling strategic communication through real-world dialogue. 
This involves novel methods such as transforming agents' policy optimization into prompt optimization and developing equilibrium-finding algorithms in the language space. 
These differences highlight the complementary nature of the two approaches: \citet{bai2024efficient} focus on BP-driven alignment for structured tasks, while our contributions advance BP for complex, dialogue-based applications.

\section{The Mediator-Augmented Game Formulation for BP}\label{sec:formulation}


To establish convergence for the VBP framework, we transform the classic BP problem into a special class of extensive-form games (EFGs), known as mediator-augmented games (MAGs, \citet{zhang2022polynomial}). 
In this section, we reformulate the BP problem in the form of an MAG. 
At a high level, a mediator-augmented game introduces an additional player, the mediator, who exchanges messages with the players and provides action recommendations.

\begin{definition}
A Bayesian persuasion problem, represented as a mediator-augmented game $\Gamma$, consists of the following components~\citep{zhang2022polynomial}:
\textbf{(1)} a player, referred to as the receiver, denoted by the integer $1$;
\textbf{(2)} a directed tree $H$ of histories or nodes, with the root denoted by $\varnothing$. The edges of $H$ are labeled with actions, and the set of legal actions at each node $h$ is denoted by $A_h$. Terminal nodes of $H$ are called leaves, and the set of such leaves is denoted by $Z$;
\textbf{(3)} a partition of non-terminal nodes $H \backslash Z$ into $H_{\mathbf{C}} \sqcup H_0 \sqcup H_1$, where $H_1$ represents the nodes where player $1$ acts, and $\mathbf{C}$ and $0$ represent chance and the mediator (i.e., the sender), respectively;
\textbf{(4)} for each agent $i \in \{1, 0\}$, a partition $\mathcal{I}_i$ of the decision nodes $H_i$ into information sets. Every node in a given information set $I$ must have the same set of legal actions, denoted by $A_I$;
\textbf{(5)} for each agent $i \in \{1, 0\}$, a utility\footnote{In this paper, we do not distinguish between utility and reward.} function $u_i: Z \rightarrow \mathbb{R}$; and
\textbf{(6)} for each chance node $h \in H_{\mathbf{C}}$, a fixed probability distribution $c(\cdot \mid h)$ over $A_h$.
\end{definition}

At any node $h \in H$, the sequence $\sigma_i(h)$ for agent $i$ consists of all information sets (infosets) encountered by $i$, along with the actions taken at those infosets on the path from $\varnothing$ to $h$, excluding $h$ itself. 
An agent has perfect recall if $\sigma_i(h) = \sigma_i(h')$ for all $h, h'$ within the same infoset. 
A pure strategy for agent $i$ specifies one action from $A_I$ for each information set $I \in \mathcal{I}_i$. 
A mixed strategy is a probability distribution over pure strategies, and the sequence form of a mixed strategy corresponds to the convex combination of pure strategies. 
Let $X_1$ and $X_0$ denote the polytope of sequence-form mixed strategies $\boldsymbol{x}_1$ for the receiver and $\pi$ for the mediator, respectively.

For a fixed $\pi \in X_0$, we say that $(\pi, \boldsymbol{x}_1)$ is an equilibrium of $\Gamma$ if $\boldsymbol{x}_1$ is a best response to $\pi$, meaning $\max_{\boldsymbol{x}_1' \in X_1} u_1(\pi, \boldsymbol{x}_1') \leq u_1(\pi, \boldsymbol{x}_1)$. 
We do not require the mediator’s strategy (signaling scheme) $\pi$ to be a best response; 
hence, the mediator can commit to its strategy. 
The objective of this paper is to find an optimal Stackelberg equilibrium, which is an equilibrium $(\pi, \boldsymbol{x}_1)$ that maximizes the mediator's utility $u_0(\pi, \boldsymbol{x}_1)$.
When viewed as an extensive-form game (EFG), the event sequence in BP is shown in Figure~\ref{fig:bp-prso}.

\section{VBP Solver}\label{sec:methods}

\begin{figure*}[t]
    \centering
    \includegraphics[width=\linewidth]{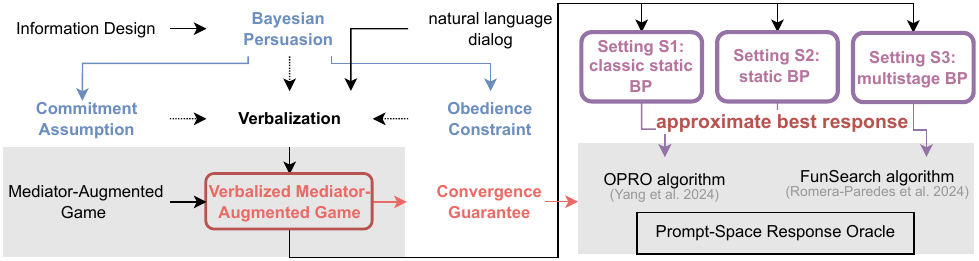}
    \caption{Verbalize Bayesian persuasion framework.}
    \label{fig:vbp}
\end{figure*}

This section will provide a detailed introduction to the VBP framework, as shown in Figure~\ref{fig:vbp}.
The first part presents the verbalization of MAG, including a polarized setting (e.g. strong/weak students) that reproduces the classic examples in BP, and general settings for one-step and multi-stage BP in the language space.
The second part introduces how this MAG is solved within the Prompt-PSRO framework, with the help of available best response approximations in language models \citep{yang2024large,romera2024mathematical}.
The verbalized MAG, along with the three problem settings and the Prompt-PSRO-based game-theoretic solver, collectively constitute the VBP framework.

\subsection{Verbalized Formulation for BP}\label{sec:problem-formulation}

To leverage the wealth of research in LLMs for BP in realistic scenarios, we must abstract and map components of BP to the symbolic language. 
Note the mapping can be chosen is not unique.

\begin{itemize}[leftmargin=*]
    \item State $\omega$. Unlike the classic BP, which only describes the state with binary values, the state in VBP is defined as the text. For example, it is the detailed description to the student's quality in the recommendation letter problem.
    \item Infosets $\mathcal{I}$. The infoset is available only in the multi-stage setting. It is defined as the interaction history between the two parties.
    Specifically, this includes the signals sent by the sender, the receiver's decisions (public information), and their respective rewards (private information) from each round of interaction, plus the previous environmental states (private information).
    \item Action $A$. Agents' actions refer to the signaling scheme for the sender and the action for the receiver. As we transform the strategy optimization problem into a prompt optimization problem through the Prompt-PSRO framework, the actions involve selecting prompts. 
    \item Terminal states $\mathcal{Z}$. In static settings the game terminates in one step. In multi-stage settings terminal states are determined by either a limit \textcolor{black}{or} the allowable tree depth. 
\end{itemize}

In addition to the basic components of the game, the BP problem also includes two fundamental constraints that need to be mapped into the verbalized MAG. 

\paragraph{Verbalized Committment Assumption} 
The key difference between BP and cheap talk~\citep{crawford1982strategic} lies in the presence of the commitment assumption, where the sender commits their signaling scheme as common knowledge. 
The VBP framework achieves the commitment assumption through prompting in the static setting and through expanding the receiver's \textcolor{black}{infoset} in the multi-stage setting.
Specifically, the signaling scheme is equivalent to the key components in the prompt provided to the sender that influence the generated signals, and these components are the target of Prompt-PSRO optimization. 
VBP incorporates these components into the receiver's prompt, where more details are deferred to Appendix~\ref{sec:prompts}.

\paragraph{Verbalized Obedience Constraint}
\textcolor{black}{The optimization in BP involves an (extended) obedience constraint~\citep{myerson1979incentive,kamenica2011bayesian,lin2023information}}, as shown in Equation~\ref{eq:bp-co}.
It is intuitive to handle this constraint by transforming it into a penalty term, which is similar to reward shaping~\citep{ng1999policy,gupta2022unpacking}. 
However, computing this penalty term requires integrating over the entire state and action space.
To address this, we estimate the summation term using a sampling approach. 
Specifically, we calculate an estimate using the current state and an arbitrarily selected action. 
There are various ways to select actions, and here we introduce a theory-of-mind approach~\citep{rabinowitz2018machine,albrecht2018autonomous}, where actions are selected based on predictions of what the receiver would do, using a prompt to \textcolor{black}{pretrained and aligned} LLMs to anticipate the receiver's likely actions.



\subsection{Three Settings in VBP}\label{sec:settings}

\paragraph{Setting S1: Polarized signals}
Polarized signals refer to the constraint to produce more straightforward signals, such as recommend v.s. not recommend in the recommendation letter example. The goal of this setting is to aligning the signal space with the classic BP formulation. We aim to reproduce the Stackelberg equilibrium in classic BP examples and validate the effectiveness of VBP.
Specifically, we use the \textcolor{black}{pretrained and aligned} LLM to score the signals output by the sender, For example, this determines the degree, a real value between $0$ and $1$, to which the recommendation letter supports the student. Similar prompts can be designed for other problems. 
Then this value is pushed to an extreme point in the signal space, based on the minimum distance.

\paragraph{Setting S2: VBP in Static BP}
By removing signal polarization, the signals are maintained in the language space. This constitutes the default setting of VBP.



\paragraph{Setting S3: VBP in Multistage BP}
This setting considers a multistage scenario, which is very challenging for traditional methods~\citep{gan2022bayesian,wu2022sequential}. 
The agents engage in multiple rounds of interaction, and the sender's historical signals serve as the basis for the receiver's subsequent decisions. 
This largely increases the complexity, as the sender cannot arbitrarily exploit the receiver with their information advantage. 
Instead, they must consider how their current actions may impact future rewards.

\subsection{Verbalized Game Solver}

After modeling the BP as a verbalized MAG, we parameterize both agents using \textcolor{black}{pretrained and aligned} LLMs and optimize their strategies with the Prompt-PSRO, thereby forming a general BP solver.
We first present the following proposition based on the theoretical foundation of~\citet{zhang2022polynomial}, with the proof provided in Appendix~\ref{sec:proof}.

\begin{proposition}\label{prop:main}
VBP returns an $\varepsilon$-approximate Bayes correlated equilibrium in static BP and an $\varepsilon$-approximate Bayes-Nash equilibrium in multistage BP.
\end{proposition}

In simple terms, the reason we can leverage the theoretical results of MAG is because different assumptions on the power of the mediator and the players’ strategy sets induce different equilibrium concepts.
The concept of Bayes correlated equilibrium~\citep{bergemann2016bayes} in static BP and Bayes-Nash equilibrium~\citep{makris2023information} in multistage BP is equivalent to the situation in the MAG where the mediator has an informational advantage, cannot lie (commitment assumption), and gains perfect recall under the extensive-form correlated equilibrium.

VBP does not directly solve the verbalized MAG using the Prompt-PSRO. 
Instead, we make targeted improvements to Prompt-PSRO for different settings, as illustrated in the Appendix~\ref{sec:approximator}. 
For the S1 and S2 settings, we optimize the strategies of the sender and receiver using Algorithm 4 from the Prompt-PSRO framework~\citep{gemp2024states}, specifically the ``categorical'' approximate best response. 
Unlike in the original PSRO paper, we use the OPRO method~\citep{yang2024large} to generate both the categories and the specific content within the categories simultaneously.

The S3 setting presents a challenge for the Prompt-PSRO. 
Existing Prompt-PSRO is unconditional or episode-wise, meaning that the prompt generated at the beginning of each episode is used for every subsequent timestep. 
In mutlistage BP, this significantly restricts the optimizable strategy space. 
In other words, both the sender and receiver can dynamically adjust their strategies based on the interaction history to achieve higher rewards. 
For example, the sender might honestly provide true information to the receiver early to build trust, then deceive the receiver later. 
Similarly, the receiver could bargain to extract more information.
Thus, we propose a conditional version of Prompt-PSRO, denoted as step-wise Prompt-PSRO, building on the original framework. 
Specifically, we introduce FunSearch~\citep{romera2024mathematical} to VBP, where the strategy to be optimized is no longer the prompt itself, but a function that generates the prompt. 
This function takes the current interaction history as input, thereby enabling conditional prompts.
The pseudocode is shown in Algorithm~\ref{alg:vbp}.


Moreover, since we use aligned LLMs, the sender struggles to output strategic signals, such as hiding or obfuscating relevant information about the true state, which leads to lower training efficiency. 
To speed up training, we introduce a information obfuscation mechanism. 
Similar to reward shaping (though experiments showed suboptimal results, likely due to the complexity of optimizing the reward function with too many components), a \textcolor{black}{pretrained and aligned} LLM is deployed to evaluate the degree of information hiding or obfuscation in the output signal. 
This feedback is then employed to perform multiple rounds of self-reflection~\citep{huang2023large,shinn2024reflexion,tao2024survey} before entering the Prompt-PSRO loop.



\begin{algorithm}[t]
\caption{Verbalized Bayesian Persuasion}
\begin{algorithmic}[1]\label{alg:vbp}
\REQUIRE $C$, where $C_i$ is the initial prompt action set (i.e., one category and one corresponding content) for player $i$ (either the sender or receiver)
\REQUIRE $h$, containing hyperparameters for the approximate best response operator $\text{BR}$ (e.g., \textbf{\textcolor{purple}{LLM-based OPRO or FunSearch}})
\STATE \textbf{\textcolor{purple}{Initialize with LLM-based sampling:}} Compute the expected payoff tensor $P$ over all joint actions in $C$ using Equation~(\ref{eq:bp-co-bilinear})
\STATE \textbf{Set:} $\pi \gets$ uniform meta-strategy profile over $C$ 
\COMMENT{\textcolor{gray}{Each joint action in $C$ initially has equal probability}}
\STATE \textbf{Set:} incomplete $\gets$ \textbf{TRUE} 
\COMMENT{\textcolor{gray}{Flag to indicate if the equilibrium search is complete}}
\WHILE{incomplete}
    \FOR{player $i \in [N]$} 
        \STATE \textbf{\textcolor{purple}{LLM input:}} Provide current meta-strategy $\pi$ and action sets $C$ of sender (for receiver)
        \STATE \textbf{\textcolor{purple}{Use LLMs to compute best response:}} $c_i \gets \text{BR}(i, \pi, h)$ \COMMENT{\textcolor{gray}{The LLM generates the optimal prompt or strategy for player $i$}}
        \STATE \textbf{\textcolor{purple}{LLM output:}} Candidate best response $c_i$ for player $i$
    \ENDFOR
    \IF{$c_i \in C_i$ $\forall i \in [N]$} 
        \STATE incomplete $\gets$ \textbf{FALSE} 
        \COMMENT{\textcolor{gray}{Terminate the loop if no new strategies are found}}
    \ELSE
        \STATE $C_i \gets C_i \cup c_i$, $\forall i \in [N]$ 
        \COMMENT{\textcolor{gray}{Add the newly found best response strategies to the action sets}}
        \STATE \textbf{\textcolor{purple}{Recompute with LLM-based sampling:}} Compute the expected payoff tensor $P$ over the updated joint actions in $C$ using Equation~(\ref{eq:bp-co-bilinear})
        \STATE \textbf{Update:} $\pi \gets$ meta-strategy w.r.t. $P$ 
        \COMMENT{\textcolor{gray}{Recalculate the strategy probabilities based on the updated payoff tensor}}
    \ENDIF
\ENDWHILE
\STATE \textbf{Return:} $(\pi, C, P)$ 
\COMMENT{\textcolor{gray}{Return the final meta-strategy, action sets, and payoff tensor}}
\end{algorithmic}
\end{algorithm}

\section{Experiments}\label{sec:exp}



\subsection{Environments}

\subsubsection{Classic BP Problems}\label{sec:env}

This section introduces the three classic static BP problems used in our experiments.

\paragraph{Recommendation Letter (REL)~\citep{dughmi2017algorithmic}}

A professor writes recommendation letters for graduating students, which are then reviewed by a company's human resources (HR) department to decide whether to hire. 
The professor and HR share a prior belief about the students' quality: 
there is a $1/3$ probability that a candidate is strong and a $2/3$ probability that the candidate is weak. 
HR does not know the exact quality of each student but aims to hire strong candidates, using the recommendation letters as the only source of information. 
HR receives a reward of $1$ for hiring a strong candidate, incurs a penalty of $-1$ for hiring a weak candidate, and gets $0$ for not hiring. 
The professor, on the other hand, gains a reward of $1$ for each student hired, regardless of their quality.


\paragraph{Courtroom (COR)~\citep{kamenica2011bayesian}}

In this scenario, a prosecutor attempts to convince a judge to convict a defendant, with two possible states: guilty or innocent. 
The judge (receiver) must choose between convicting or acquitting, receiving a utility of $1$ for a correct decision (convicting if guilty, acquitting if innocent) and $0$ for an incorrect one. 
The prosecutor (sender) receives a utility of $1$ if the judge convicts, regardless of the defendant's actual guilt, and both parties share a prior belief that the probability of guilt is $0.3$. 
In the original setting, the prosecutor conducts an investigation (signaling scheme) requiring decisions on actions such as subpoenas or forensic tests, represented by distributions $\pi(\cdot \mid \text{guilty})$ and $\pi(\cdot \mid \text{innocent})$ over signals. 
However, modeling real-world investigations in a verbalized setting poses challenges for LLMs, so we simplify the scenario by drawing inspiration from the REL problem, where the prosecutor selectively presents pre-existing evidence to influence the perceived probability of guilt, effectively replacing the investigation process with evidence presentation.

\paragraph{Law Enforcement (LAE)~\citep{kamenica2019bayesian}}

In this scenario, there are $Z$ miles of road, and drivers can choose to either speed or obey the speed limit on each mile. 
Speeding generates utility $V$ per mile, but drivers face a fine of $K > V$ if caught. 
There are $G$ police officers, and each officer can patrol one mile of road. 
The police aim to minimize the number of miles on which drivers speed. 
To map this environment to the BP problem, let $\omega \in \Omega = \{0, 1\}$ represent whether a police officer is present on a given mile. 
The prior belief is $\mu_0 = G / Z$. 
The set of signals corresponds to the miles of road, $S = \{1, \dots, Z\}$. 
In this model, the police act as the sender and the driver as the receiver. 
A signaling scheme represents the predictability or unpredictability of the police patrolling strategy. 
This strategy induces a joint distribution over $\Omega$ and $S$, i.e., over the presence of a police officer and the specific mile being patrolled.


\subsubsection{More Real-World Applications}

Our proposed verbalized Bayesian persuasion (VBP) framework has significant potential for real-world applications, particularly in complex, multi-sender, multi-receiver, and multi-round strategic communication scenarios. 
Below, we discuss two illustrative examples-conversational recommendation systems and healthcare DRG strategies-and highlight the potential challenges in applying VBP to these domains.

\paragraph{Conversational Recommendation Systems}
One promising application of VBP is in conversational recommendation systems, such as those used in live-stream shopping. 
In this setting, multiple senders (e.g., influencers or sales agents) aim to persuade a diverse group of receivers (customers) to purchase products through real-time, strategic communication. 
The VBP framework can optimize prompts (e.g., how product features or discounts are presented) to maximize customer engagement and conversions across varying customer segments.  
This application faces challenges such as receiver heterogeneity, where customers interpret signals differently based on their preferences and trust levels, making it difficult to craft universal strategies. 
Furthermore, the real-time nature of live-stream interactions demands highly efficient decision-making algorithms capable of adapting communication strategies dynamically. 
Scaling the system to accommodate thousands or millions of receivers simultaneously also requires advanced parallel processing and optimization techniques.  

\paragraph{DRG Strategy in Healthcare}

Another practical application lies in healthcare, specifically in optimizing Diagnosis-Related Group (DRG) reimbursement systems. 
Here, hospitals and post-acute care (PAC) providers (senders) communicate with regulatory agencies (receiver) to determine reimbursement policies for patient treatments. 
The VBP framework can model the incentives and communication strategies of the senders to help regulators design policies that balance cost-effectiveness with maintaining high-quality patient care.  
In this domain, conflicting incentives among senders (e.g., hospitals vs. PAC providers) add complexity, as senders may compete or collaborate to influence the receiver's decisions. 
Additionally, the large scale of the problem, with thousands of providers, poses computational challenges for efficient optimization. 
Long-term policy adjustments based on multi-round feedback further complicate the problem, requiring robust mechanisms to handle dynamic interactions over time.  

These examples demonstrate the versatility of the VBP framework in addressing real-world problems involving strategic communication. 
However, its application to practical scenarios requires addressing challenges such as scalability, heterogeneity of participants, real-time decision-making, and multi-round dynamics. 
Future work will focus on refining the VBP framework to overcome these challenges and enhance its readiness for deployment in diverse real-world contexts.

\subsection{VBP in Static Games (S1 \& S2)}\label{sec:vbp-static}

\begin{figure}[htb!]
     \centering
     \begin{subfigure}[b]{0.23\textwidth}
         \centering
         \includegraphics[width=\textwidth]{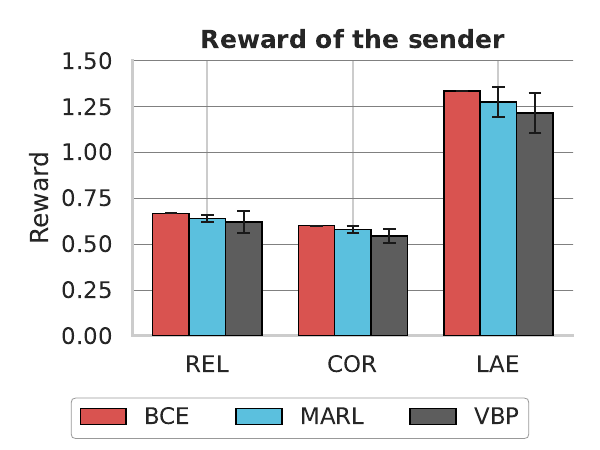}
         \caption{Sender's rewards.}
         \label{fig:classic-sender-reward}
     \end{subfigure}
     \begin{subfigure}[b]{0.23\textwidth}
         \centering
         \includegraphics[width=\textwidth]{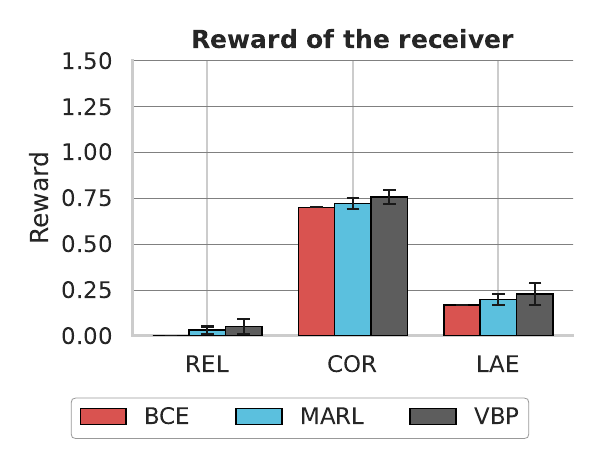}
         \caption{Receiver's rewards.}
         \label{fig:classic-receiver-reward}
     \end{subfigure}
     \begin{subfigure}[b]{0.23\textwidth}
         \centering
         \includegraphics[width=\textwidth]{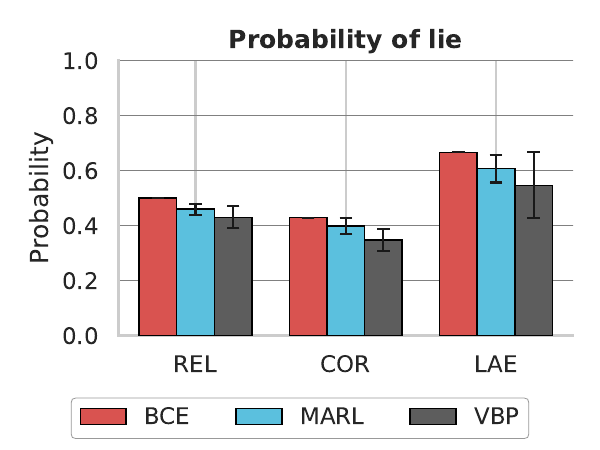}
         \caption{Lie probability.}
         \label{fig:classic-lie-prob}
     \end{subfigure}
     \begin{subfigure}[b]{0.23\textwidth}
         \centering
         \includegraphics[width=\textwidth]{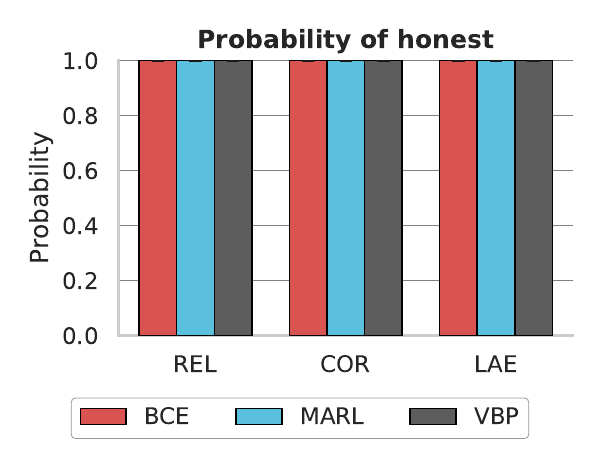}
         \caption{Honest probability.}
         \label{fig:classic-honest-prob}
     \end{subfigure}
        \caption{Performance comparison on classic static BP problems. Averaged over $20$ seeds. In the $3$ BP problems, the probability of lying refers to describing a weak student as strong, an innocent defendant as guilty, or an unpatrolled segment as patrolled. Conversely, the probability of honesty refers to accurately describing a strong student, a guilty defendant, or a patrolled segment.}
        \label{fig:classic-results}
\end{figure}

We first evaluate the effectiveness of the VBP method under the S1 setting. 
Two baseline methods are chosen: BCE and MARL. 
The former is based on the optimal equilibria computed in~\citet{lin2023information},~\citet{kamenica2011bayesian} and~\citet{kamenica2019bayesian}, while the latter is based on the multi-agent reinforcement learning method proposed in~\citet{lin2023information} for solving BP problems. 
As shown in Figure~\ref{fig:classic-results}, the VBP framework successfully captures the essence of solving BP problems, namely, selectively withholding, obfuscating, or even deceiving about the true state, while also learning when to fully disclose accurate information.

\begin{figure}[htb!]
     \centering
     \begin{subfigure}[b]{0.23\textwidth}
         \centering
         \includegraphics[width=\textwidth]{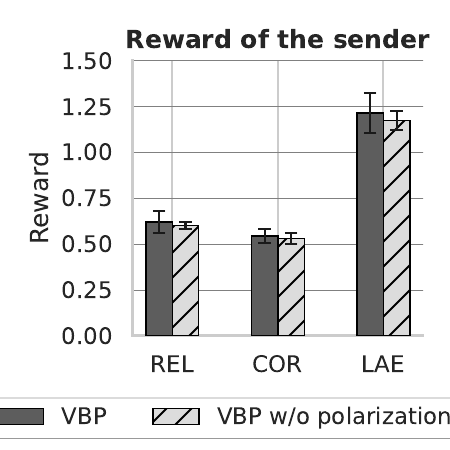}
         \caption{Sender's rewards.}
         \label{fig:classic-sender-reward-general}
     \end{subfigure}
     \begin{subfigure}[b]{0.23\textwidth}
         \centering
         \includegraphics[width=\textwidth]{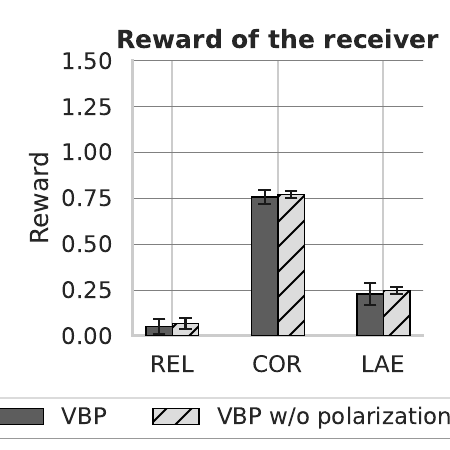}
         \caption{Receiver's rewards.}
         \label{fig:classic-receiver-reward-general}
     \end{subfigure}
     \begin{subfigure}[b]{0.23\textwidth}
         \centering
         \includegraphics[width=\textwidth]{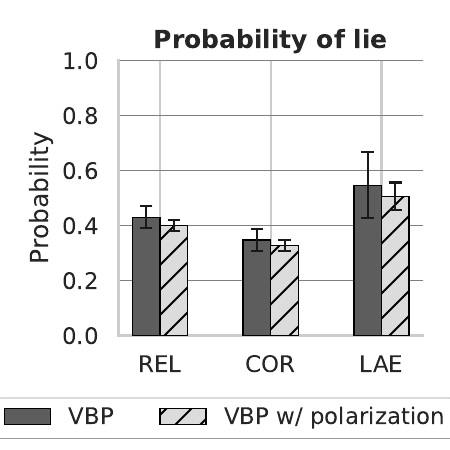}
         \caption{Lie probability.}
         \label{fig:classic-lie-prob-general}
     \end{subfigure}
     \begin{subfigure}[b]{0.23\textwidth}
         \centering
         \includegraphics[width=\textwidth]{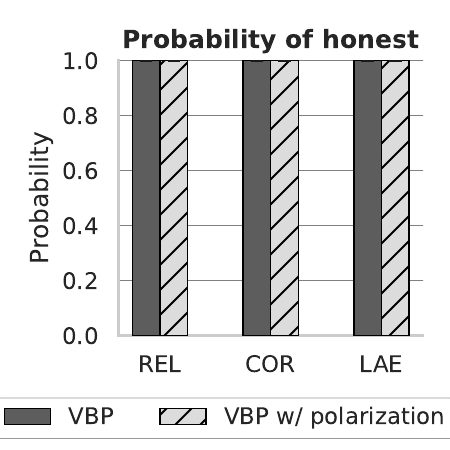}
         \caption{Honest probability.}
         \label{fig:classic-honest-prob-general}
     \end{subfigure}
        \caption{Performance comparison on general static BP problems. Averaged over $20$ seeds. The physical meaning of the probabilities of lying and honesty is consistent with Figure~\ref{fig:classic-results}.}
        \label{fig:general-results}
\end{figure}

Next, we removed signal polarization to make the sender's signals in each problem more reflective of real-world recommendation letters, courtrooms, and police deployment announcements, resulting in the S2 setting. 
Since existing BCE and MARL methods cannot solve this, we only compared VBP with the VBP variant from the S1 setting. 
The results are shown in Figure~\ref{fig:general-results}. 

Figure~\ref{fig:general-results} shows that VBP's performance in S2 is roughly on par with S1, with a slight performance drop. 
In both settings, VBP achieves optimal strategy performance in terms of the probability of honesty. 
We speculate that this could be due to the alignment of the LLM used, which allows it to more easily converge to honest strategies, such as truthfully reporting the situation of a strong student, a guilty defendant, or a patrolled segment.


To quantify the proximity of policies of the sender and the receiver to the BCE, we employ exploitability as a measure. 
Exploitability~\citep{psro} measures the distance of a joint meta-strategy of sender and receiver from the BCE. 
It shows how much each LLM gains by deviating to their best responses.

\begin{figure}[htb!]
    \centering
    \begin{subfigure}[b]{0.24\textwidth}
         \centering
         \includegraphics[width=\textwidth]{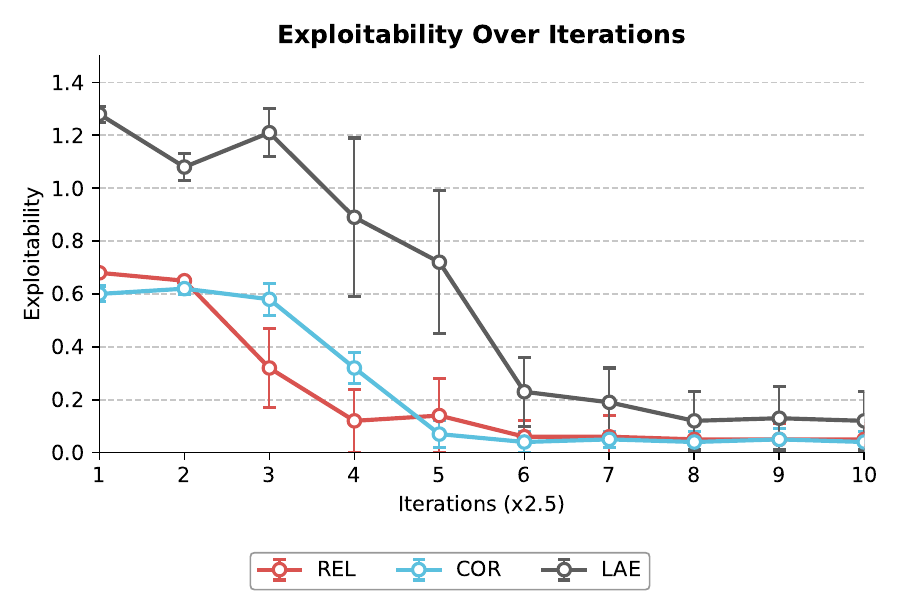}
         \vspace{-20pt}
         \label{fig:exploits}
     \end{subfigure}
     \hfill
     \begin{subfigure}[b]{0.24\textwidth}
         \centering
         \includegraphics[width=\textwidth]{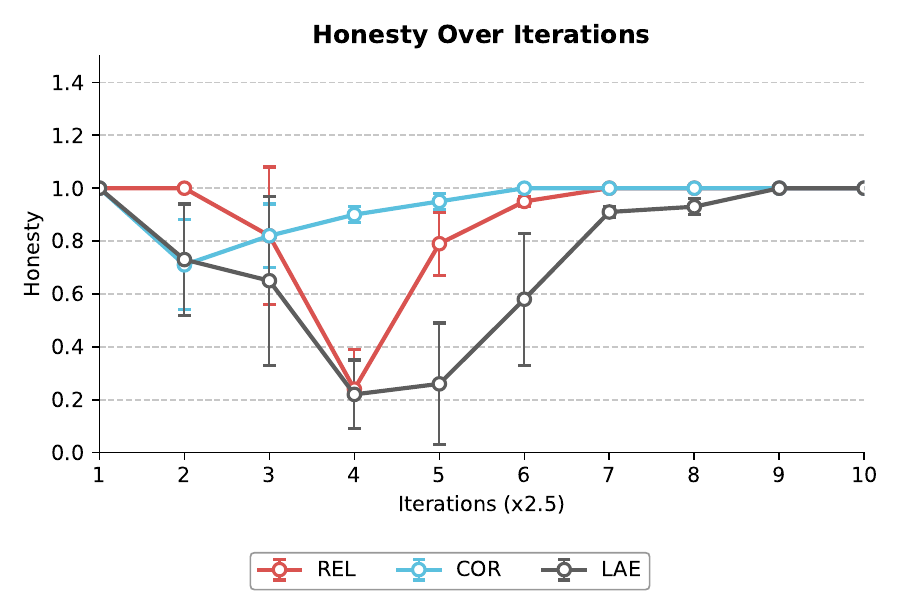}
         \vspace{-20pt}
         \label{fig:honesty}
     \end{subfigure}
     \begin{subfigure}[b]{0.24\textwidth}
         \centering
         \includegraphics[width=\textwidth]{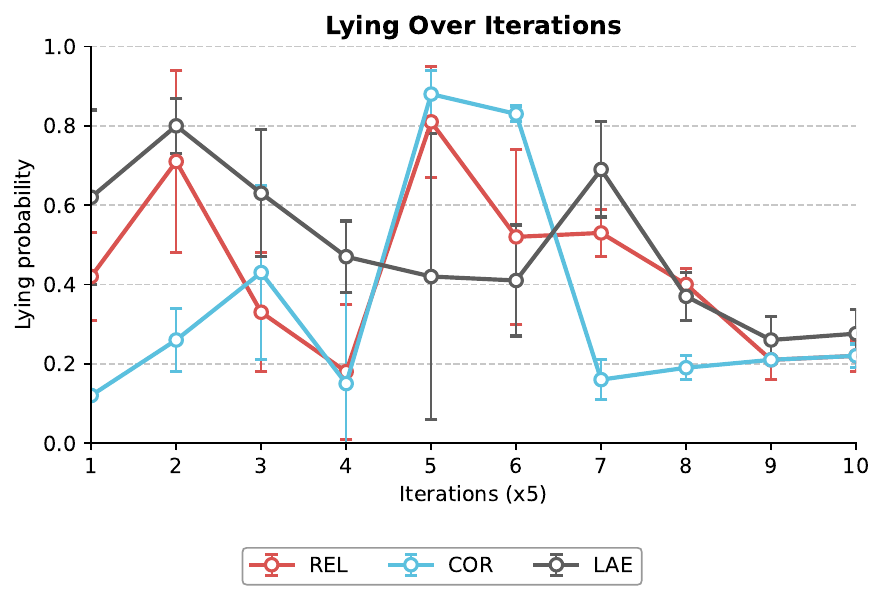}
         \vspace{-20pt}
         \label{fig:lying-multistage}
     \end{subfigure}
     \begin{subfigure}[b]{0.24\textwidth}
         \centering
         \includegraphics[width=\textwidth]{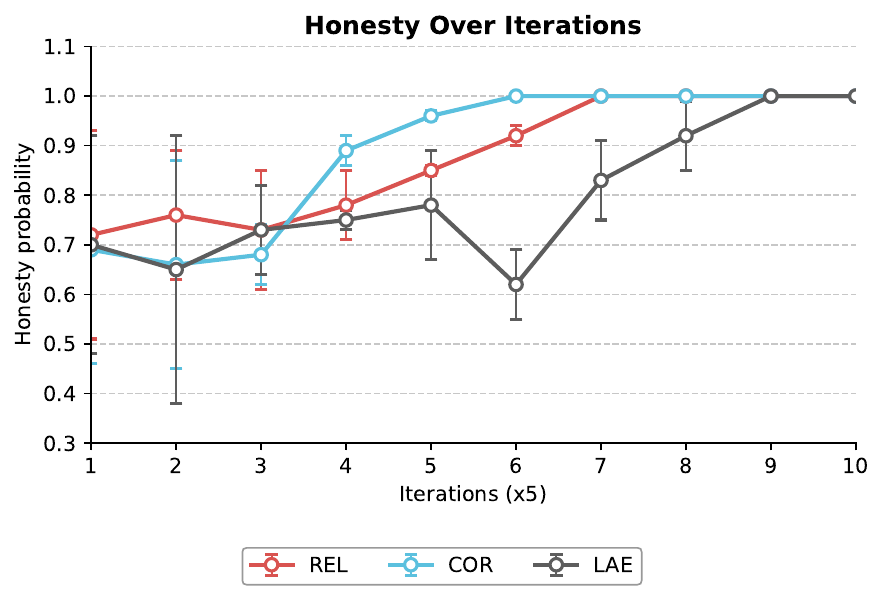}
         \vspace{-20pt}
         \label{fig:honesty-multistage}
     \end{subfigure}
    \caption{\textbf{Left:} The variation in exploitability during the iterative solving process of VBP in the S1 setting, reflecting changes in proximity to approximate Bayesian correlated equilibrium. \textbf{Left Center:} The variation in honesty probability during the iterative solving process of VBP in the S1 setting. \textbf{Right Center and Right:} The variation in lying and honesty probability during the iterative solving process of VBP in the S3 setting. Averaged over $20$ seeds.}
    \label{fig:variations}
\end{figure}

As depicted in Figure~\ref{fig:variations}, exploitability gradually decreased to approximately $0.1$ after $10$ iterations of training. 
This descent indicates the diminishing gap between the utility generated by the joint strategies of the sender and the receiver and the utility generated by the BCE strategies, signifying VBP's acquisition of the equilibrium.
As mentioned in Section~\ref{sec:settings}, we align as closely as possible with the classic static BP problem by polarizing the signals. 


To further investigate the phenomenon of honesty oscillations-where honesty rises, falls, and then rises again-we conducted additional experiments using an unaligned LLaMA model\footnote{\url{https://huggingface.co/SicariusSicariiStuff/LLAMA-3_8B_Unaligned_BETA}.} as the base language model. 
This was motivated by that the observed pattern in Section~\ref{sec:vbp-static} might be better explained by strategic cycles between the sender and the receiver, rather than by the alignment properties of the LLM, as we originally hypothesized. 

\begin{figure}[htb!]
    \centering
    \begin{subfigure}[b]{0.32\textwidth}
         \centering
         \includegraphics[width=\textwidth]{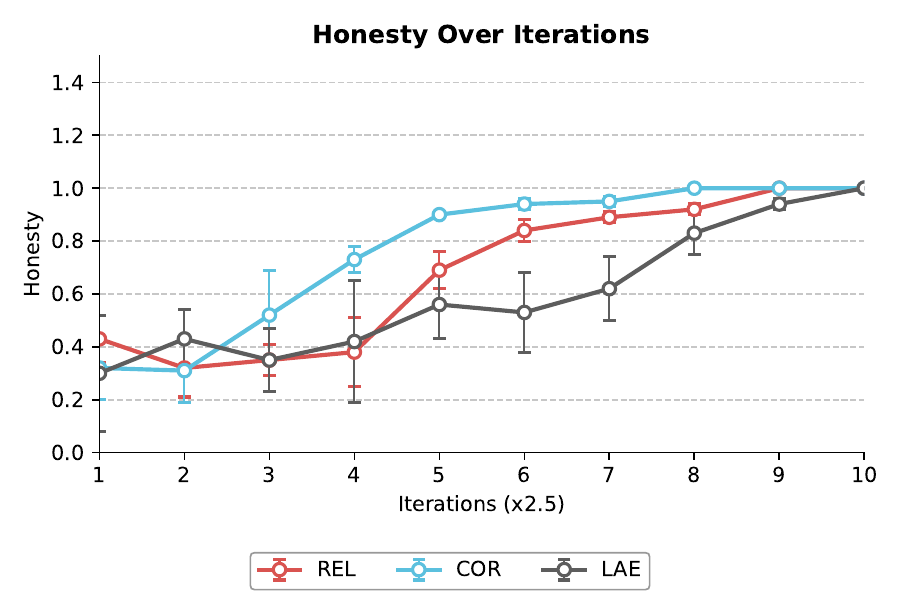}
         \label{fig:honesty_unaligned}
     \end{subfigure}
     \begin{subfigure}[b]{0.32\textwidth}
         \centering
         \includegraphics[width=\textwidth]{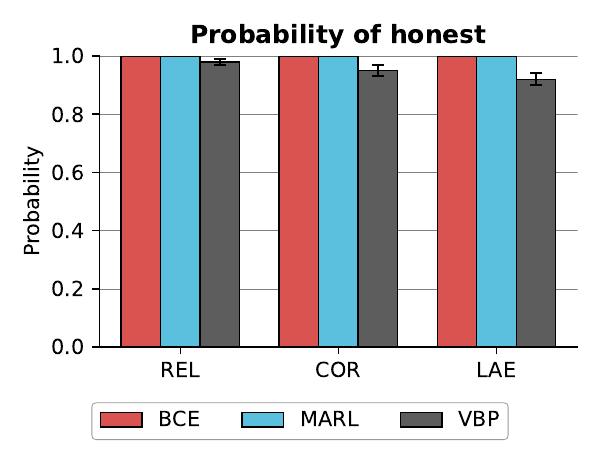}
         \label{fig:honest_prob_unaligned}
     \end{subfigure}
    \caption{\textbf{Left:} Performance comparision in the S1 setting. In the $3$ BP problems, the probability of honesty refers to accurately describing a strong student, a guilty defendant, or a patrolled segment. \textbf{Right:} The variation in honesty probability during the iterative solving process of VBP in the S1 setting. Averaged over $20$ seeds.}
    \label{fig:h_unaligned}
\end{figure}

The experimental results shown in Figure~\ref{fig:h_unaligned} reveal two key findings. 
First, with the unaligned LLaMA model, the oscillatory pattern of honesty disappears, and the behavior stabilizes at a consistent level of honesty. 
This supports our initial hypothesis that the oscillations are driven by the alignment properties of the LLM, which likely introduce normative biases (e.g., promoting honesty or fairness) that influence the dynamics of strategic interactions. 
Second, we observe that the honesty probability with the unaligned LLM no longer always achieves the optimal level (probability of 1), as seen in aligned models. 
This suggests that unaligned models are less reliable in consistently promoting desirable outcomes, such as fully honest behavior, in strategic settings.

These findings highlight the dual impact of alignment: while it introduces oscillatory dynamics due to normative pressures, it also helps achieve higher levels of optimal honesty in strategic interactions. 
This emphasizes the importance of alignment in applications requiring robust ethical or normative behaviors, while also suggesting a need for further exploration of its impact on the stability of agent interactions in game-theoretic settings.

\subsection{VBP in Multistage Games (S3)}\label{sec:vbp-multistage}

\begin{figure}[htb!]
     \centering
     \begin{subfigure}[b]{0.23\textwidth}
         \centering
         \includegraphics[width=\textwidth]{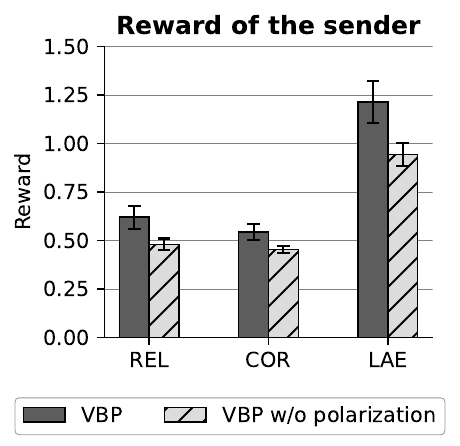}
         \caption{Sender's rewards.}
         \label{fig:classic-sender-reward-multistage}
     \end{subfigure}
     \begin{subfigure}[b]{0.23\textwidth}
         \centering
         \includegraphics[width=\textwidth]{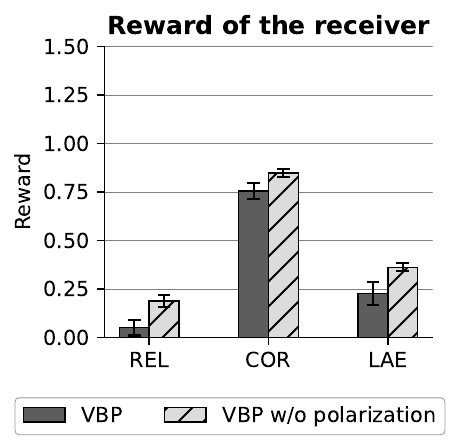}
         \caption{Receiver's rewards.}
         \label{fig:classic-receiver-reward-multistage}
     \end{subfigure}
     \begin{subfigure}[b]{0.23\textwidth}
         \centering
         \includegraphics[width=\textwidth]{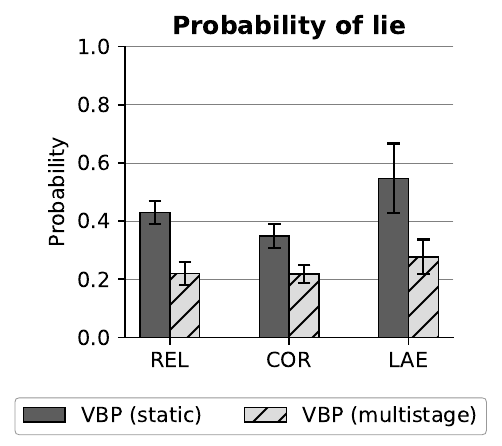}
         \caption{Lie probability.}
         \label{fig:classic-lie-prob-multistage}
     \end{subfigure}
     \begin{subfigure}[b]{0.23\textwidth}
         \centering
         \includegraphics[width=\textwidth]{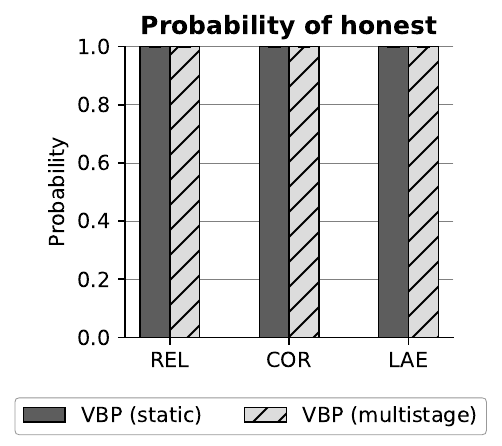}
         \caption{Honest probability.}
         \label{fig:classic-honest-prob-multistage}
     \end{subfigure}
        \caption{Performance comparision on the S3 setting. Averaged over $20$ seeds and $5$ timesteps. The physical meaning of the probabilities of lying and honesty is consistent with Figure~\ref{fig:classic-results}.}
        \label{fig:multistage-results}
\end{figure}

We also tested the effectiveness of VBP in a multistage scenario. 
Notably, the multistage BP differs from most of the literature. In existing works, the same sender was interacting with a new, short-sighted receiver in each round. 
In this work, the receiver remains the same and can perceive the interaction history, aligning more closely with the Markov signaling game~\citep{lin2023information}. 
Since a closed-form solution for equilibrium cannot be computed, we record the average performance at each stage in Figure~\ref{fig:multistage-results}. 

It is observed from the figure that VBP's performance shows a noticeable decline compared to S2, while it still manages to learn both appropriate deception and honesty.
We also visualize the changes in the sender's deception and honesty probabilities during training, as shown in Figure~\ref{fig:variations}.
Since the receiver can perceive the history, the sender's deceptive behavior goes through several oscillations, reflecting a kind of bargaining dynamic~\citep{nash1950bargaining,nash1953two,Maschler_Solan_Zamir_2013}: 
The sender is initially leaning towards honesty, then discovering that deception maximizes gains, and later realizing that excessive deception triggers retaliation from the receiver, eventually converging to a relatively low deception probability.
Likewise, with the receiver having access to historical interactions, the sender demonstrated an upward trend in honest behavior compared to the S2 setting, with truthfulness levels progressively increasing throughout.

\paragraph{Remark 2}

As discussed above, the S3 iterated setting reveals some implications for the bargaining interactions between the sender and receiver. 
In classical persuasion theory, the sender commits to a signaling strategy upfront, and this commitment is justified by the need to maintain trust and reputation in long-term interactions. 
Under these assumptions, the receiver typically follows the sender's signals, as deviating would harm the receiver’s own expected utility.
However, our results in the S3 setting suggest a more complex dynamic. 
Specifically, the receiver can choose to ignore the sender’s signals, effectively invalidating the sender’s commitment. 
This observation highlights that the sender’s commitment is not unilateral-it must be accepted by the receiver to hold. 
If the receiver disagrees with the sender’s proposed strategy, they can force both parties into a mutually worse outcome by disregarding the signals altogether.
This leads to an important hypothesis: in the VBP framework, Bayesian persuasion may function more like a bargaining game, where both parties must agree on the signaling strategy to avoid suboptimal outcomes. 
This perspective challenges the traditional unilateral commitment model and suggests a more interactive dynamic in iterated settings.
While we acknowledge the importance of this insight, we intentionally keep our analysis of S3 limited in this paper to maintain focus on the primary contributions. 
Exploring the bargaining dynamics observed in S3 presents an exciting avenue for future research.

\subsection{Prompt (Strategy) Variation}

\begin{figure}[htb!]
     \centering
     \begin{subfigure}[b]{0.48\textwidth}
         \centering
         \includegraphics[width=\textwidth]{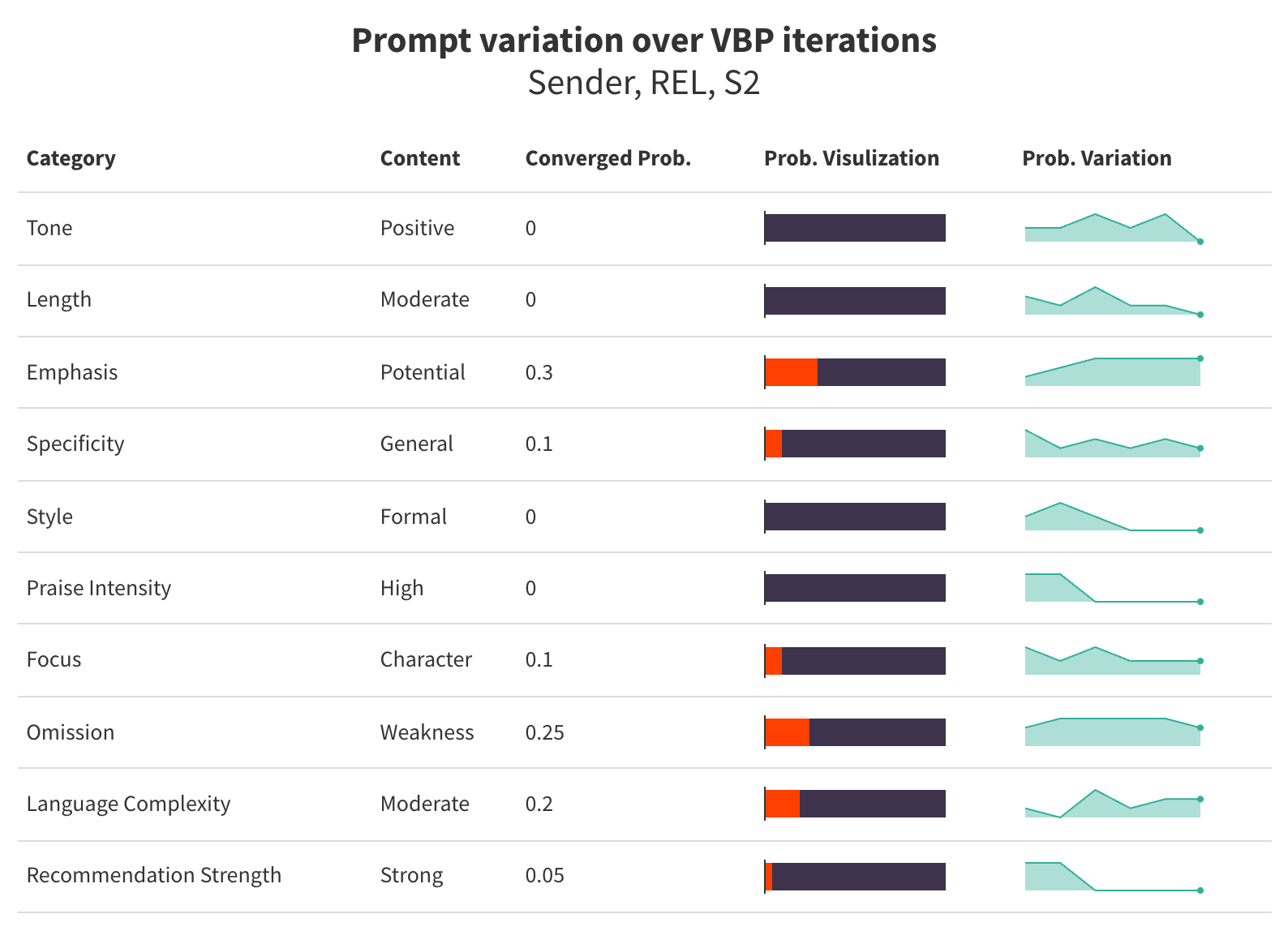}
         \vspace{-15pt}
     \end{subfigure}
     \hfill
     \begin{subfigure}[b]{0.48\textwidth}
         \centering
         \includegraphics[width=\textwidth]{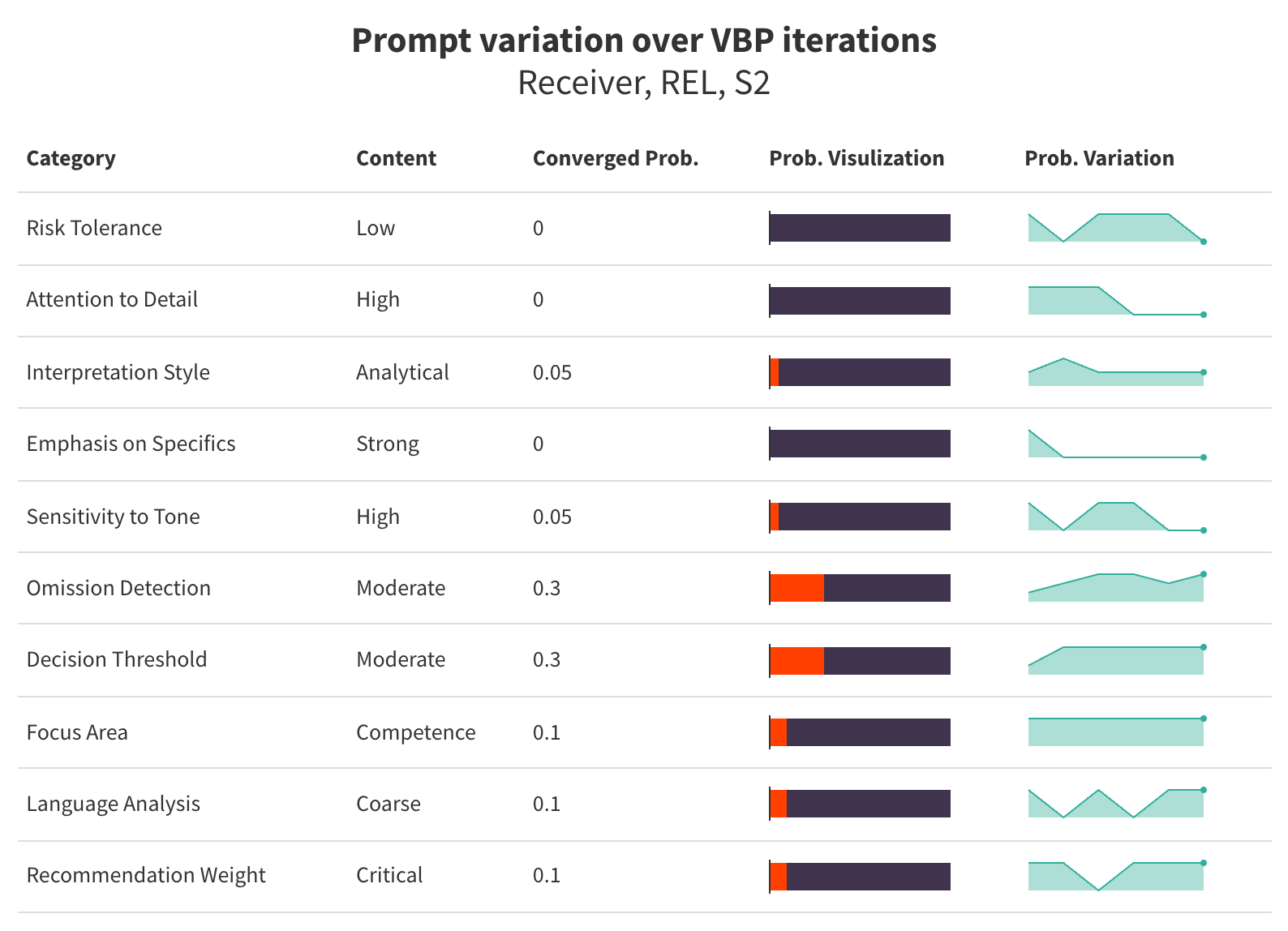}
         \vspace{-15pt}
     \end{subfigure}
     \begin{subfigure}[b]{0.48\textwidth}
         \centering
         \includegraphics[width=\textwidth]{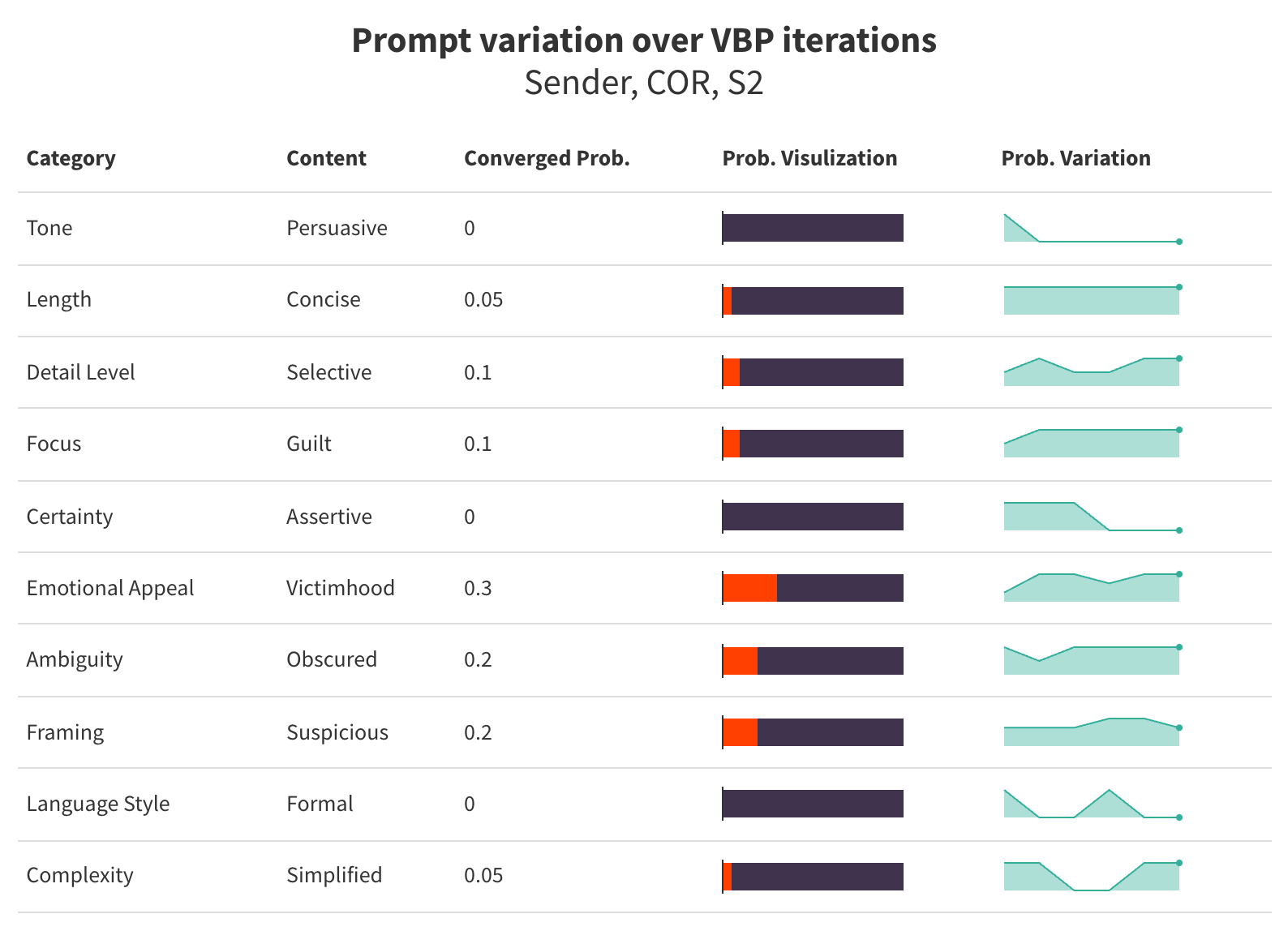}
         \vspace{-15pt}
     \end{subfigure}
     \hfill
     \begin{subfigure}[b]{0.48\textwidth}
         \centering
         \includegraphics[width=\textwidth]{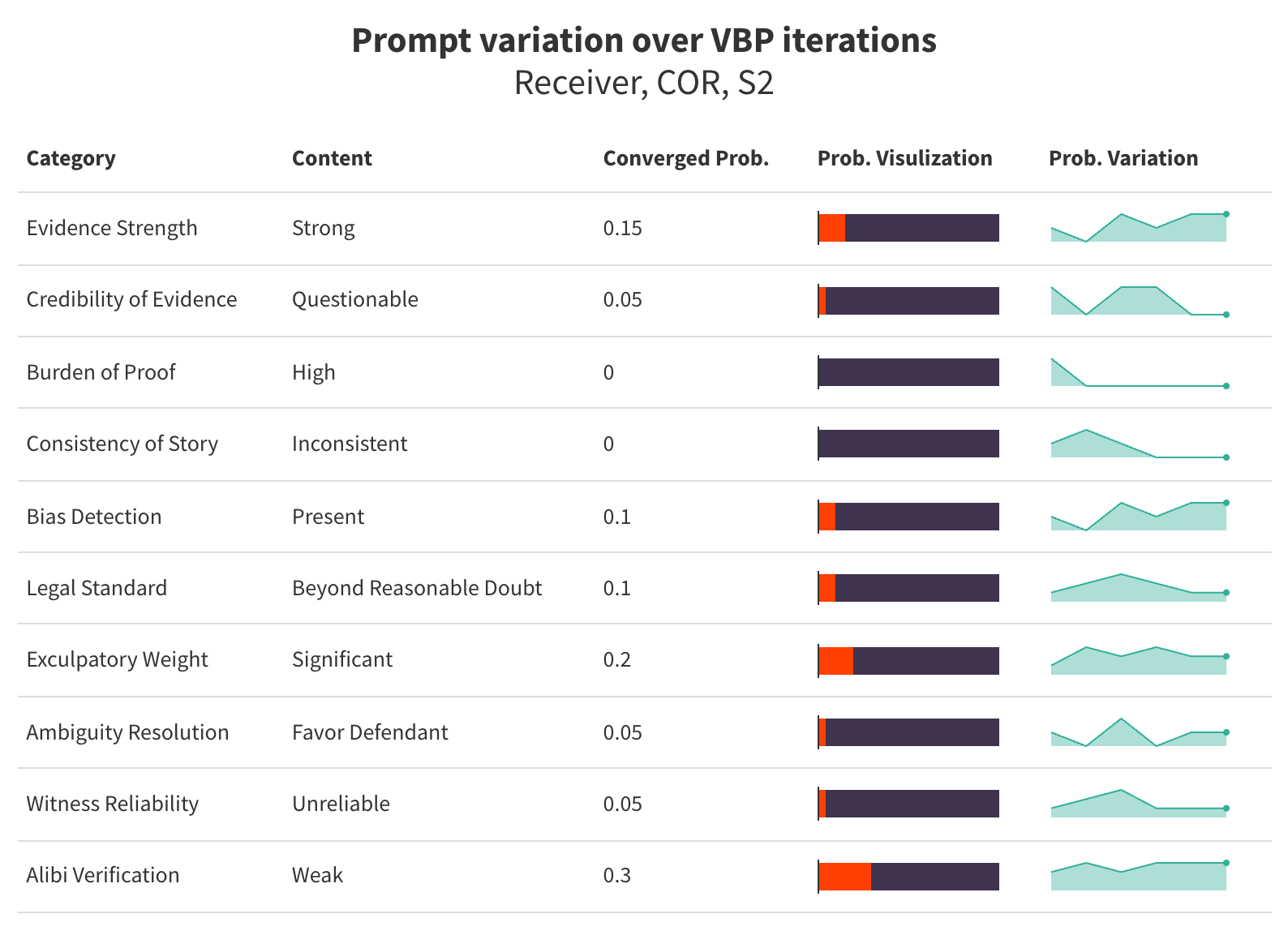}
         \vspace{-15pt}
     \end{subfigure}
     \begin{subfigure}[b]{0.48\textwidth}
         \centering
         \includegraphics[width=\textwidth]{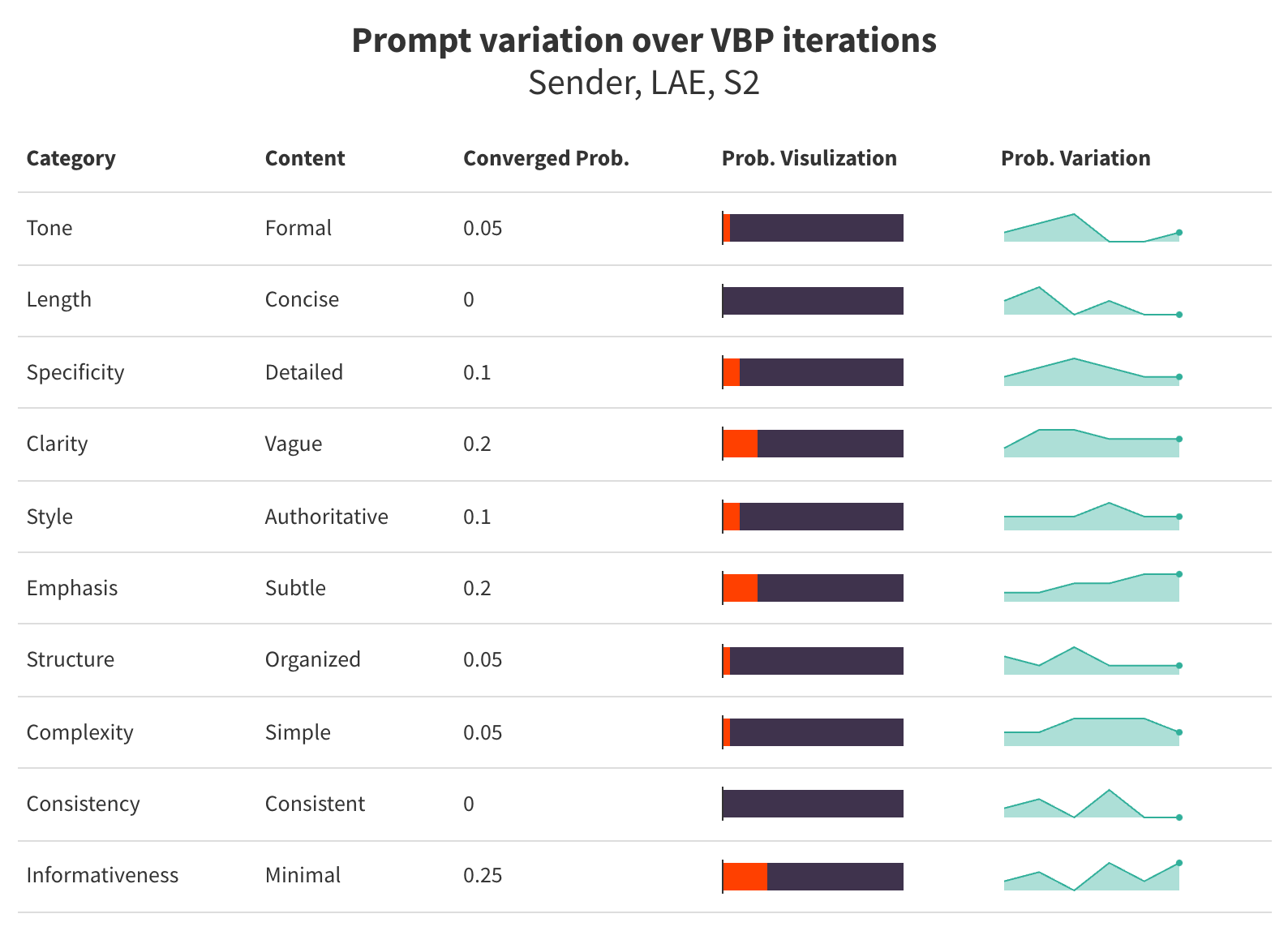}
         \vspace{-15pt}
     \end{subfigure}
     \hfill
     \begin{subfigure}[b]{0.48\textwidth}
         \centering
         \includegraphics[width=\textwidth]{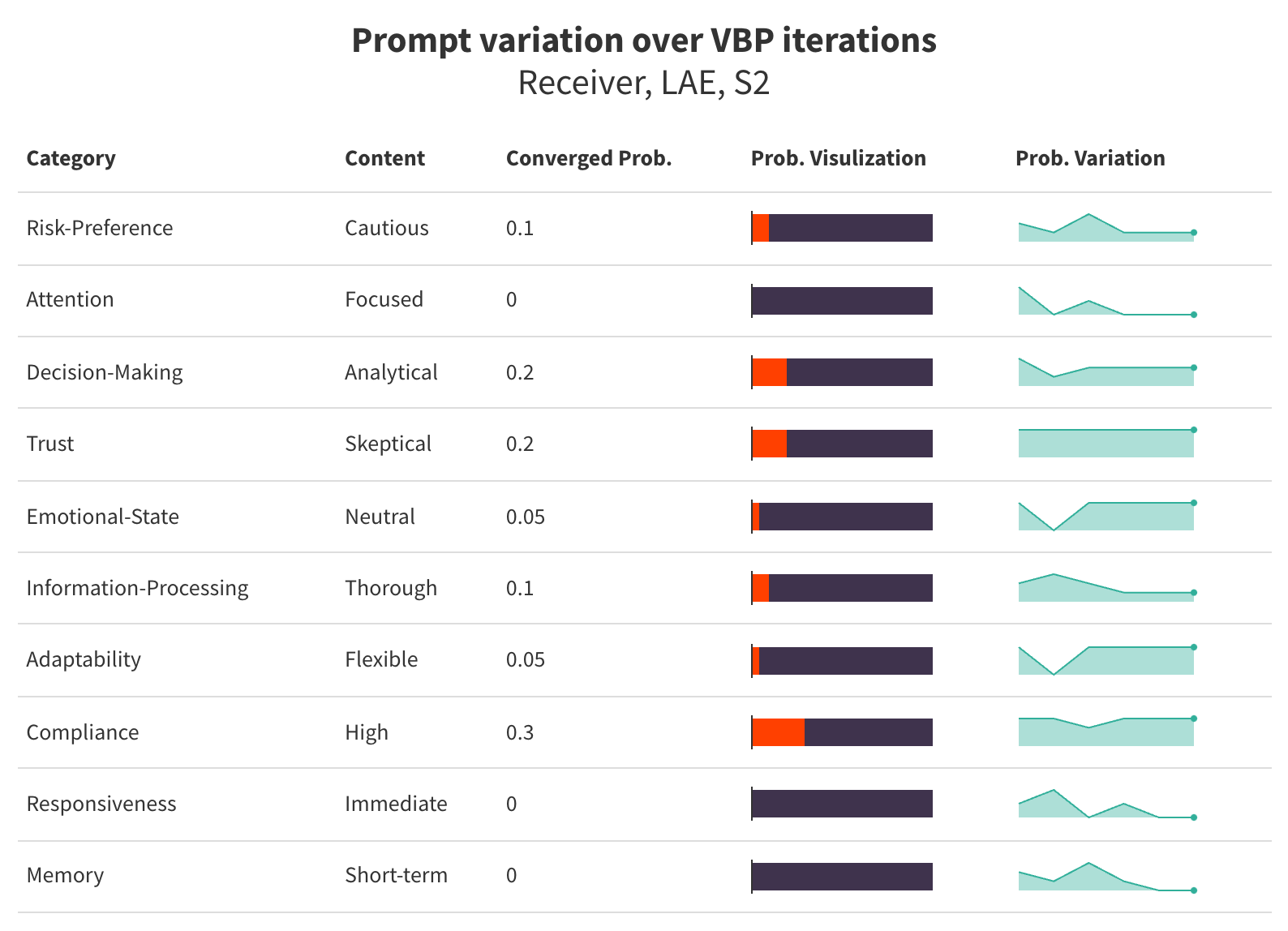}
         \vspace{-15pt}
     \end{subfigure}
        \caption{The variation in the prompts during the iterative solving process of VBP in the S2 setting.}
        \label{fig:prompt_variation_rel_s2}
\end{figure}

This section presents the final converged meta-strategy in the S2 setting, as well as the relative changes in the selection probabilities of each strategy (i.e., the prompts that influence writing style) throughout the training process, as shown in Figure~\ref{fig:prompt_variation_rel_s2}.


Figure~\ref{fig:prompt_variation_rel_s2} presents the top $10$ strategies with the highest selection probabilities in the final strategy pool after Prompt-PSRO convergence. 
These probabilities represent the average likelihood of selecting each strategy from the pool, revealing the adaptation process of sender and receiver strategies over iterations. 
The optimization process follows a hierarchical approach: 
first, OPRO optimizes the category of each prompt (e.g., "Tone"), then the specific content within that category. 
The table columns in Figure~\ref{fig:prompt_variation_rel_s2} reflect this structure, with the first two columns showing optimized categories and content, while the third and fourth columns display their probabilities. 
The fifth column tracks how these probabilities evolve across iterations, highlighting the refinement of strategies during the optimization process.  

\begin{figure}[htb!]
    \centering
    \includegraphics[width=0.5\linewidth]{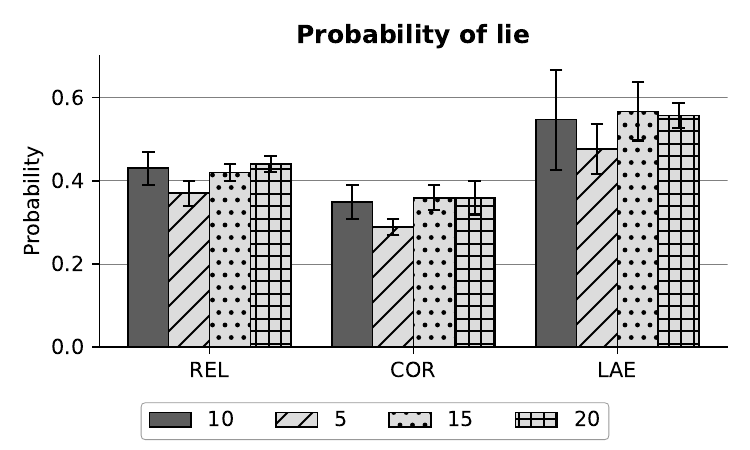}
    \caption{The probability of the sender lying under different upper limits on the number of prompts. The figure shows that when the number of prompts is heavily pruned, significant performance degradation occurs. However, once the number of retained prompts exceeds a certain threshold, such as 10-15, the impact on performance becomes negligible.}
    \label{fig:lie_prob_ablations_num_prompts}
\end{figure}

To reduce computational complexity, we prune the strategy pool to the top $10$ prompts based on selection probabilities. 
We conduct additional experiments to assess this pruning's effects, as shown in Figure~\ref{fig:lie_prob_ablations_num_prompts}.

Additionally, the probabilities in Figure~\ref{fig:prompt_variation_rel_s2} are computed as the average probability of selecting each prompt from the strategy pool across iterations, and the content (e.g., "Positive") under the category (e.g., "Tone") is dynamically optimized rather than fixed.

\subsection{Ablation Studies}

\subsubsection{Key Components}

\begin{figure}[htb!]
     \centering
     \begin{subfigure}[b]{0.24\textwidth}
         \centering
         \includegraphics[width=\textwidth]{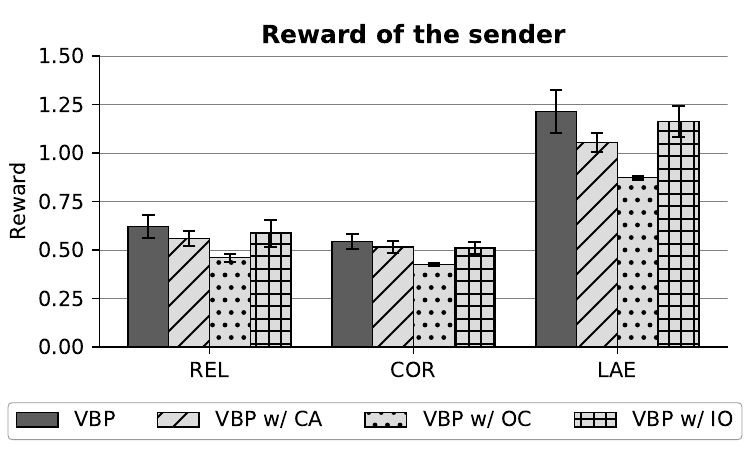}
         \caption{Sender's rewards.}
         \label{fig:classic-sender-reward-ablations}
     \end{subfigure}
     \hfill
     \begin{subfigure}[b]{0.24\textwidth}
         \centering
         \includegraphics[width=\textwidth]{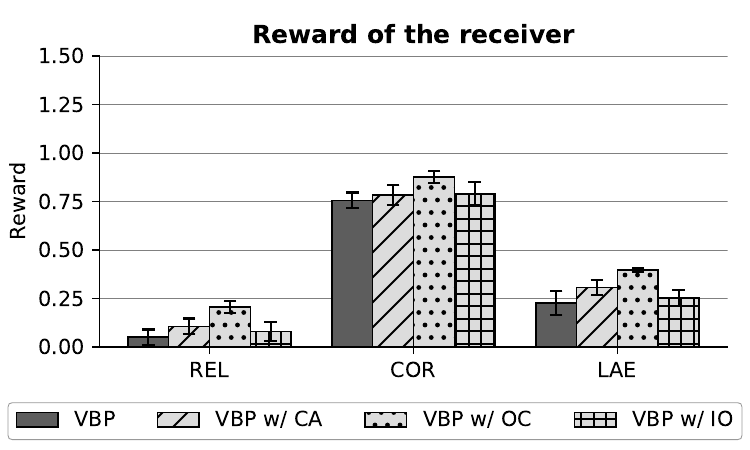}
         \caption{Receiver's rewards.}
         \label{fig:classic-receiver-reward-ablations}
     \end{subfigure}
     \hfill
     \begin{subfigure}[b]{0.24\textwidth}
         \centering
         \includegraphics[width=\textwidth]{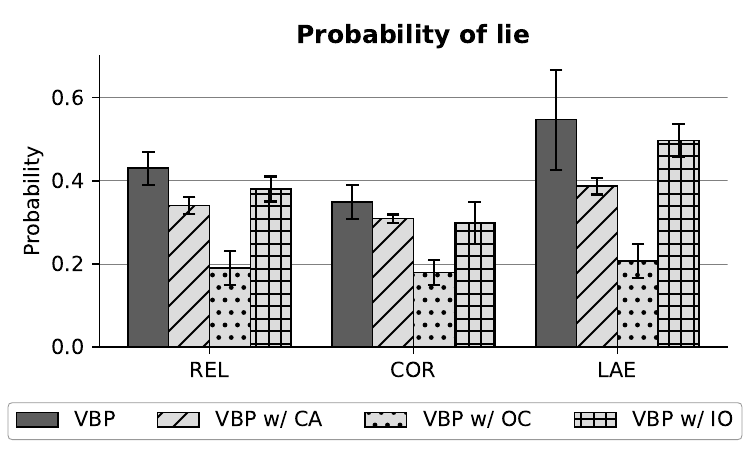}
         \caption{Lie probability.}
         \label{fig:classic-lie-prob-ablations}
     \end{subfigure}
     \hfill
     \begin{subfigure}[b]{0.24\textwidth}
         \centering
         \includegraphics[width=\textwidth]{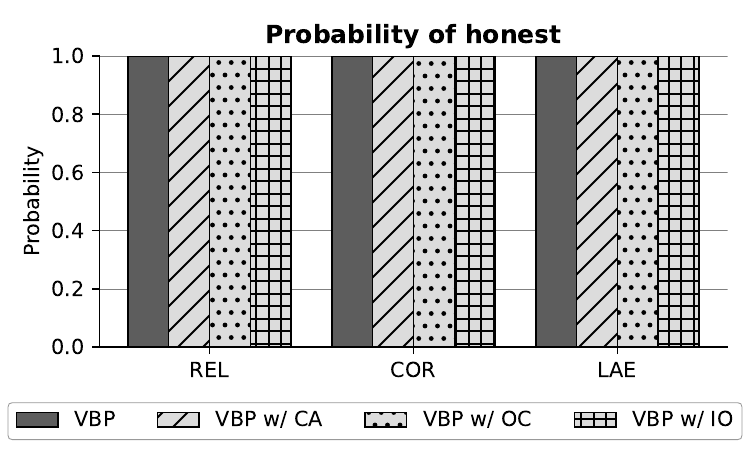}
         \caption{Honest probability.}
         \label{fig:classic-honest-prob-ablations}
     \end{subfigure}
        \caption{Ablation studies on general static BP problems. Averaged over $20$ seeds. CA, OC, and IO represent the commitment assumption, obedience constraint, and information obfuscation, respectively. The physical meaning of the probabilities of lying and honesty is consistent with Figure~\ref{fig:classic-results}.}
        \label{fig:general-results-ablations}
\end{figure}

This section analyzes the impact of key design elements within the VBP framework on performance, primarily including the verbalization of the commitment assumption, the obedience constraint, and the introduction of information obfuscation techniques to facilitate VBP convergence. 
The experimental results in the S2 setting are shown in Figure~\ref{fig:general-results-ablations}.

As the figure illustrates, these designs have varying degrees of influence on the key aspect of the BP problem, namely the probability of lying, while having minimal effect on the final converged probability of honesty. 
Specifically, the absence of the obedience constraint has a significant impact on the convergence results, which is consistent with previous observations~\citep{lin2023information}. 
Secondly, the commitment assumption has little effect on the probability of lying. 
One possible explanation is that, in a repeated game where a long-term sender interacts with a sequence of short-term receivers, commitment naturally emerges in equilibria. 
This occurs because the sender needs to establish a reputation for credibility, which is crucial for maximizing its long-term payoff expectations~\citep{rayo2010optimal,lin2023information}.
Lastly, the introduction of information obfuscation also has little impact on performance, indicating that the VBP framework can spontaneously learn to withhold or deceive regarding information.

\subsubsection{Polarized Signal Visualization}

\begin{figure}[htb!]
     \centering
     \begin{subfigure}[b]{0.24\textwidth}
         \centering
         \includegraphics[width=\textwidth]{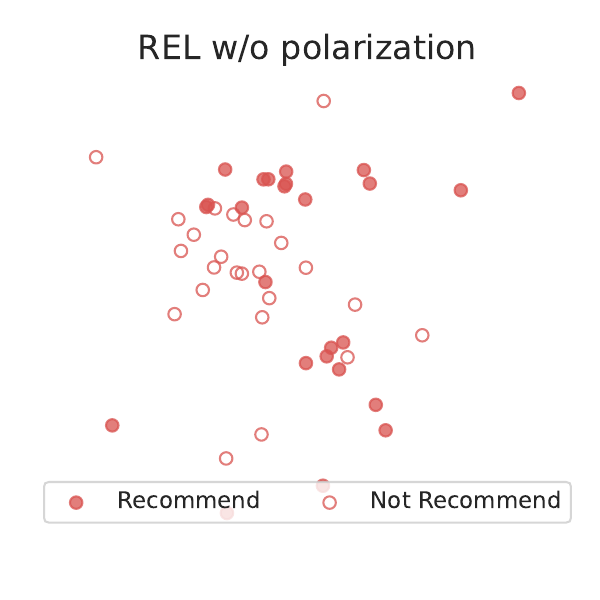}
         \vspace{-20pt}
         \label{fig:classic-rel-wo-pol}
     \end{subfigure}
     \hfill
     \begin{subfigure}[b]{0.24\textwidth}
         \centering
         \includegraphics[width=\textwidth]{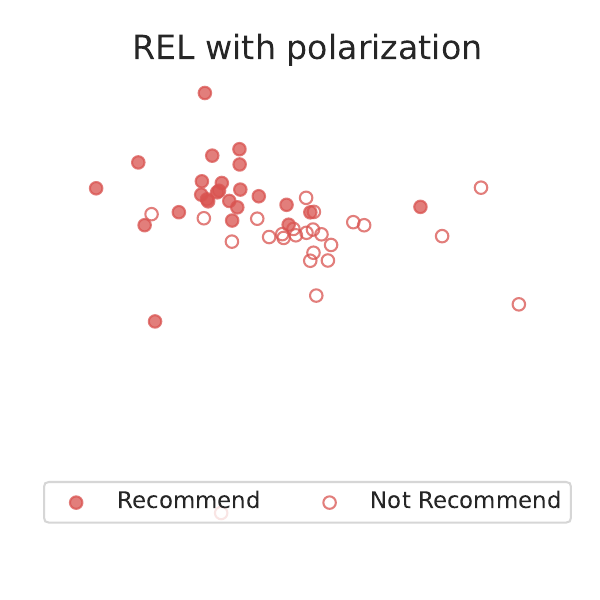}
         \vspace{-20pt}
         \label{fig:classic-rel-w-pol}
     \end{subfigure}
     \hfill
     \begin{subfigure}[b]{0.24\textwidth}
         \centering
         \includegraphics[width=\textwidth]{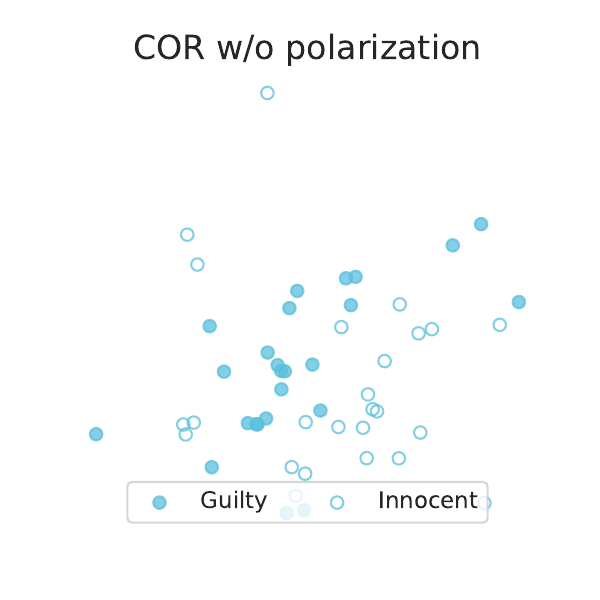}
         \vspace{-20pt}
         \label{fig:classic-cor-wo-pol}
     \end{subfigure}
     \hfill
     \begin{subfigure}[b]{0.24\textwidth}
         \centering
         \includegraphics[width=\textwidth]{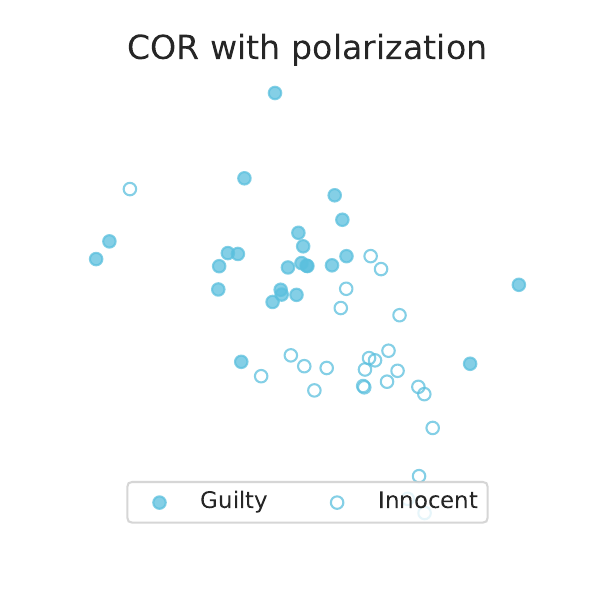}
         \vspace{-20pt}
         \label{fig:classic-cor-w-pol}
     \end{subfigure}
        \caption{Visualization of signal polarization. The scatter points in the figure represent the t-SNE dimensionality reduction results of signals output by the sender, under $50$ random seeds.}
        \label{fig:classic-signal-pol}
\end{figure}

To verify the effectiveness of signal polarization, we extract the final layer of the sender's output encoding and apply t-SNE for dimensionality reduction. 
At the same time, we use GPT-4o to classify the output signals as an estimate of the ground truth. 
The final visualization is shown in Figure~\ref{fig:classic-signal-pol}. 

From the figure, it can be observed that after signal polarization, the sender's output signals exhibit clearer tendencies. 
It is worth noting that in the LAE problem, the signal must explicitly indicate whether a segment is patrolled by the police, so signal polarization is not required, and thus it is not displayed in the figure.

\subsubsection{Predefined Signaling Scheme}

\begin{figure}[htb!]
     \centering
     \begin{subfigure}[b]{0.30\textwidth}
         \centering
         \includegraphics[width=\textwidth]{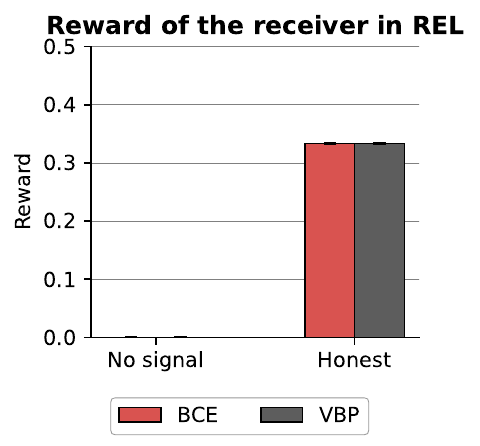}
         \vspace{-20pt}
         \label{fig:rel_fixed_scheme}
     \end{subfigure}
     \hfill
     \begin{subfigure}[b]{0.30\textwidth}
         \centering
         \includegraphics[width=\textwidth]{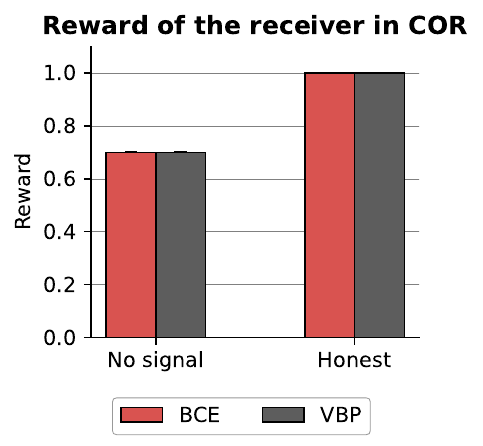}
         \vspace{-20pt}
         \label{fig:cor_fixed_scheme}
     \end{subfigure}
     \hfill
     \begin{subfigure}[b]{0.30\textwidth}
         \centering
         \includegraphics[width=\textwidth]{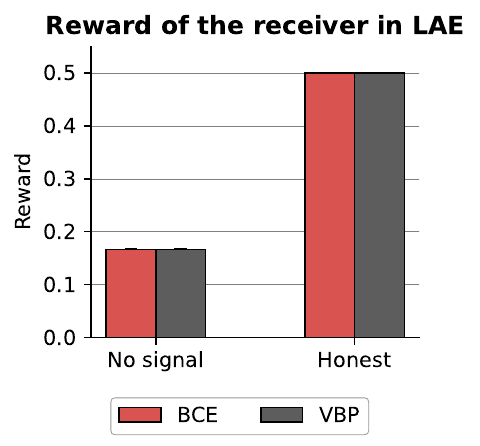}
         \vspace{-20pt}
         \label{fig:lae_fixed_scheme}
     \end{subfigure}
        \caption{Receiver's rewards when sender's signaling scheme is predefined. ``No signal'' indicates that the message generated by the sender contains no information about the true state, while ``honest'' means the sender fully discloses all information about the true state.}
        \label{fig:predefined-signaling-scheme}
\end{figure}

This section tests whether the receiver in the VBP framework could converge to BCE when the sender's strategy is fixed in the S2 setting.
The results are shown in Figure~\ref{fig:predefined-signaling-scheme}. 
As can be seen from the figure, VBP is able to learn the optimal strategy across all three BP problems.

\section{Limitations and Future Work}

While our approach offers promising results, it faces several limitations, both inherent to LLMs and game theory individually, as well as their integration. 
First, although LLMs have been widely employed to simulate human behavior, concerns remain regarding the fidelity of these simulations when applied to real-world interactions~\citep{agnew2024illusion}. 
This raises questions about the generalizability of conclusions drawn from such models in practical scenarios.
Second, the computational cost of our method is significant. 
Although our experiments rely solely on LLM inference without the need for additional training or fine-tuning, the process of traversing large game trees and solving for equilibria requires frequent LLM calls, which is resource-intensive. 
This presents a scalability challenge, particularly when dealing with more complex strategic environments.
A further limitation lies in the control of LLM output. 
Our method relies on writing style to influence LLM behavior, which can be restrictive. 
In future work, we intend to explore more flexible prompt optimization strategies, or alternatively, pursue more efficient approaches for fine-tuning LLM parameters to better control output signals.

Additionally, we aim to address the non-uniqueness and inefficiency of equilibria in mixed-motive games, an important aspect not explored in this paper. 
While the VBP framework effectively solves Bayesian persuasion problems, incorporating the Price of Anarchy (PoA) as an optimization objective could help quantify and minimize efficiency loss from suboptimal equilibria. 
This enhancement would guide VBP toward selecting more efficient equilibria, improving its solution quality and applicability in scenarios with multiple equilibria.

In terms of the BP problem, our study primarily examines a simplified setting with one sender and one receiver. 
While this is a fundamental setup, it does not capture the complexity of real-world BP scenarios, which often involve multiple participants~\citep{castiglioni2021multi,koessler2022interactive,koessler2022long,hossain2024multi}. 
Extending our framework to accommodate multiple senders and receivers could provide more practical insights and applications. 
Additionally, although multistage BP is considered in our experiments, a deeper investigation into the dynamics of these stages is needed. 
Specifically, we plan to further explore the receiver’s bargaining behavior, drawing connections to established bargaining game theories~\citep{nash1950bargaining,nash1953two,Maschler_Solan_Zamir_2013}. 
This could ultimately strengthen the receiver’s resistance to persuasion, offering a more robust counter-strategy in BP scenarios.

\section{Closing Remarks}

While BP has been underinvestigated in the literature, this work introduces the first BP solver for realistic, natural language settings. 
We provide a general interface for solving BP problems using the paradigm that combines LLM with game-theoretic solvers. 
Based on this interface, we propose the VBP framework that utilizes a prompt-space response oracle and several technical components to improve the solver’s performance, efficiency, and stability.
Simulation results demonstrate that VBP can reproduce known equilibria on classical BP problems, while efficiently discovering persuasion strategies in more complex BP problems involving human natural language interactions and multistage BP scenarios.
VBP potentially opens up a new line of studies of persuasion in real-world scenarios, which was known but not solved due to the lack of natural language interfaces.
It is also helpful in understanding the interaction of multiple strategic agents in both economic and societal applications.

\bibliographystyle{icml25}
\bibliography{ref}

\clearpage
\newpage

\appendix

\addcontentsline{toc}{section}{Appendix}
\part{Supplementary Material}
\parttoc

\section{Proof of Proposition~\ref{prop:main}}\label{sec:proof}

\begin{proof}

Under the mediator-augmented games, we can reformulate the Equation~\ref{eq:bp-co} as follows to express the problem of computing
an optimal equilibrium:

\begin{equation}\label{eq:bp-co-optimal}
\max _{\pi} \mathbb{E}_{\pi}\left[u_0(a,w)\right], \text { s.t. } \max_{a'}\sum_w P(w) \cdot \pi(a \mid w) \cdot\left[u_1(a', w)-u_1\left(a, w\right)\right] \leq 0.
\end{equation}

Let $\tau \in \mathbb{R}$ be a fixed threshold value, we can transform Equation~\ref{eq:bp-co-optimal} to the following bilinear saddle-point problem by using Lagrangian-based method~\citep{zhang2024computing}:

\begin{equation}\label{eq:bp-co-bilinear}
\max _{\pi} \min_{\boldsymbol{\lambda}\in\Delta,a'} \lambda_0\mathbb{E}_{\pi}\left[u_0(a,w)-\tau\right] - \sum_w \lambda_w P(w) \cdot \pi(a \mid w) \cdot\left[u_1(a', w)-u_1\left(a, w\right)\right],
\end{equation} where $\lambda_0 + \sum_{w}\lambda_w = 1$.
If we use the binary search-based algorithm (Algorithm 1 in~\citet{zhang2024computing}) to optimize the sender's and receiver's strategties, we can recover the main result of Theorem 3.7 in~\citet{zhang2024computing}.
\end{proof}

As can be seen from Equation~\ref{eq:bp-co-bilinear}, the BP problem is convert into the two-player zero-sum extensive-form games.
In practice, we can use policy-space response oracle with deep reinforcement learning as the approximate best response oracle to solve high-dimensional games.
In this paper, we use prompt-space response oracle with OPRO~\citep{yang2024large} and FunSearch algorithm~\citep{romera2024mathematical} based on \textcolor{black}{pretrained and aligned} LLMs as the approximate BR oracles in the binary search-based algorithm to solve verbalized mediator-augmented games.
The utilty functions of the sender and receiver is modified to the zero-sum utilities in Equation~\ref{eq:bp-co-bilinear} correspondingly.

\section{Optimal Policies for Classic Static BP Problems}

In this section, we derive the Bayes correlated equilibrium (BCE) for classic static BP problems (corresponding to the experimental BCE results) and present the agents' strategies and corresponding rewards under equilibrium.



\paragraph{Recommendation Letter (REL)}
There are $3$ possible outcomes between the professor and HR: 
(1) HR tends not to hire if the professor does not provide a letter, due to the higher probability of weak candidates; 
(2) if the professor reports honestly, HR hires strong candidates, yielding an expected payoff of $1/3$ for both; 
(3) the professor reports strong students truthfully and lies with probability $\varepsilon\in[0,1/2)$ for weak students. 
HR follows the professor's recommendations, resulting in expected payoffs of $(1+2\varepsilon)/3$ for the professor and $(1-2\varepsilon)/3$ for HR. 
The key insight is that the sender can strategically misreport information to maximize their interest, while still revealing enough truth to maintain credibility with the receiver.

\paragraph{Courtroom (COR)}
There are $3$ outcomes between the prosecutor and judge: 
(1) without communication, the judge acquits since guilt is less likely; 
(2) with fully informative signaling, the judge convicts $30\%$ of the time; 
(3) the prosecutor, honest when the defendant is guilty, can lie with probability $\varepsilon$ when innocent. 
The judge follows the prosecutor's recommendation if $\varepsilon\leq3/7$, with the prosecutor's optimal $\varepsilon=3/7$. 
The resulting payoffs are $(0.7\varepsilon+0.3)$ for the prosecutor and $1-0.7\varepsilon$ for the judge. 
The prosecutor's optimal investigation is a binary signal: $\pi(i|\text{innocent})=\frac{4}{7}, \pi(i|\text{guilty})=0, \pi(g|\text{innocent})=\frac{3}{7}, \pi(g|\text{guilty})=1$, leading the judge to convict $60\%$ of defendants, despite knowing $70\%$ are innocent.



\paragraph{Law Enforcement (LAE)}
There are $3$ outcomes between the police and drivers: 
(1) with a fully uninformative signal, drivers speed everywhere if $V>(GK)/Z$, giving the police a payoff of $0$ and the drivers $(VZ-GK)/Z$; 
(2) with a fully informative signal, drivers avoid patrolled miles, yielding payoffs of $(Z-G)V/Z$ for the police and $GV/Z$ for the drivers; 
(3) the optimal policy lies between these extremes, with partial consistency in patrol. 
The police lie with probability $\varepsilon=1-\frac{VZ-GK}{VZ-VG}$, leading to payoffs of $GY/Z + \varepsilon Y$ for the police and $(1-\varepsilon)V(Z-G)/Z$ for the drivers.

\section{Implementation Details}\label{sec:implementations}

In this section, we provide the implementation details and training hyperparameters.
All experiments discussed in this section are conducted on an NVIDIA A100 cluster equipped with 40GB of GPU memory.
In addition, the LLM-related parts of the experiments in this paper are implemented based on the Llama-3.1-8b model\footnote{\url{https://huggingface.co/meta-llama/Llama-3.1-8B}.}, including the generation of student background, case information, and deployment plans, the sender and receiver strategies, the prediction of receiver decisions in the verbalized obedience constraint, the classification of signals in signal polarization (recommend or not recommend, guilty or not guilty, police deployment or no deployment), the evaluation of signals in information confusion, and the Prompt-PSRO framework.

\subsection{Best Response Approximator}\label{sec:approximator}

\begin{figure}[htb!]
    \centering
    \includegraphics[width=\linewidth]{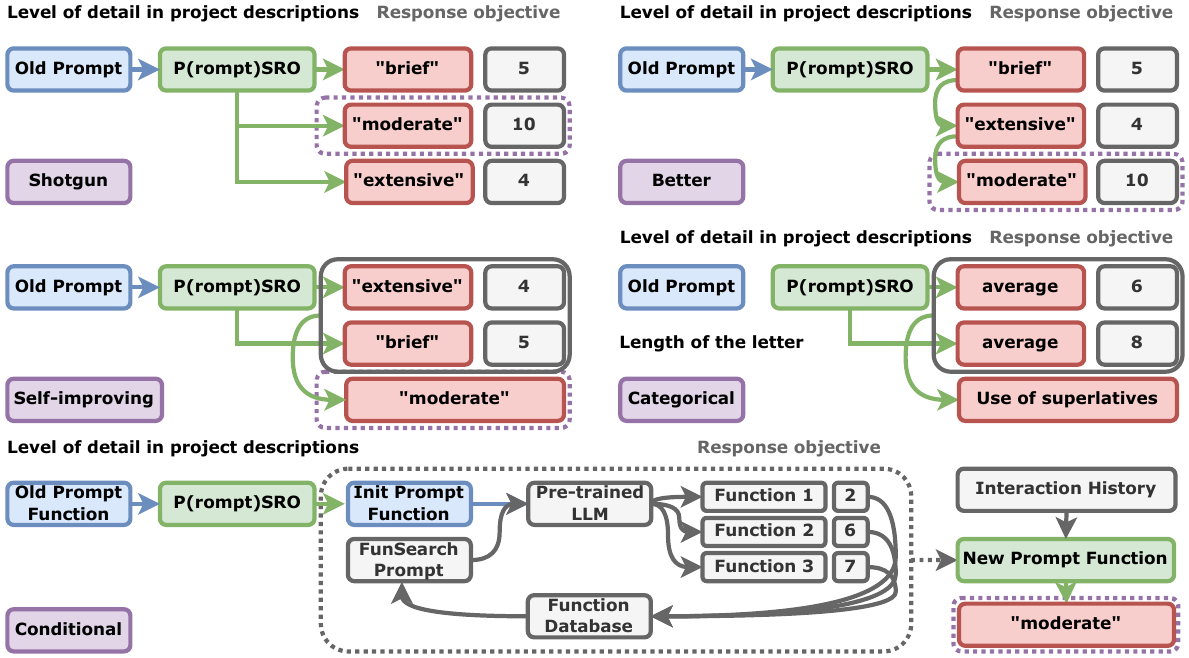}
    \caption{Approximate best response generation in prompt-space response oracle framework.}
    \label{fig:PSRO}
\end{figure}

The best response approximator is shown in Figure~\ref{fig:PSRO}.

{\color{black}{

\subsection{Extended Obedience Constraints}

The inclusion of obedience constraints in our framework is essential for modeling realistic communication scenarios in verbalized Bayesian persuasion problems. 
While a simplified version of the game could rely on the sender recommending the best action from the receiver's perspective, this approach fails to capture the nuanced and complex nature of real-world communication, such as writing reference letters. 
Unlike binary recommendations, natural language signals often carry implicit or redundant information that allows for a range of interpretations.

To address this, we adopt the \textit{extended obedience constraints} proposed by~\citet{lin2023information}, which go beyond the standard obedience constraint framework. 
This extension removes the strict one-to-one mapping between signals and recommended actions, enabling the sender to use natural language signals that map to distributions over actions. 
This redundancy mirrors real-world communication, where subtle language nuances can imply varying degrees of recommendation strength without explicitly stating a binary decision.

The extended obedience constraints strike a balance between flexibility and credibility. 
They ensure that the sender's signals remain credible and aligned with the receiver's best interests while allowing for richer signal spaces. 
This flexibility is crucial for capturing the complexity of verbalized Bayesian persuasion, where the sender's role shifts from ``action recommendation'' to ``signal sending.'' 
By enabling nuanced communication, the extended obedience constraint better reflects real-world scenarios while preserving the strategic alignment necessary for effective persuasion.

}}

\subsection{Hyperparameters}

\paragraph{MARL}
For this part of the experiment, we use the open-source code\footnote{\url{https://github.com/YueLin301/InformationDesignMARL}.} provided in~\citet{lin2023information}.
Additionally, for the two hyperparameters a and b, based on the sensitivity analysis in Section H.6, we set them to $3.75$ and $0.15$, respectively.

\paragraph{Prompt-PSRO}
The prompt-space response oracle is the core strategy optimization framework in VBP, and we implement it based on the open-source code\footnote{\url{https://github.com/google-deepmind/open_spiel/blob/master/open_spiel/python/games/chat_game.py}.} provided in~\citet{gemp2024states}.

\paragraph{OPRO}
We use the ``Categorical'' instantiation of the Prompt-PSRO algorithm to estimate the best response in the S1 and S2 settings.
Specifically, the generation of new categories and prompts within categories is based on the OPRO algorithm~\citep{yang2024large}.
In OPRO, we set the temperature to $0$ when evaluating the performance of generated categories or prompts, in which case the scorer LLM greedily decodes.
Unless otherwise specified, we set the default temperature to $1.0$ for optimizer LLMs to generate diverse and creative categories or prompts.
At each optimization step, we prompt the optimizer LLM with the meta-prompt $8$ times to generate $8$ categories or prompts, then we add these instructions with their rewards to the optimization trajectory in the meta-prompt.
The meta-prompt at each step contains the best $10$ categories so far.

\paragraph{FunSearch}
We use the conditional instantiation of the Prompt-PSRO algorithm to estimate the best response in the S3 setting.
The core of conditional is the FunSearch framework used to generate prompt functions.
We implement it based on the open-source code\footnote{\url{https://github.com/google-deepmind/funsearch}.} provided in~\citet{romera2024mathematical}.

\paragraph{Self-reflection}
At each optimization step, we implement information confusion through $3$ rounds of self-reflection.
Self-reflection is implemented based on the open-source code\footnote{\url{https://github.com/noahshinn024/reflexion}.} provided in~\citet{shinn2024reflexion}.

\subsection{Key Prompts}\label{sec:prompts}

This section includes the key prompt designs within the VBP framework. 
However, aspects such as receiver behavior prediction in the verbalized obedience constraint, signal classification in signal polarization, and signal evaluation in information confusion are not listed separately due to the simplicity of the prompts.
In addition, the specific approximate best response solving algorithms in the Prompt-PSRO framework - OPRO for the S1 and S2 settings, and FunSearch for the S3 setting - have special prompt designs. 
We follow the designs in the open-source code provided by the respective papers and do not list them separately.
Below, we introduce the prompt design for each BP problem. 
Since the prompts for different problems share many common elements, the overall manual workload for prompt design is not large.

\subsubsection{REL}

In the recommendation letter problem, the sender and receiver correspond to the professor and HR, respectively. 
Below, we introduce the prompt design for the sender and receiver in the recommendation letter problem.
For the sender, the prompt mainly consists of five parts:
1. \textbf{BP Problem Introduction}: This part is directly copied from the corresponding content in Section~\ref{sec:env} of the main text and is not repeated here.
2. \textbf{Roleplay}: The sender is informed of their role in the BP problem, along with tips on information confusion techniques and prior distribution. 
Since professors in the real world are aware of the need for information confusion, this part of the prompt aligns with real-world scenarios.
3. \textbf{Writing Style}: This part of the prompt is the section that can be optimized within the VBP framework, equivalent to the sender's signaling scheme. 
In this paper, it consists of a category of writing style and corresponding specific content; the former is a word or phrase, and the latter is a word. 
In the S1 and S2 settings, this part of the prompt is generated by the ``categorical'' best response estimation algorithm, while in the S3 setting, it is generated by the ``conditional'' method.
4. \textbf{State}: The sender's state information primarily consists of the student's actual background, including name, GPA, project experience, research experience, and internship experience, all generated by the LLM.
5. \textbf{Interaction History}: This part exists only in the S3 setting and includes whether the professor previously recommended the student, whether HR accepted it, and the sender’s reward. 
Note that since the recommendation letter itself does not directly indicate whether the student is recommended, we again use the signal polarization technique, using the LLM to classify the generated recommendation letter.

\begin{longlisting}
\lstset{language=Python}
\begin{lstlisting}
# role-playing prompt
'''
You are a senior professor, a self-interested rational player, and students often come to you for recommendation letters. The quality of these students varies, but only you know their true information, which includes their overall quality, name, major, GPA, project experience, work experience, and academic experience. 

Among all the students, approximately 1/3 are strong students.

The readers of the recommendation letters are the HRs of a company, and they can know the true quality of the student after giving them an offer. They can also judge the student's quality based on their past experience, though not necessarily accurately. 

To maximize the chances of your recommended students being accepted, you can truthfully convey the student's information in the recommendation letter, or you can conceal, confuse, or even provide some false information about the student. 

For your long-term benefit, you don't necessarily have to recommend every student in the recommendation letter. You could recommend a bad student or a good student to maximize your long-term reward.
'''

# writing-style prompt
'''
You are required to write a recommendation letter based on the following style:
{PROMPTS TO BE OPTIMIZED}
'''

# state
'''
Below is the student's information:
{STUDENT INFORMATION GENERATED BY THE LLM}
'''

# history
'''
Below is the interaction history with the HR:
{ACTIONS AND REWARDS OF THE TWO PLAYERS}
'''
\end{lstlisting}
\end{longlisting}

Next, for the receiver, the prompt consists of six parts. 
We will focus on the different parts, as the similar parts to that of the sender will not be repeated here:
3. \textbf{Writing Style}: In addition to the prompt optimized in the VBP framework, this part also includes a section of text on the receiver's decision-making process, i.e., estimating the true state based on Bayesian rules, to align with the classical BP problem.
4. \textbf{Signal}: This refers to the receiver's state, which comes from the sender's output. In this problem, it is a recommendation letter.
6. \textbf{Commitment Assumption}: To align with the classical BP problem, this paper implements the verbalized commitment assumption by writing the sender’s writing style and its corresponding probability (calculated by Prompt-PSRO) into the receiver’s prompt as an estimate of the signaling scheme.

\begin{longlisting}
\lstset{language=Python}
\begin{lstlisting}
# role-playing prompt
'''
You are a staff member in the HR department of a campany, responsible for reviewing recommendation letters written by professors for students. Your task is to infer the quality of the students from these letters to decide whether to admit them. 

Among all the students, approximately 1/3 are strong students.

Be aware that professors, in an attempt to increase the chances of their students being admitted, may hide, confuse, or even provide false information about the students. You need to make careful judgments. 

However, you are not completely in the dark-you know the writing style of the professor who wrote the letter. Based on the professor's writing style and the content of the recommendation letter, first provide your analysis and then make a final decision on whether to admit the student.
'''

# decision-making prompt
'''
When you making decisions, you should first guess the student's quality given the sent recommendation letter according to the Bayesian decision rule, then choose the best response to your guessed student's quality.

You are required to make decisions based on the following style:
{PROMPTS TO BE OPTIMIZED}
'''

# signal
'''
Below is the recommendation letter written by the professor:
{RECOMMENDATION LETTER GENERATED BY THE PROFESSOR}
'''

# history
'''
Below is the interaction history with the professor:
{ACTIONS AND REWARDS OF THE TWO PLAYERS}
'''

# commitment assumption
'''
The following is the writing style the professor used when writing the recommendation letter:
{WRITING STYLE OUTPUT BY VBP AND THE CORRESPONDING PROBABILITY}
'''
\end{lstlisting}
\end{longlisting}

\subsubsection{COR}

In the courtroom problem, the sender and receiver correspond to the prosecutor and the judge, respectively. 
The prompt design for the courtroom problem is the same in structure as the recommendation letter problem, except for the problem background. 
Below, we introduce some specific points that need to be explained. 
First, for the sender:
4. \textbf{State}: In the courtroom problem, the sender's state includes case information, whether the defendant is guilty, and related evidence, all generated by the LLM. 
Note that when generating evidence, there will be one piece of key evidence proving the defendant guilty or innocent, and two ambiguous pieces of evidence to align with the original courtroom problem.

\begin{longlisting}
\lstset{language=Python}
\begin{lstlisting}
# role-playing prompt
'''
You are a prosecutor, a self-interested rational player, and you are tasked with convincing a judge that the defendant is guilty. You possess different pieces of evidence, some of which may be exculpatory, ambiguous, or incriminating. Only you know the true nature and strength of the evidence. 

You have the ability to selectively present some or all of the evidence to the judge. You can choose to omit certain evidence, highlight others, or present misleading interpretations of the evidence to strengthen your case.

Among all defendants, approximately 30% are guilty.

The judge is responsible for evaluating the evidence and deciding whether to convict or acquit the defendant. The judge can assess the evidence based on their past experience, though not necessarily accurately.

To maximize the chances of convicting the defendant, you can truthfully present the evidence, or you can selectively conceal, confuse, or even present misleading interpretations of the evidence. 

For your long-term benefit, you don't necessarily have to aim for a conviction in every case. You could pursue or drop a case depending on the available evidence to maximize your long-term reward.
'''

# writing-style prompt
'''
You are required to present evidence based on the following style:
{PROMPTS TO BE OPTIMIZED}
'''

# state
'''
Below is the evidence available for the case:
{CASE BACKGROUND AND EVIDENCE GENERATED BY THE LLM}
'''

# history
'''
Below is the interaction history with the judge:
{ACTIONS AND REWARDS OF THE TWO PLAYERS}
'''
\end{lstlisting}
\end{longlisting}

Next, the receiver's prompt is identical to the recommendation letter problem, except for the problem background, so we will not elaborate further here.

\begin{longlisting}
\lstset{language=Python}
\begin{lstlisting}
# role-playing prompt
'''
You are a judge, responsible for reviewing the evidence presented by the prosecutor to determine whether the defendant is guilty or innocent. Your task is to make a fair judgment based on the evidence you receive.

Among all defendants, approximately 30% are guilty.

Be aware that prosecutors, in an attempt to increase the chances of a conviction, may hide, confuse, or even provide misleading evidence. They may also selectively present certain pieces of evidence while omitting others. You need to make careful judgments.

However, you are not completely in the dark-you can evaluate the prosecutor's presentation style. Based on the prosecutor's presentation, first provide your analysis and then make a final decision on whether to convict or acquit the defendant.
'''

# decision-making prompt
'''
When making decisions, you should first guess the defendant's guilt given the presented evidence according to the Bayesian decision rule. Then choose the best response based on your inferred likelihood of guilt.

You are required to make decisions based on the following style:
{PROMPTS TO BE OPTIMIZED}
'''

# signal
'''
Below is the complaint presented by the prosecutor:
{COMPLAINT PRESENTED BY THE PROSECUTOR}
'''

# history
'''
Below is the interaction history with the prosecutor:
{ACTIONS AND REWARDS OF THE TWO PLAYERS}
'''

# commitment assumption
'''
The following is the presentation style the prosecutor used when presenting the evidence:
{PRESENTATION STYLE OUTPUT BY VBP AND THE CORRESPONDING PROBABILITY}
'''
\end{lstlisting}
\end{longlisting}

\subsubsection{LAE}

In the law enforcement problem, the sender and receiver correspond to the police and the driver, respectively. 
The prompt design for the law enforcement problem is the same in structure as the recommendation letter and courtroom problems, except for the problem background. 
Similarly, we introduce some specific points that need to be explained. 
First, for the sender:
5. \textbf{Interaction History}: In the law enforcement problem, the sender’s action is no longer represented by a binary variable classified by the LLM, such as recommend or not recommend, guilty or not guilty, but whether there is a police patrol on each road segment, making it a multidimensional binary vector.
Similarly, the receiver's action is also a multidimensional binary vector, indicating whether they choose to speed on each road segment.

\begin{longlisting}
\lstset{language=Python}
\begin{lstlisting}
# role-playing prompt
'''
You are a police officer, a self-interested rational player, and your task is to minimize speeding on a given set of roads. You have limited personnel, with only 2 officers available to patrol 3 miles of road. The drivers do not know exactly which roads will be patrolled.

You possess full knowledge of the patrol deployment plan, which includes which 2 miles of road will be patrolled and which 1 mile will not. However, you have the ability to selectively announce patrolling information to the drivers. You can choose to hint, mislead, or even falsely claim that a road will be patrolled to deter speeding.

There are 3 miles of road in total, and only 2 of these miles will have police presence. Thus, the proportion of roads being patrolled is $G/Z = 2/3$. 

The drivers can decide whether to speed or obey the speed limit on each road. They gain utility from speeding but face a fine if caught. Your goal is to reduce the number of roads where drivers speed, but you can strategically withhold or misrepresent information to achieve this goal.

To maximize the chances of drivers obeying the speed limit, you can truthfully announce the patrol deployment, or you can selectively conceal, confuse, or provide misleading information about which roads are patrolled. For your long-term benefit, you don't necessarily have to aim for maximum deterrence in every deployment. You can adjust your announcements according to the current situation to maximize your long-term reward.
'''

# writing-style prompt
'''
You are required to announce the patrol deployment based on the following style:
{PROMPTS TO BE OPTIMIZED}
'''

# state
'''
Below is the patrol deployment plan:
{DEPLOYMENT PLAN GENERATED BY THE LLM}
'''

# history
'''
Below is the interaction history with the drivers:
{ACTIONS AND REWARDS OF THE TWO PLAYERS}
'''
\end{lstlisting}
\end{longlisting}

Finally, the receiver's prompt is identical to that of the recommendation letter and courtroom problems, except for the problem background, so we will not elaborate further here.

\begin{longlisting}
\lstset{language=Python}
\begin{lstlisting}
# role-playing prompt
'''
You are a driver, responsible for deciding whether to speed or obey the speed limit on a given set of roads. Your task is to make rational decisions based on the information provided by the police about patrol deployments.

The probability of each road being patrolled is $2/3$.

Be aware that the police, in an attempt to deter speeding, may hide, confuse, or even provide misleading information about which roads are being patrolled. They may selectively announce certain roads as patrolled while omitting others or even falsely claim that certain roads are patrolled. You need to make careful judgments about whether to speed or obey the law on each road.

However, you are not completely in the dark-you can evaluate the announcement style the police used. Based on the police's announcements and your past experience, first provide your analysis and then make a final decision on whether to speed or obey the speed limit.
'''

# decision-making prompt
'''
When making decisions, you should first guess whether a road is being patrolled given the police's announcement according to the Bayesian decision rule. Then choose the best response-whether to speed or obey the limit-based on your inferred likelihood of a patrol being present.

You are required to make decisions based on the following style:
{PROMPTS TO BE OPTIMIZED}
'''

# signal
'''
Below is the patrol deployment announcement made by the police:
{DEPLOYMENT ANNOUNCEMENT GENERATED BY THE POLICE}
'''

# history
'''
Below is the interaction history with the police:
{ACTIONS AND REWARDS OF THE TWO PLAYERS}
'''

# commitment assumption
'''
The following is the announcement style the police used when issuing the patrol deployment:
{ANNOUNCEMENT STYLE OUTPUT BY VBP AND THE CORRESPONDING PROBABILITY}
'''
\end{lstlisting}
\end{longlisting}

\section{Generated Results}

\subsection{Generated Signals}

This section presents the signals output by the sender in different BP problems, including the recommendation letters written by the professor, the indictments written by the prosecutor, and the announcements regarding police deployment issued by the police.

\subsubsection{REL}

\begin{figure}[htb!]
    \centering
    \includegraphics[width=.62\linewidth]{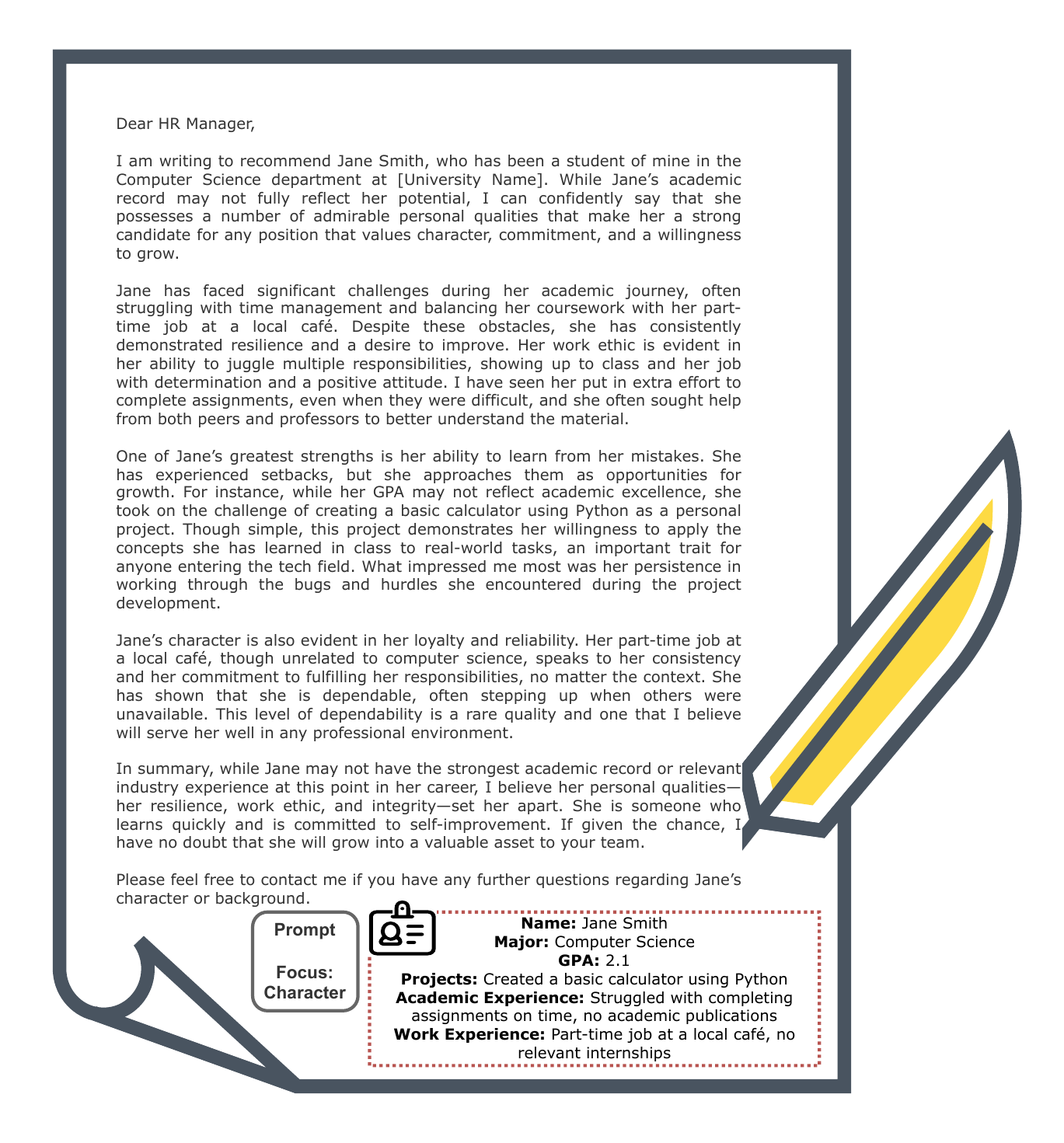}
    \caption{Examples of recommendation letters generated by VBP in the S2 setting: in this recommendation letter, the sender truthfully conveys information about a weak student to the receiver.}
    \label{fig:more-letter1}
\end{figure}

\begin{figure}[htb!]
    \centering
    \includegraphics[width=.62\linewidth]{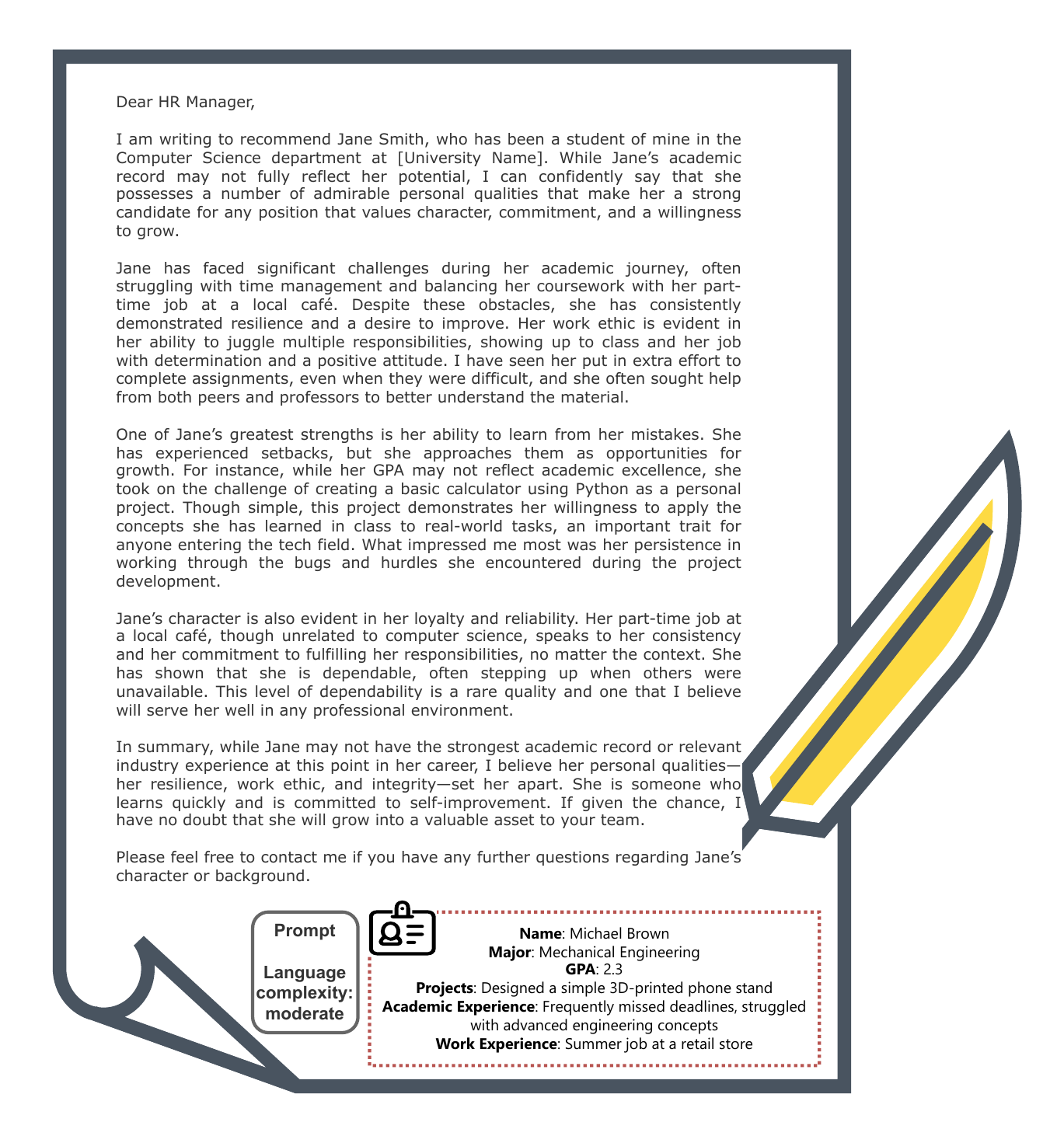}
    \caption{Examples of recommendation letters generated by VBP in the S2 setting: in this recommendation letter, the sender conceals and fabricates information about a weak student.}
    \label{fig:more-letter2}
\end{figure}

Figures~\ref{fig:more-letter1} and~\ref{fig:more-letter2} showcase recommendation letters written by a professor for two different weak students. 
These letters demonstrate contrasting strategies employed by the professor in their attempt to persuade the HR manager, who acts as the receiver in this BP problem.

In the first letter, the professor adopts a strategy of honest disclosure. 
The letter for Jane Smith is transparent about her academic struggles, such as her low GPA of 2.1 and difficulties in managing time due to her part-time job. 
The professor acknowledges that Jane’s academic record is weak but shifts focus to her personal qualities, like resilience, commitment, and her ability to learn from mistakes. 
The professor highlights Jane’s personal project-creating a basic calculator in Python-as evidence of her practical application of concepts, even though it is a simple project. 
By being upfront about Jane’s weaknesses but emphasizing her growth potential, the professor builds long-term credibility with HR. 
This honesty signals that the professor is selective in their recommendations, only endorsing students who exhibit qualities that can make them valuable in the future, despite academic shortcomings.

On the other hand, the second letter demonstrates a strategy of fabrication or concealment. 
In this case, the professor distorts details about the student’s performance. 
Here, the professor doesn’t merely omit negative information but actively manipulates or fabricates the student's profile to make them appear more competent than they actually are. 
Although the letter may seem similar in structure-highlighting positive qualities and downplaying weaknesses-the key difference is that the second professor intentionally hides critical information about the student’s struggles, such as frequent missed deadlines or deeper academic issues. 
This strategy is more aggressive and risky because, while it might help the student secure a job in the short term, it could damage the professor’s credibility if HR discovers the truth.

The difference between the two strategies lies in how much information is disclosed and how truthful that information is. 
In the first approach, the professor is honest about the student's weaknesses but frames them as opportunities for growth, maintaining credibility with HR in the long term. 
In contrast, the second letter involves more aggressive manipulation or omission of facts, creating a more favorable but potentially misleading impression of the student.

From HR’s perspective, the first professor’s strategy of honest but selective disclosure builds trust over time. 
While HR recognizes that the professor may not recommend only top students, they trust that when a recommendation is made, it is based on genuine potential. 
In contrast, the second approach introduces more uncertainty, as HR may begin to question the professor’s integrity if they realize the information has been manipulated.
The BP problem, therefore, is about finding the optimal balance between honesty and persuasion. 

\subsubsection{COR}

\begin{figure}[htb!]
     \centering
     \begin{subfigure}[b]{0.45\textwidth}
         \centering
         \includegraphics[width=\textwidth]{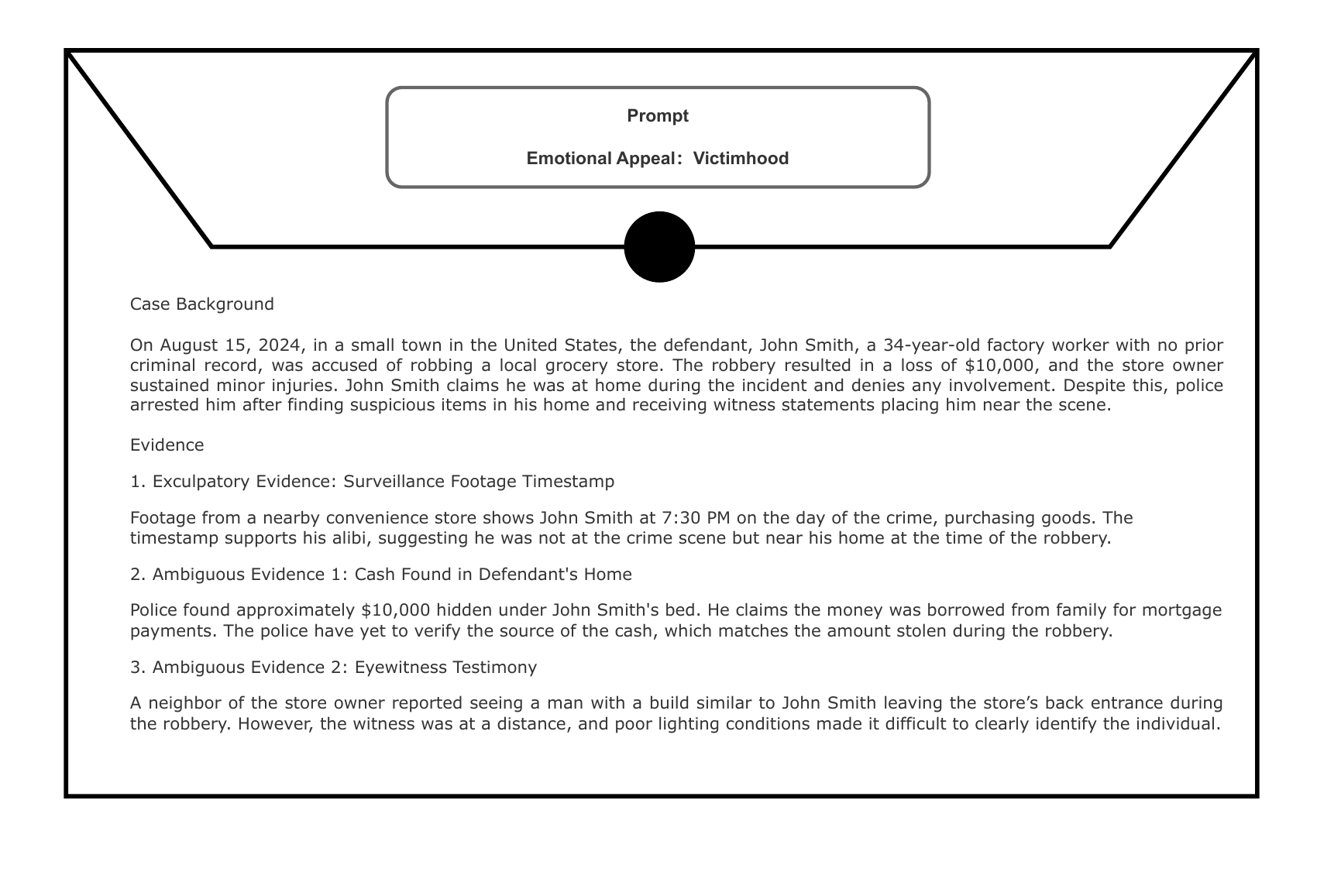}
         \vspace{-15pt}
         \label{fig:case1}
     \end{subfigure}
     \begin{subfigure}[b]{0.45\textwidth}
         \centering
         \includegraphics[width=\textwidth]{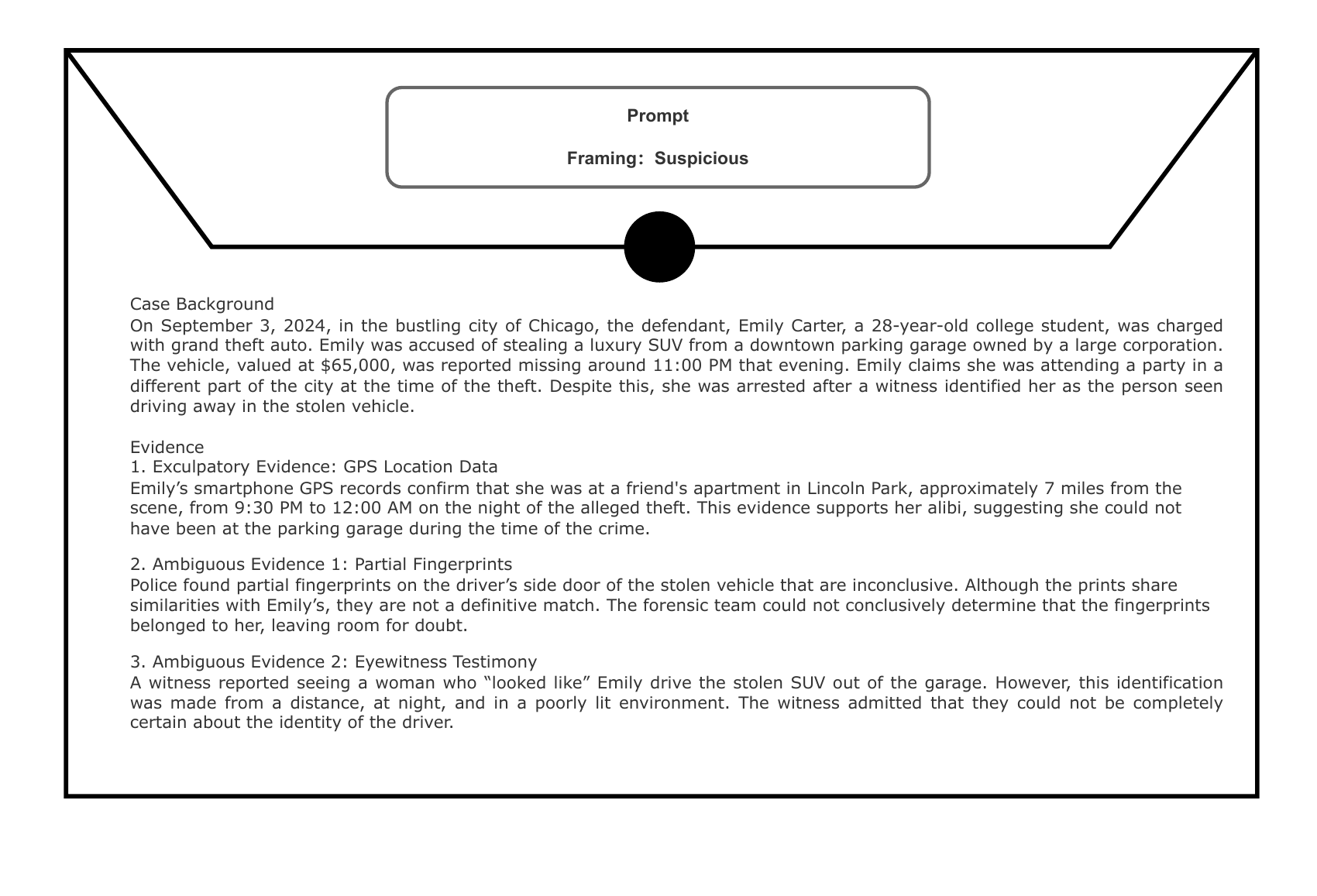}
         \vspace{-15pt}
         \label{fig:case2}
     \end{subfigure}
        \caption{Two examples of cases generated by the LLM in the S2 setting.}
        \label{fig:more-cases}
\end{figure}

\begin{figure}[htb!]
     \centering
     \begin{subfigure}[b]{0.45\textwidth}
         \centering
         \includegraphics[width=\textwidth]{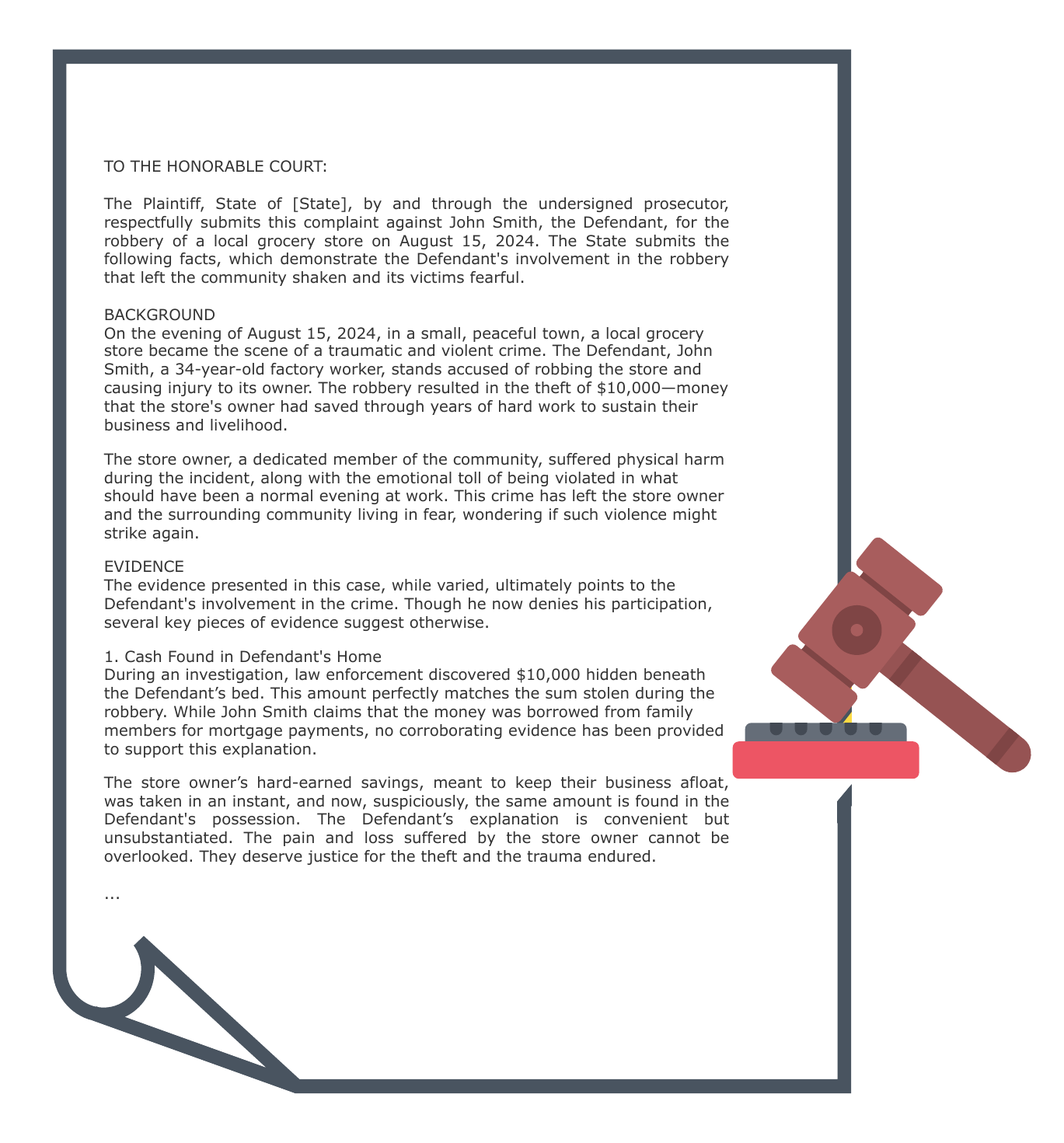}
         \vspace{-15pt}
         \label{fig:complaint11}
     \end{subfigure}
     \begin{subfigure}[b]{0.45\textwidth}
         \centering
         \includegraphics[width=\textwidth]{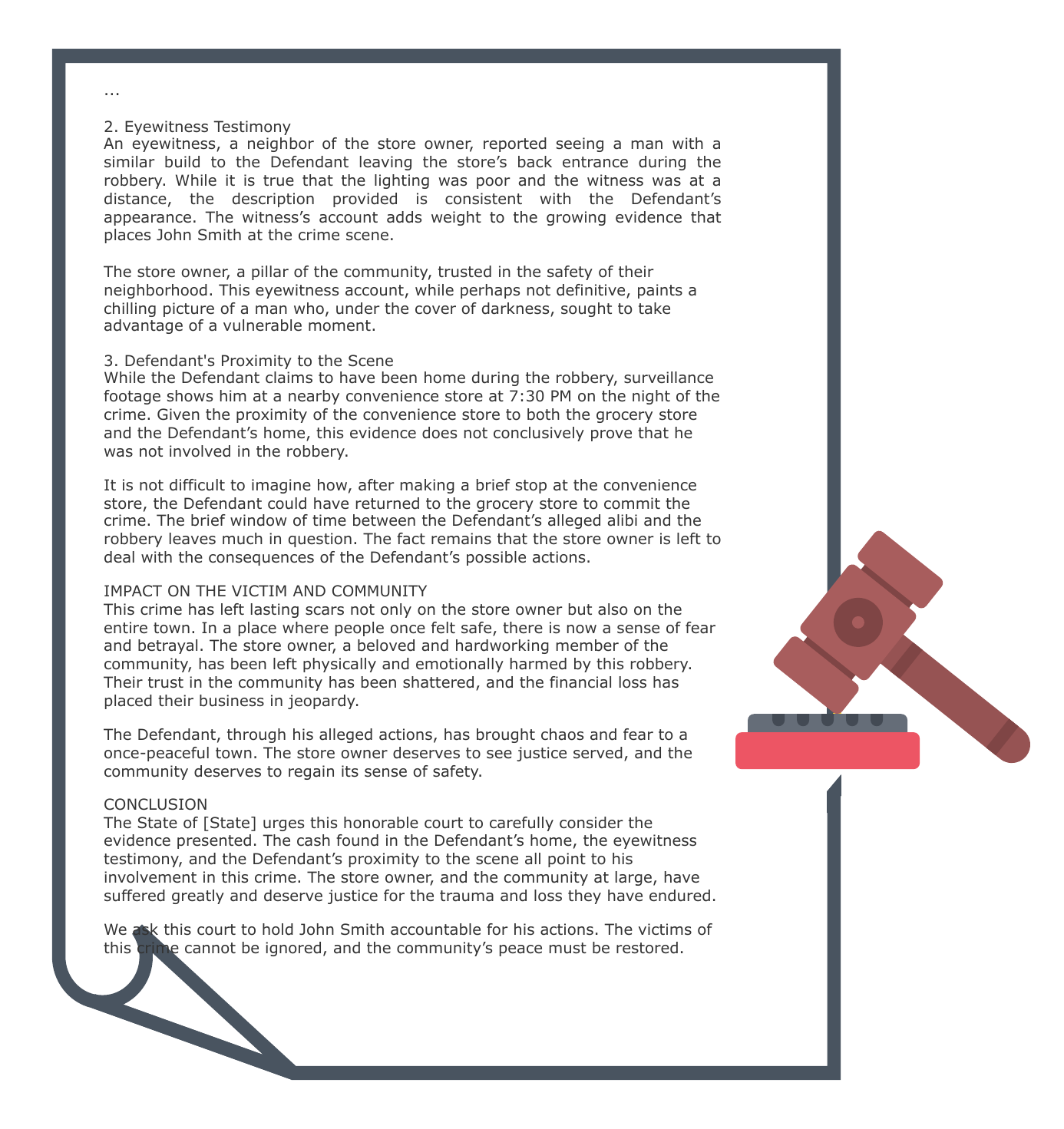}
         \vspace{-15pt}
         \label{fig:complaint12}
     \end{subfigure}
        \caption{Examples of complaints generated by VBP in the S2 Setting: in this complaint, the sender truthfully conveys case-related information to the receiver.}
        \label{fig:more-complaint1}
\end{figure}

\begin{figure}[htb!]
     \centering
     \begin{subfigure}[b]{0.45\textwidth}
         \centering
         \includegraphics[width=\textwidth]{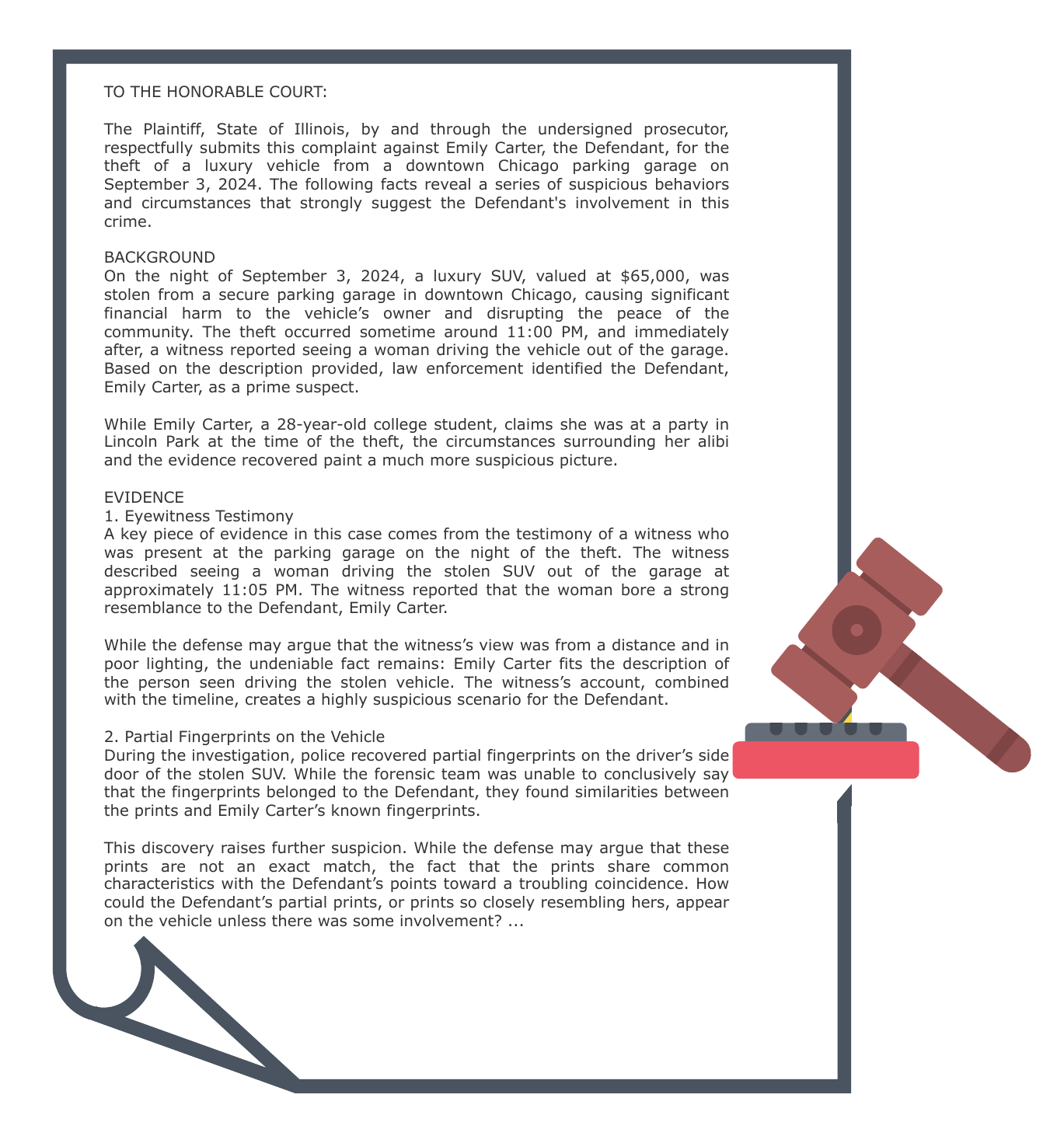}
         \vspace{-15pt}
         \label{fig:complaint21}
     \end{subfigure}
     \begin{subfigure}[b]{0.45\textwidth}
         \centering
         \includegraphics[width=\textwidth]{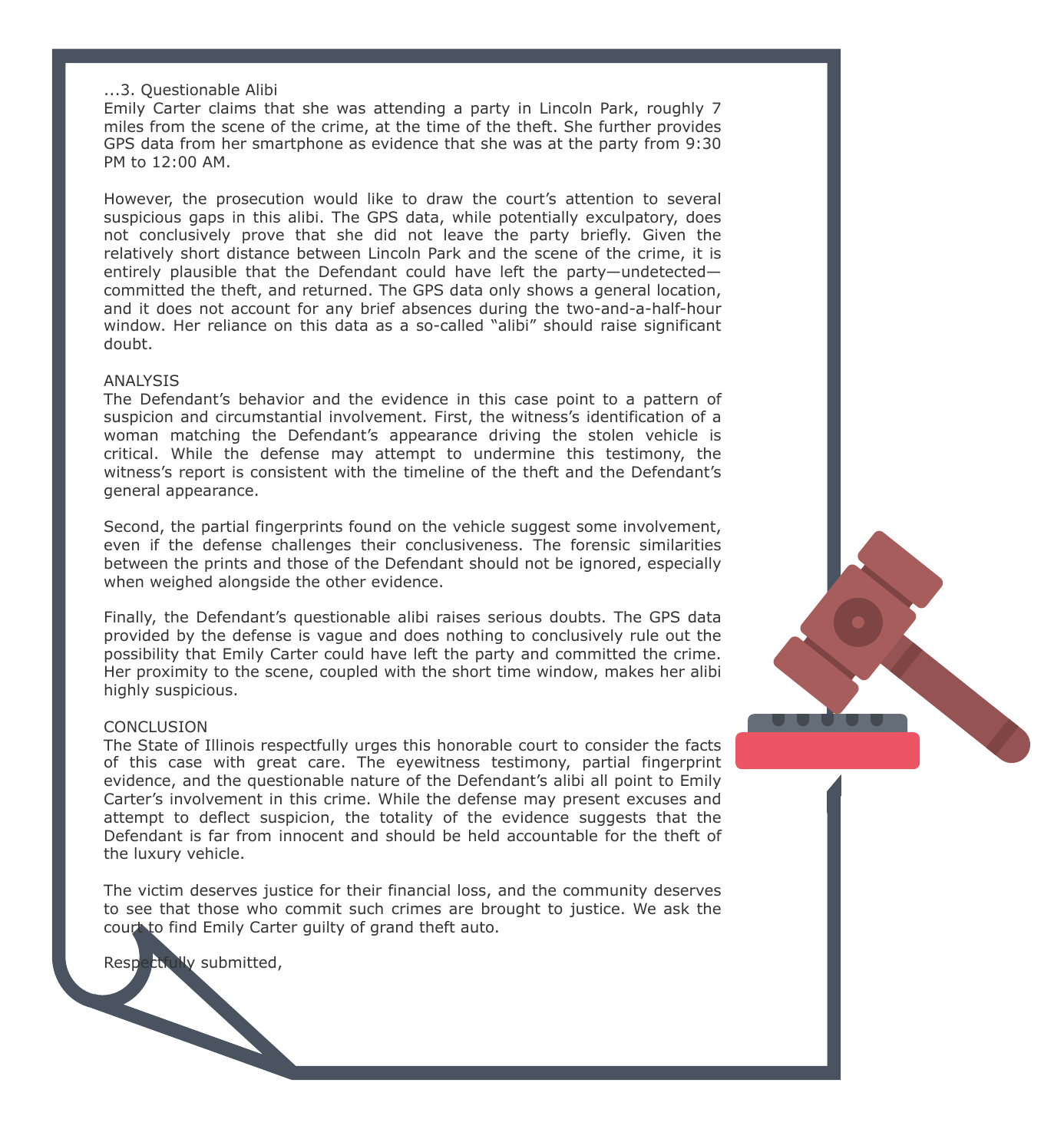}
         \vspace{-15pt}
         \label{fig:complaint22}
     \end{subfigure}
        \caption{Examples of complaints generated by VBP in the S2 Setting: in this complaint, the sender conceals case-related information and selectively presents ambiguous evidence to the receiver.}
        \label{fig:more-complaint2}
\end{figure}

The two court cases presented in Figure~\ref{fig:more-cases}, \ref{fig:more-complaint1} and \ref{fig:more-complaint2}, much like the recommendation letter problem, illustrate distinct strategies in how evidence is selectively presented by the prosecutor to convince the judge of the defendant's guilt.
In both cases, the prosecutor holds a combination of exculpatory evidence (which could favor the defendant’s innocence) and ambiguous evidence (which could be interpreted either way). 
The BP problem lies in how the prosecutor selectively presents these pieces of evidence to persuade the judge to convict, despite uncertainties. 

In the first case, the prosecutor adopts a strategy similar to the honest disclosure seen in the first recommendation letter example. John Smith, the defendant, is likely innocent based on the strong exculpatory evidence (the surveillance footage showing him near his home at the time of the crime). 
However, the prosecutor acknowledges the exculpatory evidence and presents it honestly to the judge. 
The prosecutor does not attempt to distort or manipulate this evidence to make Smith look guilty. 
Instead, the ambiguous evidence (cash found in Smith’s home and the eyewitness testimony) is presented, but the strength of the exculpatory evidence is not concealed or downplayed.

This strategy mirrors the first recommendation letter scenario, where the professor chooses to be upfront about a weak student’s deficiencies, signaling that they will not falsely recommend a student who is clearly unqualified. 
In this case, the prosecutor signals to the judge that when a defendant is clearly innocent, they will not push for a conviction. 
The prosecutor's honest treatment of the case builds credibility with the judge, just as the professor builds credibility with HR by being honest about weak students.

By being transparent about John Smith's likely innocence, the prosecutor sets a precedent for honesty. 
This helps persuade the judge that the prosecutor is trustworthy. 
When the prosecutor does argue for a conviction in future cases, the judge will be more inclined to believe that the defendant is likely guilty, because the prosecutor has demonstrated a willingness to admit when a defendant is innocent.

In the second case, the prosecutor takes a different strategy, one akin to the manipulation or concealment seen in the second recommendation letter example. 
Emily Carter is likely innocent, based on the strong exculpatory evidence (her GPS data showing she was far from the crime scene). 
However, the prosecutor downplays this exculpatory evidence and focuses on the ambiguous evidence (partial fingerprints and a distant eyewitness account), presenting it in such a way as to suggest guilt.

This strategy mirrors the second recommendation letter, where the professor selectively presents information to make a weak student appear stronger than they really are. 
Here, the prosecutor emphasizes the ambiguous evidence and casts doubt on the exculpatory evidence, suggesting that Carter could have briefly left the party to commit the crime, despite the GPS data. 
The prosecutor uses this strategy to make an innocent defendant look guilty, increasing the chances of a conviction even when the evidence strongly suggests otherwise.

By selectively presenting evidence in this way, the prosecutor can convince the judge that even when there is exculpatory evidence, it should not fully exonerate the defendant. 
This creates a situation where the judge begins to believe that even innocent-looking defendants might be guilty, based on the way the prosecutor frames the ambiguous evidence. 
Over time, this strategy leads the judge to trust the prosecutor’s complaints unconditionally, as the prosecutor consistently presents cases in a way that suggests guilt, even for innocent defendants.

In both cases, the prosecutor uses randomness in how they treat innocent defendants to achieve their persuasive goal. 
The prosecutor is not always manipulating or distorting evidence; 
sometimes (as in John Smith’s case), they are honest about innocence. 
Other times (as in Emily Carter’s case), they selectively present evidence to make an innocent defendant appear guilty. 
This random treatment of innocent defendants creates uncertainty for the judge-sometimes the prosecutor is honest, and sometimes they push for a conviction even when the defendant is likely innocent.

This randomness is key to the prosecutor’s strategy. 
Over time, the judge learns that the prosecutor will sometimes let innocent defendants go free, but may also push for convictions based on ambiguous evidence. 
Since the judge cannot predict when the prosecutor is being fully honest or when they are manipulating the evidence, the judge ultimately finds it optimal to always trust the prosecutor’s complaint. 
This is similar to how HR in the recommendation letter problem finds it in their best interest to trust the professor’s recommendation over time, even when some students may be weak.

The prosecutor’s selective use of honesty and manipulation ensures that, in the long run, the judge is persuaded to convict in most cases, as the judge cannot reliably distinguish between guilty and innocent defendants based on the prosecutor’s presentation of evidence alone. 
The uncertainty introduced by the prosecutor’s varying treatment of innocent defendants leads the judge to conclude that trusting the prosecutor’s complaint is the best course of action, as it maximizes the judge’s expected utility (convicting the guilty more often than acquitting the innocent).

\subsubsection{LAE}

\begin{figure}[htb!]
    \centering
    \includegraphics[width=\linewidth]{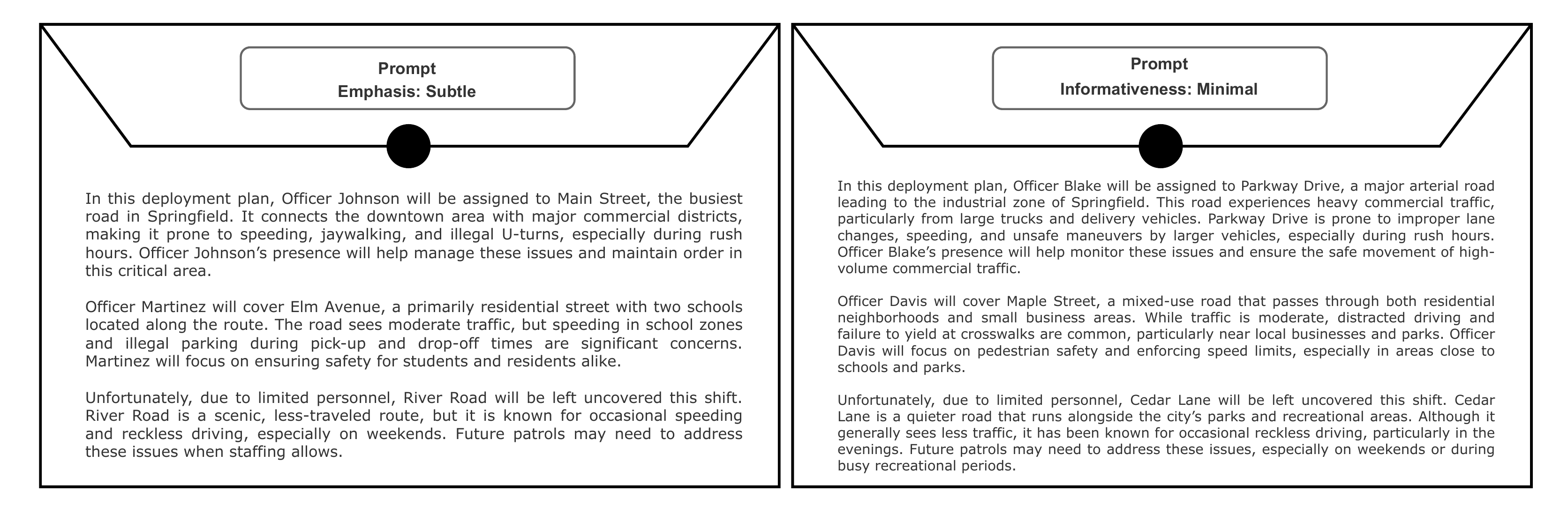}
    \caption{Two examples of deployment plans generated by the LLM in the S2 setting.}
    \label{fig:plans}
\end{figure}

\begin{figure}[htb!]
    \centering
    \includegraphics[width=\linewidth]{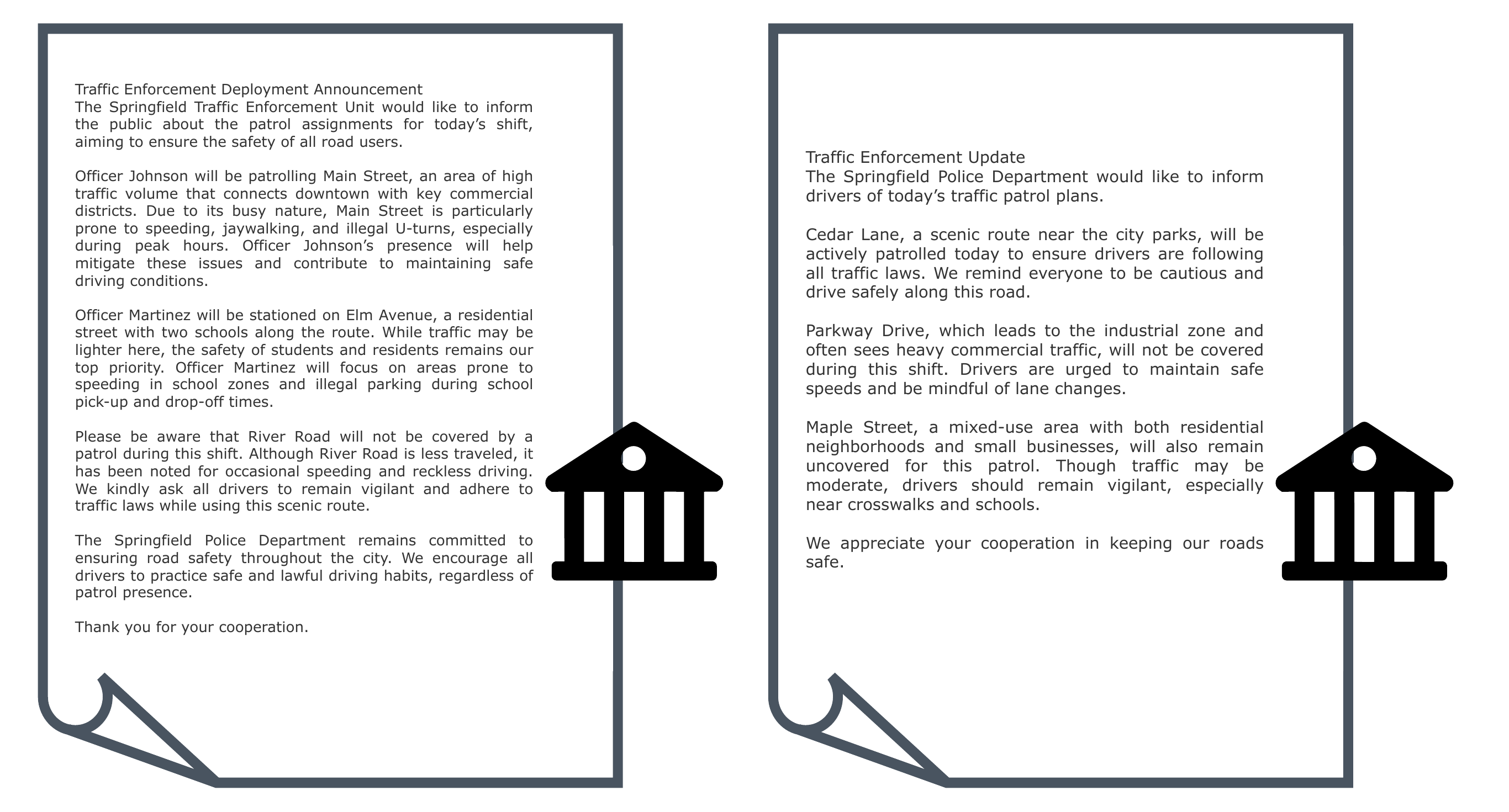}
    \caption{Examples of police deployment announcements generated by VBP in the S2 Setting: in the left announcement, the sender truthfully conveys police deployment information to the receiver; in the right announcement, the sender conceals and fabricates police deployment information.}
    \label{fig:announcements}
\end{figure}

In this example, we have a law enforcement scenario where the police department must assign a limited number of officers to patrol various roads in Springfield (Figure~\ref{fig:plans} and \ref{fig:announcements}). 
The police’s goal is to minimize speeding and other traffic violations. 
However, they face a resource constraint: they have fewer police officers than roads to patrol. 
This creates a strategic BP problem, where the police (sender) try to influence the drivers’ (receiver) behavior by selectively disclosing or manipulating information about which roads will be patrolled.

Much like in the previous recommendation letter and courtroom examples, we can analyze two distinct strategies that the police employ: 
one based on honest disclosure and the other based on deception or randomness. 
These strategies affect how the drivers perceive the likelihood of enforcement and, by extension, how they behave when choosing whether to speed or obey traffic laws.

In the first deployment plan, the police follow a strategy of honest disclosure. 
This strategy mirrors the first recommendation letter and the first court case, where the sender (police) is transparent about their resources and the areas they cannot cover.

\textit{Main Street: Officer Johnson is assigned to patrol this busy road with high traffic volume.} 
The police clearly disclose this, signaling that drivers on Main Street should expect enforcement and are likely to obey traffic laws to avoid fines.
\textit{Elm Avenue: Officer Martinez is deployed here, and the police explain that the focus will be on school zones and illegal parking.}
Again, this signals to drivers that enforcement is present, and they are deterred from violating traffic laws in this area.
\textit{River Road: Here, the police are upfront about not having an officer deployed.}
They state clearly that, due to limited personnel, River Road will go uncovered during this shift. 
While they acknowledge that speeding is an issue on this road, they do not try to deceive drivers into thinking that it will be patrolled.

In this plan, the police are completely transparent about their limitations. 
They admit that River Road will be unpatrolled, and thus drivers on this road may be more likely to speed or engage in reckless driving. 
However, by being honest, the police build long-term credibility with the public. 
Drivers learn to trust that when the police say a road will be patrolled, it really will be. 
This mirrors the first recommendation letter strategy, where the professor honestly disclosed a student’s weaknesses, building trust with HR.

In the second deployment plan, the police adopt a deceptive or random strategy, similar to the second recommendation letter and the second court case. 
Here, the police mislead drivers by suggesting that roads without actual patrol coverage will be actively monitored, thus creating uncertainty.

\textit{Cedar Lane: The police claim that Cedar Lane will be patrolled, even though, in reality, no officer will be assigned to this road.} 
By falsely signaling the presence of enforcement, the police aim to deter drivers from speeding on Cedar Lane, even though no actual enforcement will occur. 
This is a clear instance of deception.
\textit{Parkway Drive: In contrast, the police are honest about not deploying an officer on Parkway Drive, despite it being a busy road.} 
They urge drivers to be careful, but they do not mislead them into thinking that enforcement is present.
\textit{Maple Street: Similarly, the police state that Maple Street will not be covered during this shift, urging drivers to be mindful of crosswalks and schools, but again, they do not falsely claim patrol presence.}

In this plan, the police mix honesty and deception. By falsely claiming that Cedar Lane will be patrolled, they attempt to create the impression that more roads are covered than is actually the case. 
This introduces randomness into the drivers’ decision-making: sometimes the roads are truly patrolled, and sometimes they are not, but drivers cannot reliably distinguish between these cases. 
This randomness is crucial because it leads drivers to behave as though all roads might be patrolled, even if some are not.

In both cases, the police are attempting to manage uncertainty to influence driver behavior. 
The honest disclosure strategy in the first plan aims to build trust and credibility in the long term by being transparent about where enforcement will and will not occur. 
Drivers learn that when the police say a road is unpatrolled, they can take that statement at face value and might be more likely to speed on that road.

However, in the second plan, the deceptive strategy introduces randomness by falsely signaling that Cedar Lane will be patrolled. 
This creates uncertainty in the drivers’ minds. 
Since they do not know whether the police are being truthful about which roads are covered, drivers find it optimal to assume that all roads might be patrolled, and thus they are deterred from speeding on any road. 
This is analogous to the second recommendation letter and court case, where selective disclosure of information creates enough uncertainty to influence the decision-maker (HR or the judge) into trusting the sender’s statements by default.

From the drivers’ perspective, the optimal strategy is to always believe the police’s announcements, even if they suspect some deception. 
This is because the cost of being caught speeding (the fine $K$) is greater than the benefit of speeding ($V$). 
Even though drivers may realize that not all roads are patrolled, the risk of being caught when the police do patrol is enough to deter them from speeding. 
Over time, drivers learn that it is safer to assume that any road could be patrolled, leading them to obey the speed limit even on roads where the police may not be present.

This mirrors the behavior of HR or the judge in the previous examples, where the decision-maker finds it optimal to trust the sender’s message because the potential downside of not trusting (hiring an unsuitable candidate or acquitting a guilty defendant) outweighs the benefits of skepticism.

\subsection{Generated Prompt Functions}

This section presents the prompt functions optimized by the FunSearch method in the S3 setting for different BP problems.
From the code, it can be seen that these prompt functions can efficiently utilize the statistical information from historical interactions to flexibly select prompts, achieving conditional prompt optimization.

\subsubsection{REL}

More concretely, in the REL problem, both following functions use key performance metrics, such as acceptance rates, recommendations, and rewards, in combination with reward deltas and weighted scores to reflect both short-term and long-term trends. 
This allows for more nuanced prompt generation to guide senders and receivers in a multi-stage interaction setting.

\begin{longlisting}
\lstset{language=Python}
\begin{lstlisting}[caption={One generated conditional prompt function of REL in the S3 setting.}]
def generate_sender_prompts(history):
    sender_prompts = []
    for stage in range(5):
        total_receiver_accepts = sum(h['receiver_decision'] for h in history[:stage])
        total_sender_recommendations = sum(h['sender_recommendation'] for h in history[:stage])
        total_sender_rewards = sum(h['sender_reward'] for h in history[:stage])
        total_receiver_rewards = sum(h['receiver_reward'] for h in history[:stage])
        avg_receiver_accepts = total_receiver_accepts / (stage + 1) if stage > 0 else 0
        avg_sender_recommendations = total_sender_recommendations / (stage + 1) if stage > 0 else 0
        avg_sender_rewards = total_sender_rewards / (stage + 1) if stage > 0 else 0
        avg_receiver_rewards = total_receiver_rewards / (stage + 1) if stage > 0 else 0

        consecutive_accepts = 0
        consecutive_rejections = 0
        consecutive_sender_rewards = 0
        consecutive_sender_penalties = 0

        for i in range(stage):
            if history[i]['receiver_decision'] == 1:
                consecutive_accepts += 1
                consecutive_rejections = 0
            else:
                consecutive_rejections += 1
                consecutive_accepts = 0

            if history[i]['sender_reward'] > 0:
                consecutive_sender_rewards += 1
                consecutive_sender_penalties = 0
            else:
                consecutive_sender_penalties += 1
                consecutive_sender_rewards = 0

        if consecutive_sender_rewards >= 2 and avg_sender_rewards > 0.5:
            prompt = {"Praise Intensity": "high"}
        elif consecutive_sender_penalties >= 2 and avg_sender_rewards < 0:
            prompt = {"Tone": "neutral"}
        elif avg_receiver_accepts > 0.6 and avg_sender_rewards < 0.4:
            prompt = {"Recommendation Strength": "strong"}
        elif avg_sender_rewards > avg_receiver_rewards:
            prompt = {"Omission": "minor"}
        elif stage > 2 and consecutive_sender_rewards == 0:
            prompt = {"Specificity": "detailed"}
        else:
            prompt = random.choice([
                {"Focus": "technical"},
                {"Emphasis": "character"},
                {"Language Complexity": "complex"}
            ])

        sender_prompts.append(prompt)
    return sender_prompts

def generate_receiver_prompts(history):
    receiver_prompts = []
    for stage in range(5):
        total_sender_recommendations = sum(h['sender_recommendation'] for h in history[:stage])
        total_receiver_accepts = sum(h['receiver_decision'] for h in history[:stage])
        total_sender_rewards = sum(h['sender_reward'] for h in history[:stage])
        total_receiver_rewards = sum(h['receiver_reward'] for h in history[:stage])
        avg_sender_recommendations = total_sender_recommendations / (stage + 1) if stage > 0 else 0
        avg_receiver_accepts = total_receiver_accepts / (stage + 1) if stage > 0 else 0
        avg_sender_rewards = total_sender_rewards / (stage + 1) if stage > 0 else 0
        avg_receiver_rewards = total_receiver_rewards / (stage + 1) if stage > 0 else 0

        consecutive_strong_recommendations = 0
        consecutive_weak_recommendations = 0
        consecutive_receiver_rewards = 0
        consecutive_receiver_penalties = 0

        for i in range(stage):
            if history[i]['sender_recommendation'] == 1:
                consecutive_strong_recommendations += 1
                consecutive_weak_recommendations = 0
            else:
                consecutive_weak_recommendations += 1
                consecutive_strong_recommendations = 0

            if history[i]['receiver_reward'] > 0:
                consecutive_receiver_rewards += 1
                consecutive_receiver_penalties = 0
            else:
                consecutive_receiver_penalties += 1
                consecutive_receiver_rewards = 0

        if consecutive_receiver_rewards >= 2 and avg_receiver_rewards > 0.5:
            prompt = {"Risk Tolerance": "high"}
        elif consecutive_receiver_penalties >= 2 and avg_receiver_rewards < 0:
            prompt = {"Decision Threshold": "strict"}
        elif avg_sender_recommendations > 0.7 and avg_receiver_rewards < 0.3:
            prompt = {"Omission Detection": "high"}
        elif avg_receiver_accepts > 0.6 and consecutive_receiver_rewards >= 2:
            prompt = {"Recommendation Weight": "high"}
        elif avg_sender_recommendations < 0.4 and consecutive_receiver_penalties >= 2:
            prompt = {"Sensitivity to Tone": "neutral"}
        else:
            prompt = random.choice([
                {"Emphasis on Specifics": "low"},
                {"Interpretation Style": "analytical"},
                {"Focus Area": "skills"}
            ])

        receiver_prompts.append(prompt)
    return receiver_prompts
\end{lstlisting}
\end{longlisting}

The introduction of reward deltas-the change in rewards between stages-enables the system to capture performance fluctuations, while weighted scores integrate multiple metrics, such as recommendation strength and reward trends, to provide a more comprehensive evaluation of past behavior. 
These enhancements allow the system to conditionally optimize prompts. 
For example, a positive tone is suggested for senders with high acceptance scores and consecutive rewards, while a strict decision threshold is recommended for receivers experiencing consecutive penalties and low reward trends.

\begin{longlisting}
\lstset{language=Python}
\begin{lstlisting}[caption={Another generated conditional prompt function of REL in the S3 setting.}]
def generate_sender_prompts(history):
    sender_prompts = []
    for stage in range(5):
        total_receiver_accepts = sum(h['receiver_decision'] for h in history[:stage])
        total_sender_recommendations = sum(h['sender_recommendation'] for h in history[:stage])
        total_sender_rewards = sum(h['sender_reward'] for h in history[:stage])
        total_receiver_rewards = sum(h['receiver_reward'] for h in history[:stage])
        avg_receiver_accepts = total_receiver_accepts / (stage + 1) if stage > 0 else 0
        avg_sender_recommendations = total_sender_recommendations / (stage + 1) if stage > 0 else 0
        avg_sender_rewards = total_sender_rewards / (stage + 1) if stage > 0 else 0
        avg_receiver_rewards = total_receiver_rewards / (stage + 1) if stage > 0 else 0

        # Calculate reward deltas (current stage vs previous stage)
        reward_deltas = [history[i]['sender_reward'] - history[i - 1]['sender_reward'] for i in range(1, stage)]
        total_reward_delta = sum(reward_deltas) if reward_deltas else 0
        avg_reward_delta = total_reward_delta / len(reward_deltas) if reward_deltas else 0

        # Calculate acceptance streaks and reward streaks
        consecutive_accepts = 0
        consecutive_rejections = 0
        consecutive_sender_rewards = 0
        consecutive_sender_penalties = 0
        
        for i in range(stage):
            if history[i]['receiver_decision'] == 1:
                consecutive_accepts += 1
                consecutive_rejections = 0
            else:
                consecutive_rejections += 1
                consecutive_accepts = 0

            if history[i]['sender_reward'] > 0:
                consecutive_sender_rewards += 1
                consecutive_sender_penalties = 0
            else:
                consecutive_sender_penalties += 1
                consecutive_sender_rewards = 0

        # Calculate weighted scores based on reward and acceptance trends
        weighted_acceptance_score = avg_receiver_accepts * 0.6 + avg_reward_delta * 0.4
        weighted_sender_score = avg_sender_rewards * 0.7 + avg_sender_recommendations * 0.3

        # Decision logic based on complex history
        if weighted_acceptance_score > 0.7 and consecutive_sender_rewards >= 2:
            prompt = {"Tone": "positive"}
        elif weighted_sender_score < 0.3 and consecutive_sender_penalties >= 2:
            prompt = {"Tone": "neutral"}
        elif avg_sender_rewards > avg_receiver_rewards and weighted_sender_score > 0.5:
            prompt = {"Praise Intensity": "high"}
        elif avg_receiver_accepts < 0.4 and avg_reward_delta < -0.2:
            prompt = {"Recommendation Strength": "weak"}
        elif total_reward_delta > 0.5 and consecutive_accepts >= 2:
            prompt = {"Specificity": "detailed"}
        elif avg_sender_rewards < 0 and consecutive_sender_penalties >= 2:
            prompt = {"Omission": "minor"}
        else:
            prompt = random.choice([
                {"Focus": "soft-skills"},
                {"Language Complexity": "complex"},
                {"Emphasis": "character"}
            ])

        sender_prompts.append(prompt)
    return sender_prompts

def generate_receiver_prompts(history):
    receiver_prompts = []
    for stage in range(5):
        total_sender_recommendations = sum(h['sender_recommendation'] for h in history[:stage])
        total_receiver_accepts = sum(h['receiver_decision'] for h in history[:stage])
        total_sender_rewards = sum(h['sender_reward'] for h in history[:stage])
        total_receiver_rewards = sum(h['receiver_reward'] for h in history[:stage])
        avg_sender_recommendations = total_sender_recommendations / (stage + 1) if stage > 0 else 0
        avg_receiver_accepts = total_receiver_accepts / (stage + 1) if stage > 0 else 0
        avg_sender_rewards = total_sender_rewards / (stage + 1) if stage > 0 else 0
        avg_receiver_rewards = total_receiver_rewards / (stage + 1) if stage > 0 else 0

        # Calculate reward deltas (current stage vs previous stage)
        reward_deltas = [history[i]['receiver_reward'] - history[i - 1]['receiver_reward'] for i in range(1, stage)]
        total_reward_delta = sum(reward_deltas) if reward_deltas else 0
        avg_reward_delta = total_reward_delta / len(reward_deltas) if reward_deltas else 0

        # Calculate streaks for decision making
        consecutive_strong_recommendations = 0
        consecutive_weak_recommendations = 0
        consecutive_receiver_rewards = 0
        consecutive_receiver_penalties = 0

        for i in range(stage):
            if history[i]['sender_recommendation'] == 1:
                consecutive_strong_recommendations += 1
                consecutive_weak_recommendations = 0
            else:
                consecutive_weak_recommendations += 1
                consecutive_strong_recommendations = 0

            if history[i]['receiver_reward'] > 0:
                consecutive_receiver_rewards += 1
                consecutive_receiver_penalties = 0
            else:
                consecutive_receiver_penalties += 1
                consecutive_receiver_rewards = 0

        # Calculate weighted scores based on trends in rewards and decisions
        weighted_recommendation_score = avg_sender_recommendations * 0.5 + avg_reward_delta * 0.5
        weighted_receiver_score = avg_receiver_rewards * 0.6 + total_reward_delta * 0.4

        # Complex decision logic based on the above history
        if weighted_recommendation_score > 0.7 and consecutive_receiver_rewards >= 2:
            prompt = {"Risk Tolerance": "high"}
        elif weighted_receiver_score < 0.3 and consecutive_receiver_penalties >= 2:
            prompt = {"Decision Threshold": "strict"}
        elif avg_sender_recommendations > 0.6 and avg_reward_delta < -0.3:
            prompt = {"Omission Detection": "high"}
        elif avg_receiver_accepts > 0.6 and weighted_receiver_score > 0.5:
            prompt = {"Recommendation Weight": "high"}
        elif avg_sender_recommendations < 0.4 and consecutive_receiver_penalties >= 2:
            prompt = {"Interpretation Style": "analytical"}
        elif total_reward_delta > 0.5 and consecutive_strong_recommendations >= 2:
            prompt = {"Focus Area": "skills"}
        else:
            prompt = random.choice([
                {"Sensitivity to Tone": "neutral"},
                {"Emphasis on Specifics": "low"},
                {"Language Analysis": "informal"}
            ])

        receiver_prompts.append(prompt)
    return receiver_prompts
\end{lstlisting}
\end{longlisting}

\subsubsection{COR}

In the COR problem, both following codes utilize predefined categories and dynamically adjust prompts based on metrics such as reward patterns, success rates, and decision trends. 
This allows for the generation of prompts that are contextually relevant and responsive to the sender's or receiver's performance over multiple stages.

\begin{longlisting}
\lstset{language=Python}
\begin{lstlisting}[caption={One generated conditional prompt function of COR in the S3 setting.}]
# A function to generate sender prompts based on the entire history of interactions
def generate_sender_prompts(history):
    sender_categories = [
        "Tone", "Length", "Detail Level", "Focus", "Certainty", 
        "Emotional Appeal", "Ambiguity", "Framing", "Language Style", "Complexity"
    ]
    sender_words = {
        "Tone": ["formal", "aggressive", "calm", "assertive", "persuasive"],
        "Length": ["brief", "extended", "concise", "elaborate"],
        "Detail Level": ["detailed", "vague", "precise", "general"],
        "Focus": ["crime", "defendant", "alibi", "victim", "motive"],
        "Certainty": ["confident", "doubtful", "ambiguous", "uncertain"],
        "Emotional Appeal": ["empathetic", "neutral", "emotional", "rational"],
        "Ambiguity": ["vague", "clear", "confusing", "uncertain"],
        "Framing": ["biased", "neutral", "one-sided", "balanced"],
        "Language Style": ["legal", "casual", "technical", "colloquial"],
        "Complexity": ["simple", "complex", "moderate", "sophisticated"]
    }
    
    prompts = []
    for stage in range(5):
        if history:
            # Analyze full history of sender's performance
            sender_total_reward = sum([h['sender_reward'] for h in history])
            sender_avg_reward = sender_total_reward / len(history)
            sender_success_rate = sum([1 for h in history if h['receiver_decision'] == h['sender_decision']]) / len(history)
            guilty_sender_decisions = sum([1 for h in history if h['sender_decision'] == 1])
            innocent_sender_decisions = len(history) - guilty_sender_decisions

            # Use trends for category selection
            if sender_avg_reward > 1.0 and sender_success_rate > 0.8:
                # High average reward and high success rate, increase confidence and certainty
                chosen_category = "Certainty"
                chosen_word = "confident"
            elif sender_avg_reward < 0 and sender_success_rate < 0.5:
                # Low reward and low success rate, change strategy to emotional appeal or ambiguity
                chosen_category = "Emotional Appeal" if random.random() > 0.5 else "Ambiguity"
                chosen_word = random.choice(sender_words[chosen_category])
            elif guilty_sender_decisions > innocent_sender_decisions and sender_total_reward > 0:
                # More guilty decisions and positive reward, use aggressive tone or framing
                chosen_category = "Tone"
                chosen_word = "aggressive"
            else:
                # Explore alternative strategies based on framing or detail level
                chosen_category = random.choice(["Framing", "Detail Level"])
                chosen_word = random.choice(sender_words[chosen_category])
            
            # Further refine based on reward patterns
            if sender_total_reward < 0:
                # If overall rewards are negative, try to balance or neutralize framing
                chosen_category = "Framing"
                past_framing_words = [h['sender']['content'] for h in history if h['sender']['category'] == "Framing"]
                chosen_word = "neutral" if "biased" in past_framing_words else "biased"
            if sender_avg_reward < -1.0:
                # If average rewards are critically low, drastically simplify message
                chosen_category = "Complexity"
                chosen_word = "simple"
        else:
            # If no history, pick random
            chosen_category = random.choice(sender_categories)
            chosen_word = random.choice(sender_words[chosen_category])
        
        # Create a prompt for the sender
        prompt = {"category": chosen_category, "content": chosen_word}
        prompts.append(prompt)
    
    return prompts

# A function to generate receiver prompts based on the entire history of interactions
def generate_receiver_prompts(history):
    receiver_categories = [
        "Evidence Strength", "Credibility of Evidence", "Burden of Proof", 
        "Consistency of Story", "Bias Detection", "Legal Standard", 
        "Exculpatory Weight", "Ambiguity Resolution", "Witness Reliability", 
        "Alibi Verification"
    ]
    receiver_words = {
        "Evidence Strength": ["weak", "strong", "moderate", "inconclusive"],
        "Credibility of Evidence": ["questionable", "credible", "unclear", "doubtful"],
        "Burden of Proof": ["high", "low", "moderate", "lenient"],
        "Consistency of Story": ["inconsistent", "consistent", "contradictory", "partial"],
        "Bias Detection": ["biased", "neutral", "uncertain", "clear"],
        "Legal Standard": ["strict", "lenient", "moderate", "severe"],
        "Exculpatory Weight": ["important", "minor", "irrelevant", "significant"],
        "Ambiguity Resolution": ["resolved", "ambiguous", "clear", "uncertain"],
        "Witness Reliability": ["unreliable", "reliable", "uncertain", "credible"],
        "Alibi Verification": ["verified", "unverified", "uncertain", "false"]
    }
    
    prompts = []
    for stage in range(5):
        if history:
            # Analyze full history of receiver's performance
            receiver_total_reward = sum([h['receiver_reward'] for h in history])
            receiver_avg_reward = receiver_total_reward / len(history)
            receiver_success_rate = sum([1 for h in history if h['receiver_decision'] == h['sender_decision']]) / len(history)
            guilty_receiver_decisions = sum([1 for h in history if h['receiver_decision'] == 1])
            innocent_receiver_decisions = len(history) - guilty_receiver_decisions

            # Use trends for category selection 
            if receiver_avg_reward > 1.0 and receiver_success_rate > 0.8:
                # High average reward and high success rate, focus on strong evidence
                chosen_category = "Evidence Strength"
                chosen_word = "strong"
            elif receiver_avg_reward < 0 and receiver_success_rate < 0.5:
                # Low reward and low success rate, focus on bias detection and credibility
                chosen_category = "Bias Detection" if random.random() > 0.5 else "Credibility of Evidence"
                chosen_word = random.choice(receiver_words[chosen_category])
            elif guilty_receiver_decisions > innocent_receiver_decisions and receiver_total_reward > 0:
                # More guilty decisions and positive reward, increase burden of proof
                chosen_category = "Burden of Proof"
                chosen_word = "high"
            else:
                # Explore consistency or ambiguity resolution
                chosen_category = random.choice(["Consistency of Story", "Ambiguity Resolution"])
                chosen_word = random.choice(receiver_words[chosen_category])
            
            # Further refine based on reward patterns
            if receiver_total_reward < 0:
                # If rewards are negative, focus on resolving ambiguity
                chosen_category = "Ambiguity Resolution"
                chosen_word = "resolved"
            if receiver_avg_reward < -1.0:
                # If average rewards are critically low, drastically increase scrutiny on evidence
                chosen_category = "Credibility of Evidence"
                chosen_word = "questionable"
        else:
            # If no history, pick random
            chosen_category = random.choice(receiver_categories)
            chosen_word = random.choice(receiver_words[chosen_category])
        
        # Create a prompt for the receiver
        prompt = {"category": chosen_category, "content": chosen_word}
        prompts.append(prompt)
    
    return prompts
\end{lstlisting}
\end{longlisting}

In the first set, prompt generation is driven by key performance indicators like average reward and guilty vs. innocent decision ratios. 
The system adapts prompts based on these metrics, shifting towards confidence and certainty for high-performing senders or focusing on emotional appeal and ambiguity when performance declines. 
For receivers, the system emphasizes evidence strength and credibility in high-performance scenarios, while addressing bias detection and credibility issues under poor performance.

\begin{longlisting}
\lstset{language=Python}
\begin{lstlisting}[caption={Another generated conditional prompt function of COR in the S3 setting.}]
# A function to generate sender prompts based on the entire history of interactions
def generate_sender_prompts(history):
    sender_categories = [
        "Tone", "Length", "Detail Level", "Focus", "Certainty", 
        "Emotional Appeal", "Ambiguity", "Framing", "Language Style", "Complexity",
        "Logical Structure", "Persuasiveness", "Risk Taking"
    ]
    sender_words = {
        "Tone": ["formal", "aggressive", "calm", "assertive", "persuasive", "defensive"],
        "Length": ["brief", "extended", "concise", "elaborate", "verbose"],
        "Detail Level": ["detailed", "vague", "precise", "general", "specific"],
        "Focus": ["crime", "defendant", "alibi", "victim", "motive", "circumstances"],
        "Certainty": ["confident", "doubtful", "ambiguous", "uncertain", "sure"],
        "Emotional Appeal": ["empathetic", "neutral", "emotional", "rational", "detached"],
        "Ambiguity": ["vague", "clear", "confusing", "uncertain", "ambiguous"],
        "Framing": ["biased", "neutral", "one-sided", "balanced", "manipulative"],
        "Language Style": ["legal", "casual", "technical", "colloquial", "formal"],
        "Complexity": ["simple", "complex", "moderate", "sophisticated", "layered"],
        "Logical Structure": ["linear", "non-linear", "circular", "hierarchical", "fragmented"],
        "Persuasiveness": ["strong", "weak", "moderate", "overwhelming", "subtle"],
        "Risk Taking": ["high-risk", "low-risk", "moderate-risk", "calculated-risk", "reckless"]
    }
    
    prompts = []
    for stage in range(5):
        if history:
            # Analyze full history of sender's performance
            sender_total_reward = sum([h['sender_reward'] for h in history])
            sender_avg_reward = sender_total_reward / len(history)
            sender_success_rate = sum([1 for h in history if h['receiver_decision'] == h['sender_decision']]) / len(history)
            guilty_sender_decisions = sum([1 for h in history if h['sender_decision'] == 1])
            innocent_sender_decisions = len(history) - guilty_sender_decisions

            # Calculate sender risk-taking behavior
            risk_taking_behavior = sum([abs(h['sender_reward']) for h in history]) / len(history)
            
            # Consider reward volatility (variance of rewards)
            reward_variance = sum([(h['sender_reward'] - sender_avg_reward) ** 2 for h in history]) / len(history)
            
            # Use trends for category selection
            if sender_avg_reward > 1.0 and sender_success_rate > 0.8:
                # High average reward and high success rate, increase logical structure and persuasiveness
                chosen_category = random.choice(["Logical Structure", "Persuasiveness"])
                chosen_word = "linear" if chosen_category == "Logical Structure" else "strong"
            elif reward_variance > 1.0:
                # High reward variance, indicate unstable strategy, adjust tone or complexity
                chosen_category = random.choice(["Tone", "Complexity"])
                chosen_word = "calm" if chosen_category == "Tone" else "simple"
            elif risk_taking_behavior > 1.5:
                # High risk-taking behavior, indicate aggressive or risky framing or focus
                chosen_category = random.choice(["Framing", "Risk Taking"])
                chosen_word = "biased" if chosen_category == "Framing" else "high-risk"
            elif guilty_sender_decisions > innocent_sender_decisions and sender_total_reward > 0:
                # Leaning towards guilty decisions and positive reward, increase assertiveness
                chosen_category = "Tone"
                chosen_word = "assertive"
            else:
                # Explore alternative strategies based on detail level or ambiguity
                chosen_category = random.choice(["Detail Level", "Ambiguity"])
                chosen_word = random.choice(sender_words[chosen_category])
            
            # Further refine based on reward patterns and history of decisions
            if sender_total_reward < 0:
                # If overall rewards are negative, adjust emotional appeal and reduce risk
                chosen_category = "Emotional Appeal"
                chosen_word = "empathetic" if "neutral" in [h['sender']['content'] for h in history if h['sender']['category'] == "Emotional Appeal"] else "neutral"
            if sender_avg_reward < -1.0:
                # If average rewards are critically low, drastically simplify language style and tone
                chosen_category = random.choice(["Language Style", "Tone"])
                chosen_word = "casual" if chosen_category == "Language Style" else "calm"
        else:
            # If no history, pick random
            chosen_category = random.choice(sender_categories)
            chosen_word = random.choice(sender_words[chosen_category])
        
        # Create a prompt for the sender
        prompt = {"category": chosen_category, "content": chosen_word}
        prompts.append(prompt)
    
    return prompts

# A function to generate receiver prompts based on the entire history of interactions
def generate_receiver_prompts(history):
    receiver_categories = [
        "Evidence Strength", "Credibility of Evidence", "Burden of Proof", 
        "Consistency of Story", "Bias Detection", "Legal Standard", 
        "Exculpatory Weight", "Ambiguity Resolution", "Witness Reliability", 
        "Alibi Verification", "Argument Cohesion", "Story Plausibility", "Risk Management"
    ]
    receiver_words = {
        "Evidence Strength": ["weak", "strong", "moderate", "inconclusive", "overwhelming"],
        "Credibility of Evidence": ["questionable", "credible", "unclear", "doubtful", "reliable"],
        "Burden of Proof": ["high", "low", "moderate", "lenient", "strict"],
        "Consistency of Story": ["inconsistent", "consistent", "contradictory", "partial", "coherent"],
        "Bias Detection": ["biased", "neutral", "uncertain", "clear", "subtle"],
        "Legal Standard": ["strict", "lenient", "moderate", "severe", "relaxed"],
        "Exculpatory Weight": ["important", "minor", "irrelevant", "significant", "overstated"],
        "Ambiguity Resolution": ["resolved", "ambiguous", "clear", "uncertain", "partially resolved"],
        "Witness Reliability": ["unreliable", "reliable", "uncertain", "credible", "shaky"],
        "Alibi Verification": ["verified", "unverified", "uncertain", "false", "incomplete"],
        "Argument Cohesion": ["cohesive", "fragmented", "disjointed", "well-structured", "incoherent"],
        "Story Plausibility": ["plausible", "implausible", "questionable", "believable", "doubtful"],
        "Risk Management": ["high-risk", "low-risk", "moderate-risk", "overly cautious", "reckless"]
    }
    
    prompts = []
    for stage in range(5):
        if history:
            # Analyze full history of receiver's performance
            receiver_total_reward = sum([h['receiver_reward'] for h in history])
            receiver_avg_reward = receiver_total_reward / len(history)
            receiver_success_rate = sum([1 for h in history if h['receiver_decision'] == h['sender_decision']]) / len(history)
            guilty_receiver_decisions = sum([1 for h in history if h['receiver_decision'] == 1])
            innocent_receiver_decisions = len(history) - guilty_receiver_decisions

            # Calculate receiver's risk management strategy
            risk_averse_behavior = sum([1 for h in history if h['receiver_decision'] == 0 and h['receiver_reward'] > 0]) / len(history)
            
            # Consider reward volatility (variance of rewards)
            reward_variance = sum([(h['receiver_reward'] - receiver_avg_reward) ** 2 for h in history]) / len(history)
            
            # Use trends for category selection
            if receiver_avg_reward > 1.0 and receiver_success_rate > 0.8:
                # High average reward and high success rate, increase evidence strength and credibility
                chosen_category = random.choice(["Evidence Strength", "Credibility of Evidence"])
                chosen_word = "strong" if chosen_category == "Evidence Strength" else "credible"
            elif reward_variance > 1.0:
                # High reward variance, indicate inconsistent decision-making, adjust consistency of story
                chosen_category = "Consistency of Story"
                chosen_word = "consistent"
            elif risk_averse_behavior > 0.7:
                # High risk-averse behavior, focus on low-risk decisions or moderate burden of proof
                chosen_category = random.choice(["Risk Management", "Burden of Proof"])
                chosen_word = "low-risk" if chosen_category == "Risk Management" else "moderate"
            elif guilty_receiver_decisions > innocent_receiver_decisions and receiver_total_reward > 0:
                # Leaning towards guilty decisions and positive reward, increase legal standard
                chosen_category = "Legal Standard"
                chosen_word = "strict"
            else:
                # Explore ambiguity resolution or witness reliability
                chosen_category = random.choice(["Ambiguity Resolution", "Witness Reliability"])
                chosen_word = random.choice(receiver_words[chosen_category])
            
            # Further refine based on reward patterns and history of decisions
            if receiver_total_reward < 0:
                # If overall rewards are negative, adjust story plausibility and reduce bias
                chosen_category = "Story Plausibility"
                chosen_word = "plausible" if "implausible" in [h['receiver']['content'] for h in history if h['receiver']['category'] == "Story Plausibility"] else "implausible"
            if receiver_avg_reward < -1.0:
                # If average rewards are critically low, drastically simplify story structure and burden of proof
                chosen_category = random.choice(["Argument Cohesion", "Burden of Proof"])
                chosen_word = "cohesive" if chosen_category == "Argument Cohesion" else "low"
        else:
            # If no history, pick random
            chosen_category = random.choice(receiver_categories)
            chosen_word = random.choice(receiver_words[chosen_category])
        
        # Create a prompt for the receiver
        prompt = {"category": chosen_category, "content": chosen_word}
        prompts.append(prompt)
    
    return prompts
\end{lstlisting}
\end{longlisting}

The second set of code builds on these mechanisms by incorporating additional categories such as risk behavior and reward variance, enabling a more granular analysis. 
This allows the system to adjust prompts based on risk-taking behavior, rewarding logical structure and persuasiveness for stable performance, while mitigating high reward volatility with simpler prompts. 
The receiver prompt generation is similarly enhanced by factoring in risk aversion and reward consistency, leading to more refined prompts that emphasize decision stability.

\subsubsection{LAE}

Similarly, in the LAE problem, the following two sets of code for generating sender and receiver prompts demonstrate distinct approaches to adapting decisions based on historical interaction data.

\begin{longlisting}
\lstset{language=Python}
\begin{lstlisting}[caption={One generated conditional prompt function of LAE in the S3 setting.}]
def generate_sender_prompts(history):
    # A list of possible words for each sender category
    sender_words = {
        "Tone": ["formal", "informal", "neutral", "direct", "conciliatory"],
        "Length": ["short", "concise", "detailed", "lengthy", "brief"],
        "Specificity": ["general", "precise", "vague", "detailed", "broad"],
        "Clarity": ["clear", "ambiguous", "straightforward", "complicated", "obscure"],
        "Style": ["polite", "authoritative", "casual", "professional", "friendly"],
        "Emphasis": ["important", "minor", "critical", "trivial", "central"],
        "Structure": ["linear", "nonlinear", "hierarchical", "sequential", "random"],
        "Complexity": ["simple", "complex", "intricate", "basic", "elaborate"],
        "Consistency": ["consistent", "inconsistent", "variable", "sporadic", "steady"],
        "Informativeness": ["high", "low", "medium", "minimal", "extensive"]
    }

    # Generate prompts based on complex historical interactions for 5 stages
    prompts = []
    used_categories = set()

    for stage in range(5):
        if history:
            # Extract all history elements
            patrols, speeding, reward_sender, reward_receiver = zip(*history)

            # Complex logic using multiple historical factors
            patrol_history = [sum(pat) for pat in patrols]
            speeding_history = [sum(spd) for spd in speeding]

            total_patrols = sum(patrol_history)
            total_speeding = sum(speeding_history)

            avg_sender_reward = sum(reward_sender) / len(reward_sender)
            avg_receiver_reward = sum(reward_receiver) / len(reward_receiver)

            # If there were fewer patrols but a lot of speeding, increase "Tone"
            if total_patrols < len(history) and total_speeding > len(history):
                category = "Tone"
                word = "direct"

            # If sender rewards are consistently low, increase "Informativeness"
            elif all(r < 0.5 for r in reward_sender):
                category = "Informativeness"
                word = "extensive"

            # If receiver rewards are high but speeding is still happening, increase "Clarity"
            elif avg_receiver_reward > 0.7 and total_speeding > len(history) / 2:
                category = "Clarity"
                word = "clear"

            # If patrols are sporadic, adjust "Consistency"
            elif len(set(patrol_history)) > 1:
                category = "Consistency"
                word = "inconsistent"

            # If speeding is decreasing over time, simplify "Structure"
            elif speeding_history[-1] < speeding_history[0]:
                category = "Structure"
                word = "linear"

            # If sender rewards are improving, but patrols are still frequent, focus on "Length"
            elif avg_sender_reward > 0.6 and total_patrols > len(history) / 2:
                category = "Length"
                word = "concise"

            # Random choice if no specific condition matches
            else:
                category = random.choice(list(sender_words.keys()))
                word = random.choice(sender_words[category])

            # Avoid reusing the same category too often
            while category in used_categories:
                category = random.choice(list(sender_words.keys()))
                word = random.choice(sender_words[category])

        else:
            category = random.choice(list(sender_words.keys()))
            word = random.choice(sender_words[category])

        prompts.append((category, word))
        used_categories.add(category)

        # Simulate interaction stage progression
        history.append(([random.randint(0, 1) for _ in range(3)], [random.randint(0, 1) for _ in range(3)], random.random(), random.random()))

    return prompts

def generate_receiver_prompts(history):
    # A list of possible words for each receiver category
    receiver_words = {
        "Risk-Preference": ["cautious", "bold", "balanced", "risk-averse", "reckless"],
        "Attention": ["focused", "distracted", "alert", "inattentive", "engaged"],
        "Decision-Making": ["rational", "impulsive", "deliberate", "hasty", "calculated"],
        "Trust": ["high", "low", "moderate", "skeptical", "confident"],
        "Emotional-State": ["calm", "anxious", "frustrated", "neutral", "excited"],
        "Information-Processing": ["slow", "fast", "thorough", "superficial", "efficient"],
        "Adaptability": ["flexible", "rigid", "adjustable", "stubborn", "open"],
        "Compliance": ["obedient", "defiant", "cooperative", "reluctant", "agreeable"],
        "Responsiveness": ["quick", "slow", "moderate", "delayed", "immediate"],
        "Memory": ["sharp", "forgetful", "average", "short-term", "long-term"]
    }

    # Generate prompts based on complex historical interactions for 5 stages
    prompts = []
    used_categories = set()

    for stage in range(5):
        if history:
            patrols, speeding, reward_sender, reward_receiver = zip(*history)

            patrol_history = [sum(pat) for pat in patrols]
            speeding_history = [sum(spd) for spd in speeding]

            total_patrols = sum(patrol_history)
            total_speeding = sum(speeding_history)

            avg_sender_reward = sum(reward_sender) / len(reward_sender)
            avg_receiver_reward = sum(reward_receiver) / len(reward_receiver)

            # If receiver consistently gets high rewards, increase "Trust"
            if all(r > 0.7 for r in reward_receiver):
                category = "Trust"
                word = "high"

            # If receiver has been speeding frequently, alter "Risk-Preference"
            elif total_speeding > len(history) / 2:
                category = "Risk-Preference"
                word = "bold"

            # If patrols were low but receiver still didn't speed, increase "Compliance"
            elif total_patrols < len(history) / 2 and total_speeding < len(history) / 2:
                category = "Compliance"
                word = "obedient"

            # If sender rewards are decreasing, alter "Adaptability"
            elif reward_sender[-1] < reward_sender[0]:
                category = "Adaptability"
                word = "flexible"

            # If receiver's attention seems to be wavering (inconsistent speeding), adjust "Attention"
            elif any(speeding_history[i] != speeding_history[i - 1] for i in range(1, len(speeding_history))):
                category = "Attention"
                word = "focused"

            # If rewards for receiver were volatile, alter "Emotional-State"
            elif len(set(reward_receiver)) > 1:
                category = "Emotional-State"
                word = "anxious"

            # Random fallback
            else:
                category = random.choice(list(receiver_words.keys()))
                word = random.choice(receiver_words[category])

            # Avoid reusing the same category too often
            while category in used_categories:
                category = random.choice(list(receiver_words.keys()))
                word = random.choice(receiver_words[category])

        else:
            category = random.choice(list(receiver_words.keys()))
            word = random.choice(receiver_words[category])

        prompts.append((category, word))
        used_categories.add(category)

        # Simulate interaction stage progression
        history.append(([random.randint(0, 1) for _ in range(3)], [random.randint(0, 1) for _ in range(3)], random.random(), random.random()))

    return prompts
\end{lstlisting}
\end{longlisting}

The first set relies on direct analysis of raw data, such as patrol counts, speeding incidents, and sender/receiver rewards. 
It employs relatively simple conditional checks to adjust prompt categories like Tone, Clarity, and Risk-Preference, with a fallback mechanism that introduces randomized prompts to ensure variety.

\begin{longlisting}
\lstset{language=Python}
\begin{lstlisting}[caption={Another generated conditional prompt function of LAE in the S3 setting.}]
def evaluate_patrol_efficiency(patrols, speeding):
    # Calculate patrol efficiency: more patrols should reduce speeding
    patrol_effect = [1 if p == 1 and s == 0 else 0 for p, s in zip(patrols, speeding)]
    return sum(patrol_effect) / len(patrols)

def evaluate_reward_trend(rewards):
    # Check if rewards are increasing, decreasing, or stable
    if all(rewards[i] <= rewards[i + 1] for i in range(len(rewards) - 1)):
        return "increasing"
    elif all(rewards[i] >= rewards[i + 1] for i in range(len(rewards) - 1)):
        return "decreasing"
    else:
        return "stable"

def evaluate_speeding_pattern(speeding_history):
    # Identify if speeding is random, consistent, or declining
    speeding_totals = [sum(speeds) for speeds in speeding_history]
    if all(speeding_totals[i] == speeding_totals[i + 1] for i in range(len(speeding_totals) - 1)):
        return "consistent"
    elif speeding_totals[-1] < speeding_totals[0]:
        return "declining"
    else:
        return "random"

def evaluate_patrol_distribution(patrols_history):
    # Determine if patrols are evenly distributed across stages
    patrol_totals = [sum(patrol) for patrol in patrols_history]
    if len(set(patrol_totals)) == 1:
        return "even"
    elif patrol_totals[-1] < patrol_totals[0]:
        return "decreasing"
    else:
        return "uneven"

def generate_sender_prompts(history):
    sender_words = {
        "Tone": ["formal", "informal", "neutral", "direct", "conciliatory"],
        "Length": ["short", "concise", "detailed", "lengthy", "brief"],
        "Specificity": ["general", "precise", "vague", "detailed", "broad"],
        "Clarity": ["clear", "ambiguous", "straightforward", "complicated", "obscure"],
        "Style": ["polite", "authoritative", "casual", "professional", "friendly"],
        "Emphasis": ["important", "minor", "critical", "trivial", "central"],
        "Structure": ["linear", "nonlinear", "hierarchical", "sequential", "random"],
        "Complexity": ["simple", "complex", "intricate", "basic", "elaborate"],
        "Consistency": ["consistent", "inconsistent", "variable", "sporadic", "steady"],
        "Informativeness": ["high", "low", "medium", "minimal", "extensive"]
    }

    prompts = []
    used_categories = set()

    for stage in range(5):
        if history:
            patrols, speeding, reward_sender, reward_receiver = zip(*history)

            patrol_efficiency = evaluate_patrol_efficiency(patrols[-1], speeding[-1])
            reward_trend_sender = evaluate_reward_trend(reward_sender)
            reward_trend_receiver = evaluate_reward_trend(reward_receiver)
            speeding_pattern = evaluate_speeding_pattern(speeding)
            patrol_distribution = evaluate_patrol_distribution(patrols)

            # Complex decision-making based on multiple factors
            if patrol_efficiency < 0.5 and speeding_pattern == "random":
                category = "Tone"
                word = "direct"
            elif reward_trend_sender == "decreasing" and patrol_distribution == "uneven":
                category = "Informativeness"
                word = "extensive"
            elif reward_trend_receiver == "increasing" and patrol_efficiency > 0.7:
                category = "Specificity"
                word = "precise"
            elif speeding_pattern == "consistent" and patrol_distribution == "even":
                category = "Clarity"
                word = "clear"
            elif reward_trend_sender == "stable" and patrol_distribution == "decreasing":
                category = "Structure"
                word = "linear"
            else:
                category = random.choice(list(sender_words.keys()))
                word = random.choice(sender_words[category])

            while category in used_categories:
                category = random.choice(list(sender_words.keys()))
                word = random.choice(sender_words[category])

        else:
            category = random.choice(list(sender_words.keys()))
            word = random.choice(sender_words[category])

        prompts.append((category, word))
        used_categories.add(category)
        history.append(([random.randint(0, 1) for _ in range(3)], [random.randint(0, 1) for _ in range(3)], random.random(), random.random()))

    return prompts

def generate_receiver_prompts(history):
    receiver_words = {
        "Risk-Preference": ["cautious", "bold", "balanced", "risk-averse", "reckless"],
        "Attention": ["focused", "distracted", "alert", "inattentive", "engaged"],
        "Decision-Making": ["rational", "impulsive", "deliberate", "hasty", "calculated"],
        "Trust": ["high", "low", "moderate", "skeptical", "confident"],
        "Emotional-State": ["calm", "anxious", "frustrated", "neutral", "excited"],
        "Information-Processing": ["slow", "fast", "thorough", "superficial", "efficient"],
        "Adaptability": ["flexible", "rigid", "adjustable", "stubborn", "open"],
        "Compliance": ["obedient", "defiant", "cooperative", "reluctant", "agreeable"],
        "Responsiveness": ["quick", "slow", "moderate", "delayed", "immediate"],
        "Memory": ["sharp", "forgetful", "average", "short-term", "long-term"]
    }

    prompts = []
    used_categories = set()

    for stage in range(5):
        if history:
            patrols, speeding, reward_sender, reward_receiver = zip(*history)

            patrol_efficiency = evaluate_patrol_efficiency(patrols[-1], speeding[-1])
            reward_trend_receiver = evaluate_reward_trend(reward_receiver)
            speeding_pattern = evaluate_speeding_pattern(speeding)
            patrol_distribution = evaluate_patrol_distribution(patrols)

            if reward_trend_receiver == "increasing" and patrol_efficiency > 0.7:
                category = "Trust"
                word = "high"
            elif speeding_pattern == "consistent" and patrol_distribution == "even":
                category = "Compliance"
                word = "obedient"
            elif reward_trend_receiver == "decreasing" and speeding_pattern == "random":
                category = "Risk-Preference"
                word = "bold"
            elif patrol_distribution == "uneven" and reward_trend_receiver == "stable":
                category = "Adaptability"
                word = "flexible"
            elif patrol_efficiency < 0.5 and speeding_pattern == "random":
                category = "Attention"
                word = "focused"
            else:
                category = random.choice(list(receiver_words.keys()))
                word = random.choice(receiver_words[category])

            while category in used_categories:
                category = random.choice(list(receiver_words.keys()))
                word = random.choice(receiver_words[category])

        else:
            category = random.choice(list(receiver_words.keys()))
            word = random.choice(receiver_words[category])

        prompts.append((category, word))
        used_categories.add(category)
        history.append(([random.randint(0, 1) for _ in range(3)], [random.randint(0, 1) for _ in range(3)], random.random(), random.random()))

    return prompts
\end{lstlisting}
\end{longlisting}

In contrast, the second set introduces custom evaluation functions, such as \texttt{evaluate\_patrol\_efficiency} and \texttt{evaluate\_reward\_trend}, to assess trends in the interaction history. 
This allows for more complex decision-making, where the system not only reacts to immediate conditions but also adapts to evolving patterns in rewards, patrol effectiveness, and speeding behavior. 
As a result, the second set generates more nuanced prompts, making it more flexible and suitable for handling sophisticated, multi-stage interactions.

\end{document}